
\def\unlock{\catcode`@=11} 
\def\lock{\catcode`@=12} 
\unlock
%
%
%
%
%

\font\fourteenrm=cmr10 scaled\magstep2
\font\twelverm=cmr10 scaled\magstep1
\font\ninerm=cmr9          \font\sixrm=cmr6

\font\fourteenbf=cmbx10 scaled\magstep2
\font\twelvebf=cmbx10 scaled\magstep1

\font\seventeeni=cmmi10 scaled\magstep3     \skewchar\seventeeni='177
\font\fourteeni=cmmi10 scaled\magstep2      \skewchar\fourteeni='177
\font\twelvei=cmmi10 scaled\magstep1        \skewchar\twelvei='177
\font\ninei=cmmi9                           \skewchar\ninei='177
\font\sixi=cmmi6                            \skewchar\sixi='177
\font\seventeensy=cmsy10 scaled\magstep3    \skewchar\seventeensy='60
\font\fourteensy=cmsy10 scaled\magstep2     \skewchar\fourteensy='60
\font\twelvesy=cmsy10 scaled\magstep1       \skewchar\twelvesy='60
\font\ninesy=cmsy9                          \skewchar\ninesy='60
\font\sixsy=cmsy6                           \skewchar\sixsy='60

\font\fourteenex=cmex10 scaled\magstep2
\font\twelveex=cmex10 scaled\magstep1

\font\fourteensl=cmsl10 scaled\magstep2
\font\twelvesl=cmsl10 scaled\magstep1

\font\fourteenit=cmti10 scaled\magstep2
\font\twelveit=cmti10 scaled\magstep1
\font\twelvett=cmtt10 scaled\magstep1
\font\twelvecp=cmcsc10 scaled\magstep1
\font\tencp=cmcsc10
\newfam\cpfam
%
%
\newcount\f@ntkey            \f@ntkey=0
\def\samef@nt{\relax \ifcase\f@ntkey \rm \or\oldstyle \or\or
         \or\it \or\sl \or\bf \or\tt \or\caps \fi }
\def\fourteenpoint{\relax
    \textfont0=\fourteenrm          \scriptfont0=\tenrm
    \scriptscriptfont0=\sevenrm
     \def\rm{\fam0 \fourteenrm \f@ntkey=0 }\relax
    \textfont1=\fourteeni           \scriptfont1=\teni
    \scriptscriptfont1=\seveni
     \def\oldstyle{\fam1 \fourteeni\f@ntkey=1 }\relax
    \textfont2=\fourteensy          \scriptfont2=\tensy
    \scriptscriptfont2=\sevensy
    \textfont3=\fourteenex     \scriptfont3=\fourteenex
    \scriptscriptfont3=\fourteenex
    \def\it{\fam\itfam \fourteenit\f@ntkey=4 }\textfont\itfam=\fourteenit
    \def\sl{\fam\slfam \fourteensl\f@ntkey=5 }\textfont\slfam=\fourteensl
    \scriptfont\slfam=\tensl
    \def\bf{\fam\bffam \fourteenbf\f@ntkey=6 }\textfont\bffam=\fourteenbf
    \scriptfont\bffam=\tenbf     \scriptscriptfont\bffam=\sevenbf
    \def\tt{\fam\ttfam \twelvett \f@ntkey=7 }\textfont\ttfam=\twelvett
    \h@big=11.9\p@{} \h@Big=16.1\p@{} \h@bigg=20.3\p@{} \h@Bigg=24.5\p@{}
    \def\caps{\fam\cpfam \twelvecp \f@ntkey=8 }\textfont\cpfam=\twelvecp
    \setbox\strutbox=\hbox{\vrule height 12pt depth 5pt width\z@}
    \samef@nt}
\def\twelvepoint{\relax
    \textfont0=\twelverm          \scriptfont0=\ninerm
    \scriptscriptfont0=\sixrm
     \def\rm{\fam0 \twelverm \f@ntkey=0 }\relax
    \textfont1=\twelvei           \scriptfont1=\ninei
    \scriptscriptfont1=\sixi
     \def\oldstyle{\fam1 \twelvei\f@ntkey=1 }\relax
    \textfont2=\twelvesy          \scriptfont2=\ninesy
    \scriptscriptfont2=\sixsy
    \textfont3=\twelveex          \scriptfont3=\twelveex
    \scriptscriptfont3=\twelveex
    \def\it{\fam\itfam \twelveit \f@ntkey=4 }\textfont\itfam=\twelveit
    \def\sl{\fam\slfam \twelvesl \f@ntkey=5 }\textfont\slfam=\twelvesl
    \scriptfont\slfam=\ninerm
    \def\bf{\fam\bffam \twelvebf \f@ntkey=6 }\textfont\bffam=\twelvebf
    \scriptfont\bffam=\ninerm     \scriptscriptfont\bffam=\sixrm
    \def\tt{\fam\ttfam \twelvett \f@ntkey=7 }\textfont\ttfam=\twelvett
    \h@big=10.2\p@{}
    \h@Big=13.8\p@{}
    \h@bigg=17.4\p@{}
    \h@Bigg=21.0\p@{}
    \def\caps{\fam\cpfam \twelvecp \f@ntkey=8 }\textfont\cpfam=\twelvecp
    \setbox\strutbox=\hbox{\vrule height 10pt depth 4pt width\z@}
    \samef@nt}
\def\tenpoint{\relax
    \textfont0=\tenrm          \scriptfont0=\sevenrm
    \scriptscriptfont0=\fiverm
    \def\rm{\fam0 \tenrm \f@ntkey=0 }\relax
    \textfont1=\teni           \scriptfont1=\seveni
    \scriptscriptfont1=\fivei
    \def\oldstyle{\fam1 \teni \f@ntkey=1 }\relax
    \textfont2=\tensy          \scriptfont2=\sevensy
    \scriptscriptfont2=\fivesy
    \textfont3=\tenex          \scriptfont3=\tenex
    \scriptscriptfont3=\tenex
    \def\it{\fam\itfam \tenit \f@ntkey=4 }\textfont\itfam=\tenit
    \def\sl{\fam\slfam \tensl \f@ntkey=5 }\textfont\slfam=\tensl
    \def\bf{\fam\bffam \tenbf \f@ntkey=6 }\textfont\bffam=\tenbf
    \scriptfont\bffam=\sevenbf     \scriptscriptfont\bffam=\fivebf
    \def\tt{\fam\ttfam \tentt \f@ntkey=7 }\textfont\ttfam=\tentt
    \def\caps{\fam\cpfam \tencp \f@ntkey=8 }\textfont\cpfam=\tencp
    \setbox\strutbox=\hbox{\vrule height 8.5pt depth 3.5pt width\z@}
    \samef@nt}
%
%
%
%
\newdimen\h@big  \h@big=8.5\p@
\newdimen\h@Big  \h@Big=11.5\p@
\newdimen\h@bigg  \h@bigg=14.5\p@
\newdimen\h@Bigg  \h@Bigg=17.5\p@
\def\big#1{{\hbox{$\left#1\vbox to\h@big{}\right.\n@space$}}}
\def\Big#1{{\hbox{$\left#1\vbox to\h@Big{}\right.\n@space$}}}
\def\bigg#1{{\hbox{$\left#1\vbox to\h@bigg{}\right.\n@space$}}}
\def\Bigg#1{{\hbox{$\left#1\vbox to\h@Bigg{}\right.\n@space$}}}
%
%
%
\normalbaselineskip = 20pt plus 0.2pt minus 0.1pt
\normallineskip = 1.5pt plus 0.1pt minus 0.1pt
\normallineskiplimit = 1.5pt
\newskip\normaldisplayskip
\normaldisplayskip = 18pt plus 4pt minus 8pt
\newskip\normaldispshortskip
\normaldispshortskip = 5pt plus 4pt
\newskip\normalparskip
\normalparskip = 6pt plus 2pt minus 1pt
\newskip\skipregister
\skipregister = 5pt plus 2pt minus 1.5pt
\newif\ifsingl@    \newif\ifdoubl@
\newif\iftwelv@    \twelv@true
\def\singlespace{\singl@true\doubl@false\spaces@t}
\def\doublespace{\singl@false\doubl@true\spaces@t}
\def\normalspace{\singl@false\doubl@false\spaces@t}
\def\Tenpoint{\tenpoint\twelv@false\spaces@t}
\def\Twelvepoint{\twelvepoint\twelv@true\spaces@t}
\def\spaces@t{\relax%
 \iftwelv@ \ifsingl@\subspaces@t3:4;\else\subspaces@t1:1;\fi%
 \else \ifsingl@\subspaces@t3:5;\else\subspaces@t4:5;\fi \fi%
 \ifdoubl@ \multiply\baselineskip by 5%
 \divide\baselineskip by 4 \fi \unskip}
\def\subspaces@t#1:#2;{%
      \baselineskip = \normalbaselineskip%
      \multiply\baselineskip by #1 \divide\baselineskip by #2%
      \lineskip = \normallineskip%
      \multiply\lineskip by #1 \divide\lineskip by #2%
      \lineskiplimit = \normallineskiplimit%
      \multiply\lineskiplimit by #1 \divide\lineskiplimit by #2%
      \parskip = \normalparskip%
      \multiply\parskip by #1 \divide\parskip by #2%
      \abovedisplayskip = \normaldisplayskip%
      \multiply\abovedisplayskip by #1 \divide\abovedisplayskip by #2%
      \belowdisplayskip = \abovedisplayskip%
      \abovedisplayshortskip = \normaldispshortskip%
      \multiply\abovedisplayshortskip by #1%
        \divide\abovedisplayshortskip by #2%
      \belowdisplayshortskip = \abovedisplayshortskip%
      \advance\belowdisplayshortskip by \belowdisplayskip%
      \divide\belowdisplayshortskip by 2%
      \smallskipamount = \skipregister%
      \multiply\smallskipamount by #1 \divide\smallskipamount by #2%
      \medskipamount = \smallskipamount \multiply\medskipamount by 2%
      \bigskipamount = \smallskipamount \multiply\bigskipamount by 4 }
\def\normalbaselines{ \baselineskip=\normalbaselineskip%
   \lineskip=\normallineskip \lineskiplimit=\normallineskip%
   \iftwelv@\else \multiply\baselineskip by 4 \divide\baselineskip by 5%
     \multiply\lineskiplimit by 4 \divide\lineskiplimit by 5%
     \multiply\lineskip by 4 \divide\lineskip by 5 \fi }
\Twelvepoint  
\interlinepenalty=50
\interfootnotelinepenalty=5000
\predisplaypenalty=9000
\postdisplaypenalty=500
\hfuzz=1pt
\vfuzz=0.2pt
%
%
%
\def\pagecontents{%
   \ifvoid\topins\else\unvbox\topins\vskip\skip\topins\fi
   \dimen@ = \dp255 \unvbox255
   \ifvoid\footins\else\vskip\skip\footins\footrule\unvbox\footins\fi
   \ifr@ggedbottom \kern-\dimen@ \vfil \fi }
\def\makeheadline{\vbox to 0pt{ \skip@=\topskip
      \advance\skip@ by -12pt \advance\skip@ by -2\normalbaselineskip
      \vskip\skip@ \line{\vbox to 12pt{}\the\headline} \vss
      }\nointerlineskip}
\def\makefootline{\baselineskip = 1.5\normalbaselineskip
                 \line{\the\footline}}
\newif\iffrontpage
\newif\ifletterstyle
\newif\ifp@genum
\def\nopagenumbers{\p@genumfalse}
\def\pagenumbers{\p@genumtrue}
\pagenumbers
\newtoks\paperheadline
\newtoks\letterheadline
\newtoks\letterfrontheadline
\newtoks\lettermainheadline
\newtoks\paperfootline
\newtoks\letterfootline
\newtoks\date
\footline={\ifletterstyle\the\letterfootline\else\the\paperfootline\fi}
\paperfootline={\hss\iffrontpage\else\ifp@genum\tenrm\folio\hss\fi\fi}
\letterfootline={\hfil}
\headline={\ifletterstyle\the\letterheadline\else\the\paperheadline\fi}
\paperheadline={\hfil}
\letterheadline{\iffrontpage\the\letterfrontheadline
     \else\the\lettermainheadline\fi}
\lettermainheadline={\rm\ifp@genum page \ \folio\fi\hfil\the\date}
\def\monthname{\relax\ifcase\month 0/\or January\or February\or
   March\or April\or May\or June\or July\or August\or September\or
   October\or November\or December\else\number\month/\fi}
\date={\monthname\ \number\day, \number\year}
\countdef\pagenumber=1  \pagenumber=1
\def\advancepageno{\global\advance\pageno by 1
   \ifnum\pagenumber<0 \global\advance\pagenumber by -1
    \else\global\advance\pagenumber by 1 \fi \global\frontpagefalse }
\def\folio{\ifnum\pagenumber<0 \romannumeral-\pagenumber
           \else \number\pagenumber \fi }
\def\footrule{\dimen@=\prevdepth\nointerlineskip
   \vbox to 0pt{\vskip -0.25\baselineskip \hrule width 0.35\hsize \vss}
   \prevdepth=\dimen@ }
\newtoks\foottokens
\foottokens={\Tenpoint\singlespace}
\newdimen\footindent
\footindent=24pt
\def\vfootnote#1{\insert\footins\bgroup  \the\foottokens
   \interlinepenalty=\interfootnotelinepenalty \floatingpenalty=20000
   \splittopskip=\ht\strutbox \boxmaxdepth=\dp\strutbox
   \leftskip=\footindent \rightskip=\z@skip
   \parindent=0.5\footindent \parfillskip=0pt plus 1fil
   \spaceskip=\z@skip \xspaceskip=\z@skip
   \Textindent{$ #1 $}\footstrut\futurelet\next\fo@t}
\def\Textindent#1{\noindent\llap{#1\enspace}\ignorespaces}
\def\footnote#1{\attach{#1}\vfootnote{#1}}

\let\footsymbol=\star
\newcount\lastf@@t           \lastf@@t=-1
\newcount\footsymbolcount    \footsymbolcount=0
\newif\ifPhysRev
\def\footsymbolgen{\relax \ifPhysRev \iffrontpage \NPsymbolgen\else
      \PRsymbolgen\fi \else \NPsymbolgen\fi
   \global\lastf@@t=\pageno \footsymbol }
\def\NPsymbolgen{\ifnum\footsymbolcount<0 \global\footsymbolcount=0\fi
   {\iffrontpage \else \advance\lastf@@t by 1 \fi
    \ifnum\lastf@@t<\pageno \global\footsymbolcount=0
     \else \global\advance\footsymbolcount by 1 \fi }
   \ifcase\footsymbolcount \fd@f\star\or \fd@f\dagger\or \fd@f\ddagger\or
    \fd@f\ast\or \fd@f\natural\or \fd@f\diamond\or \fd@f\bullet\or
    \fd@f\nabla\else \fd@f\dagger\global\footsymbolcount=0 \fi }
\def\fd@f#1{\xdef\footsymbol{#1}}
\def\PRsymbolgen{\ifnum\footsymbolcount>0 \global\footsymbolcount=0\fi
      \global\advance\footsymbolcount by -1
      \xdef\footsymbol{\sharp\number-\footsymbolcount} }
\def\space@ver#1{\let\@sf=\empty \ifmmode #1\else \ifhmode
   \edef\@sf{\spacefactor=\the\spacefactor}\unskip${}#1$\relax\fi\fi}
\def\attach#1{\space@ver{\strut^{\mkern 2mu #1} }\@sf\ }
%
%
%
\newcount\chapternumber      \chapternumber=0
\newcount\sectionnumber      \sectionnumber=0
\newcount\equanumber         \equanumber=0
\let\chapterlabel=0
\newtoks\chapterstyle        \chapterstyle={\Number}
\newskip\chapterskip         \chapterskip=\bigskipamount
\newskip\sectionskip         \sectionskip=\medskipamount
\newskip\headskip            \headskip=8pt plus 3pt minus 3pt
\newdimen\chapterminspace    \chapterminspace=15pc
\newdimen\sectionminspace    \sectionminspace=10pc
\newdimen\referenceminspace  \referenceminspace=25pc
\def\chapterreset{\global\advance\chapternumber by 1
   \ifnum\equanumber<0 \else\global\equanumber=0\fi
   \sectionnumber=0 \makel@bel}
\def\makel@bel{\xdef\chapterlabel{%
\the\chapterstyle{\the\chapternumber}.}}
\def\sectionlabel{\number\sectionnumber \quad }
\def\alphabetic#1{\count255='140 \advance\count255 by #1\char\count255}
\def\Alphabetic#1{\count255='100 \advance\count255 by #1\char\count255}
\def\Roman#1{\uppercase\expandafter{\romannumeral #1}}
\def\roman#1{\romannumeral #1}
\def\Number#1{\number #1}
\def\unnumberedchapters{\let\makel@bel=\relax \let\chapterlabel=\relax
\let\sectionlabel=\relax \equanumber=-1 }
\def\titlestyle#1{\par\begingroup \interlinepenalty=9999
     \leftskip=0.02\hsize plus 0.23\hsize minus 0.02\hsize
     \rightskip=\leftskip \parfillskip=0pt
     \hyphenpenalty=9000 \exhyphenpenalty=9000
     \tolerance=9999 \pretolerance=9000
     \spaceskip=0.333em \xspaceskip=0.5em
     \iftwelv@\twelvepoint\fourteenrm\else\twelvepoint\fi
   \noindent #1\par\endgroup }
\def\spacecheck#1{\dimen@=\pagegoal\advance\dimen@ by -\pagetotal
   \ifdim\dimen@<#1 \ifdim\dimen@>0pt \vfil\break \fi\fi}
\def\chapter#1{\par \penalty-300 \vskip\chapterskip
   \spacecheck\chapterminspace
   \chapterreset \titlestyle{\chapterlabel \ #1}
   \nobreak\vskip\headskip \penalty 30000
   \wlog{\string\chapter\ \chapterlabel} }

\def\section#1{\par \ifnum\the\lastpenalty=30000\else
   \penalty-200\vskip\sectionskip \spacecheck\sectionminspace\fi
   \wlog{\string\section\ \chapterlabel \the\sectionnumber}
   \global\advance\sectionnumber by 1  \noindent
   {\caps\enspace\chapterlabel \sectionlabel #1}\par
   \nobreak\vskip\headskip \penalty 30000 }
\def\subsection#1{\par
   \ifnum\the\lastpenalty=30000\else \penalty-100\smallskip \fi
   \noindent\undertext{#1}\enspace \vadjust{\penalty5000}}

\def\undertext#1{\vtop{\hbox{#1}\kern 1pt \hrule}}
\def\APPENDIX#1#2{\par\penalty-300\vskip\chapterskip
   \spacecheck\chapterminspace \chapterreset \xdef\chapterlabel{#1}
   \titlestyle{APPENDIX #2} \nobreak\vskip\headskip \penalty 30000
   \wlog{\string\Appendix\ \chapterlabel} }
\def\Appendix#1{\APPENDIX{#1}{#1}}
\def\appendix{\APPENDIX{A}{}}
%
%
%

\newif\ifdraftmode
\draftmodefalse

\def\eqname#1{\relax \ifnum\equanumber<0
     \xdef#1{{(\number-\equanumber)}}\global\advance\equanumber by -1
    \else \global\advance\equanumber by 1
      \xdef#1{{(\chapterlabel \number\equanumber)}} \fi}
%



%
\def\eqinsert#1{\noalign{\dimen@=\prevdepth \nointerlineskip
   \setbox0=\hbox to\displaywidth{\hfil #1}
   \vbox to 0pt{\vss\hbox{$\!\box0\!$}\kern-0.5\baselineskip}
   \prevdepth=\dimen@}}
%

%

%

%
%
\def\GENITEM#1;#2{\par \hangafter=0 \hangindent=#1
    \Textindent{$ #2 $}\ignorespaces}
\outer\def\newitem#1=#2;{\gdef#1{\GENITEM #2;}}
\newdimen\itemsize                \itemsize=30pt
\newitem\item=1\itemsize;
\newitem\sitem=1.75\itemsize;     
\newitem\ssitem=2.5\itemsize;     
\outer\def\newlist#1=#2&#3&#4;{\toks0={#2}\toks1={#3}%
   \count255=\escapechar \escapechar=-1
   \alloc@0\list\countdef\insc@unt\listcount     \listcount=0
   \edef#1{\par
      \countdef\listcount=\the\allocationnumber
      \advance\listcount by 1
      \hangafter=0 \hangindent=#4
      \Textindent{\the\toks0{\listcount}\the\toks1}}
   \expandafter\expandafter\expandafter
    \edef\c@t#1{begin}{\par
      \countdef\listcount=\the\allocationnumber \listcount=1
      \hangafter=0 \hangindent=#4
      \Textindent{\the\toks0{\listcount}\the\toks1}}
   \expandafter\expandafter\expandafter
    \edef\c@t#1{con}{\par \hangafter=0 \hangindent=#4 \noindent}
   \escapechar=\count255}
\def\c@t#1#2{\csname\string#1#2\endcsname}
\newlist\point=\Number&.&1.0\itemsize;
\newlist\subpoint=(\alphabetic&)&1.75\itemsize;
\newlist\subsubpoint=(\roman&)&2.5\itemsize;
\let\spoint=\subpoint

%
%
%
\newcount\referencecount     \referencecount=0
\newif\ifreferenceopen       \newwrite\referencewrite
\newtoks\rw@toks
\def\NPrefmark#1{\attach{\scriptscriptstyle [ #1 ] }}
\def\APrefmark#1{[#1]}
\let\PRrefmark=\attach
\def\refmark#1{\relax\ifPhysRev\PRrefmark{#1}\else\NPrefmark{#1}\fi}
\def\refend{\refmark{\number\referencecount}}
\newcount\lastrefsbegincount \lastrefsbegincount=0
\def\refsend{\refmark{\count255=\referencecount
   \advance\count255 by-\lastrefsbegincount
   \ifcase\count255 \number\referencecount
   \or \number\lastrefsbegincount,\number\referencecount
   \else \number\lastrefsbegincount-\number\referencecount \fi}}
\def\refch@ck{\chardef\rw@write=\referencewrite
   \ifreferenceopen \else \referenceopentrue
   \immediate\openout\referencewrite=referenc.tex \fi}
%
{\catcode`\^^M=\active 
  \gdef\obeyendofline{\catcode`\^^M\active \let^^M\ }}%
%
{\catcode`\^^M=\active 
  \gdef\ignoreendofline{\catcode`\^^M=5}}
{\obeyendofline\gdef\rw@start#1{\def\t@st{#1} \ifx\t@st\blankend%
\endgroup \@sf \relax \else \ifx\t@st\bl@nkend \endgroup \@sf \relax%
\else \rw@begin#1
\backtotext
\fi \fi } }
{\obeyendofline\gdef\rw@begin#1
{\def\n@xt{#1}\rw@toks={#1}\relax%
\rw@next}}
\def\blankend{}
{\obeylines\gdef\bl@nkend{
}}
\newif\iffirstrefline  \firstreflinetrue
\def\rwr@teswitch{\ifx\n@xt\blankend \let\n@xt=\rw@begin %
 \else\iffirstrefline \global\firstreflinefalse%
\immediate\write\rw@write{\noexpand\obeyendofline \the\rw@toks}%
\let\n@xt=\rw@begin%
      \else\ifx\n@xt\rw@@d \def\n@xt{\immediate\write\rw@write{%
        \noexpand\ignoreendofline}\endgroup \@sf}%
             \else \immediate\write\rw@write{\the\rw@toks}%
             \let\n@xt=\rw@begin\fi\fi \fi}
\def\rw@next{\rwr@teswitch\n@xt}
\def\rw@@d{\backtotext} \let\rw@end=\relax
\let\backtotext=\relax

\newdimen\refindent     \refindent=30pt
\def\refitem#1{\par \hangafter=0 \hangindent=\refindent \Textindent{#1}}
\def\REFNUM#1{\space@ver{}\refch@ck \firstreflinetrue%
 \global\advance\referencecount by 1 \xdef#1{\the\referencecount}}
\def\refnum#1{\space@ver{}\refch@ck \firstreflinetrue%
 \global\advance\referencecount by 1 \xdef#1{\the\referencecount}\refend}

\def\REF#1{\REFNUM#1%
 \immediate\write\referencewrite{%
            \noexpand\refitem{\ifdraftmode{\sevenrm
               \noexpand\string\string#1\ }\fi#1.}}%
      \begingroup\obeyendofline\rw@start}
\def\ref{\refnum\?%
 \immediate\write\referencewrite{\noexpand\refitem{\?.}}%
\begingroup\obeyendofline\rw@start}
\def\Ref#1{\refnum#1%
 \immediate\write\referencewrite{\noexpand\refitem{#1.}}%
\begingroup\obeyendofline\rw@start}
\def\REFS#1{\REFNUM#1\global\lastrefsbegincount=\referencecount
\immediate\write\referencewrite{\noexpand\refitem{#1.}}%
\begingroup\obeyendofline\rw@start}
%

%
%

%
\newcount\figurecount     \figurecount=0
\newif\iffigureopen       \newwrite\figurewrite
\def\figch@ck{\chardef\rw@write=\figurewrite \iffigureopen\else
   \immediate\openout\figurewrite=figures.aux
   \figureopentrue\fi}
\def\FIGNUM#1{\space@ver{}\figch@ck \firstreflinetrue%
 \global\advance\figurecount by 1 \xdef#1{\the\figurecount}}
\def\FIG#1{\FIGNUM#1
   \immediate\write\figurewrite{\noexpand\refitem{#1.}}%
   \begingroup\obeyendofline\rw@start}
\def\fig{\FIGNUM\? fig.~\?%
\immediate\write\figurewrite{\noexpand\refitem{\?.}}%
\begingroup\obeyendofline\rw@start}
\def\figure{\FIGNUM\? figure~\?
   \immediate\write\figurewrite{\noexpand\refitem{\?.}}%
   \begingroup\obeyendofline\rw@start}
\def\Fig{\FIGNUM\? Fig.~\?%
\immediate\write\figurewrite{\noexpand\refitem{\?.}}%
\begingroup\obeyendofline\rw@start}
\def\Figure{\FIGNUM\? Figure~\?%
\immediate\write\figurewrite{\noexpand\refitem{\?.}}%
\begingroup\obeyendofline\rw@start}
\newcount\tablecount     \tablecount=0
\newif\iftableopen       \newwrite\tablewrite
\def\tabch@ck{\chardef\rw@write=\tablewrite \iftableopen\else
   \immediate\openout\tablewrite=tables.aux
   \tableopentrue\fi}
\def\TABNUM#1{\space@ver{}\tabch@ck \firstreflinetrue%
 \global\advance\tablecount by 1 \xdef#1{\the\tablecount}}
\def\TABLE#1{\TABNUM#1
   \immediate\write\tablewrite{\noexpand\refitem{#1.}}%
   \begingroup\obeyendofline\rw@start}
\def\Table{\TABNUM\? Table~\?%
\immediate\write\tablewrite{\noexpand\refitem{\?.}}%
\begingroup\obeyendofline\rw@start}
\newif\ifsymbolopen       \newwrite\symbolwrite
\def\symch@ck{\ifsymbolopen\else
   \immediate\openout\symbolwrite=symbols.aux
   \symbolopentrue\fi}
\def\symdef#1#2{\def#1{#2}%
      \symch@ck%
      \immediate\write\symbolwrite{$$ \hbox{\noexpand\string\string#1}
                \noexpand\qquad\noexpand\longrightarrow\noexpand\qquad
                           \string#1 $$}}
%
%

%
%
%
\def\masterreset{\global\pagenumber=1 \global\chapternumber=0
   \global\equanumber=0 \global\sectionnumber=0
   \global\referencecount=0 \global\figurecount=0 \global\tablecount=0 }
\def\FRONTPAGE{\ifvoid255\else\vfill\penalty-2000\fi
      \masterreset\global\frontpagetrue
      \global\lastf@@t=0 \global\footsymbolcount=0}

\def\paperstyle{\letterstylefalse\normalspace\papersize}
\def\letterstyle{\letterstyletrue\singlespace\lettersize}
\def\papersize{\hsize=35pc\vsize=50pc\hoffset=1pc\voffset=6pc
               \skip\footins=\bigskipamount}
\def\lettersize{\hsize=6.5in\vsize=8.5in\hoffset=0in\voffset=1in
   \skip\footins=\smallskipamount \multiply\skip\footins by 3 }
\paperstyle   
%
%
\def\MEMO{\letterstyle\FRONTPAGE \letterfrontheadline={\hfil}
    \line{\quad\fourteenrm NTC MEMORANDUM\hfil\twelverm\the\date\quad}
    \medskip \memod@f}

\def\memit@m#1{\smallskip \hangafter=0 \hangindent=1in
      \Textindent{\caps #1}}
\def\memod@f{\xdef\to{\memit@m{To:}}\xdef\from{\memit@m{From:}}%
     \xdef\topic{\memit@m{Topic:}}\xdef\subject{\memit@m{Subject:}}%
     \xdef\rule{\bigskip\hrule height 1pt\bigskip}}
\memod@f
\newskip\lettertopfil
\lettertopfil = 0pt plus 1.5in minus 0pt
\newskip\letterbottomfil
\letterbottomfil = 0pt plus 2.3in minus 0pt
\newskip\spskip \setbox0\hbox{\ } \spskip=-1\wd0
\def\addressee#1{\medskip\rightline{\the\date\hskip 30pt} \bigskip
   \vskip\lettertopfil
   \ialign to\hsize{\strut ##\hfil\tabskip 0pt plus \hsize \cr #1\crcr}
   \medskip\noindent\hskip\spskip}
\newskip\signatureskip       \signatureskip=40pt
\def\signed#1{\par \penalty 9000 \bigskip \dt@pfalse
  \everycr={\noalign{\ifdt@p\vskip\signatureskip\global\dt@pfalse\fi}}
  \setbox0=\vbox{\singlespace \halign{\tabskip 0pt \strut ##\hfil\cr
   \noalign{\global\dt@ptrue}#1\crcr}}
  \line{\hskip 0.5\hsize minus 0.5\hsize \box0\hfil} \medskip }

\def\endletter{\ifnum\pagenumber=1 \vskip\letterbottomfil\supereject
\else \vfil\supereject \fi}
\newbox\letterb@x
\def\lettertext{\par\unvcopy\letterb@x\par}
\def\multiletter{\setbox\letterb@x=\vbox\bgroup
      \everypar{\vrule height 1\baselineskip depth 0pt width 0pt }
      \singlespace \topskip=\baselineskip }
\def\letterend{\par\egroup}
%
%
%
\newskip\frontpageskip
\newtoks\pubtype
\newtoks\Pubnum
\newtoks\pubnum
\newif\ifp@bblock  \p@bblocktrue
\def\PH@SR@V{\doubl@true \baselineskip=24.1pt plus 0.2pt minus 0.1pt
             \parskip= 3pt plus 2pt minus 1pt }
\def\PHYSREV{\paperstyle\PhysRevtrue\PH@SR@V}
\def\titlepage{\FRONTPAGE\paperstyle\ifPhysRev\PH@SR@V\fi
   \ifp@bblock\p@bblock\fi}
\def\nopubblock{\p@bblockfalse}

\frontpageskip=1\medskipamount plus .5fil
\pubtype={\tensl Preliminary Version}
\Pubnum={$\caps SLAC - PUB - \the\pubnum $}
\pubnum={0000}
\def\p@bblock{\begingroup \tabskip=\hsize minus \hsize
   \baselineskip=1.5\ht\strutbox \topspace-2\baselineskip
   \halign to\hsize{\strut ##\hfil\tabskip=0pt\crcr
   \the\Pubnum\cr \the\date\cr \the\pubtype\cr}\endgroup}
\def\title#1{\vskip\frontpageskip \titlestyle{#1} \vskip\headskip }
\def\author#1{\vskip\frontpageskip\titlestyle{\twelvecp #1}\nobreak}

\def\address#1{\par\kern 5pt\titlestyle{\twelvepoint\it #1}}
\def\andaddress{\par\kern 5pt \centerline{\sl and} \address}

\def\abstract{\vskip\frontpageskip\centerline{\fourteenrm ABSTRACT}
              \vskip\headskip }

%
%
%
\def\ie{\hbox{\it i.e.}}     
\def\eg{\hbox{\it e.g.}}     

\def\\{\relax\ifmmode\backslash\else$\backslash$\fi}
\def\globaleqnumbers{\relax\if\equanumber<0\else\global\equanumber=-1\fi}

\def\journal#1&#2(#3){\unskip, \sl #1~\bf #2 \rm (19#3) }

\def\topspace{\hrule height 0pt depth 0pt \vskip}

\let\int=\intop         
\def\prop{\mathrel{{\mathchoice{\pr@p\scriptstyle}{\pr@p\scriptstyle}{
                \pr@p\scriptscriptstyle}{\pr@p\scriptscriptstyle} }}}
\def\pr@p#1{\setbox0=\hbox{$\cal #1 \char'103$}
   \hbox{$\cal #1 \char'117$\kern-.4\wd0\box0}}
\def\lsim{\mathrel{\mathpalette\@versim<}}
\def\gsim{\mathrel{\mathpalette\@versim>}}
\def\@versim#1#2{\lower0.5ex\vbox{\baselineskip\z@skip\lineskip-.1ex
  \lineskiplimit\z@\ialign{$\m@th#1\hfil##\hfil$\crcr#2\crcr\sim\crcr}}}
%
%
%
\let\sec@nt=\sec
\def\sec{\relax\ifmmode\let\n@xt=\sec@nt\else\let\n@xt\section\fi\n@xt}
\def\obsolete#1{\message{Macro \string #1 is obsolete.}}
\def\firstsec#1{\obsolete\firstsec \section{#1}}
\def\firstsubsec#1{\obsolete\firstsubsec \subsection{#1}}
\def\thispage#1{\obsolete\thispage \global\pagenumber=#1\frontpagefalse}
\def\thischapter#1{\obsolete\thischapter \global\chapternumber=#1}
\def\nextequation#1{\obsolete\nextequation \global\equanumber=#1
   \ifnum\the\equanumber>0 \global\advance\equanumber by 1 \fi}
\def\BOXITEM{\afterassigment\B@XITEM\setbox0=}
\def\B@XITEM{\par\hangindent\wd0 \noindent\box0 }
%

%
%
%
%
%
\lock
\message{   }
%
%
%












\newbox\figbox
\newdimen\zero  \zero=0pt
\newdimen\figmove
\newdimen\figwidth
\newdimen\figheight
\newdimen\textwidth
\newtoks\figtoks
\newcount\figcounta
\newcount\figcountb
\newcount\figlines
\def\figreset{\global\figcounta=-1 \global\figcountb=-1
\global\figmove=\baselineskip
\global\figlines=1 \global\figtoks={ } }
\def\picture#1by#2:#3{\global\setbox\figbox=\vbox{\vskip #1
\hbox{\vbox{\hsize=#2 \noindent #3}}}
\global\setbox\figbox=\vbox{\kern 10pt
\hbox{\kern 10pt \box\figbox \kern 10pt }\kern 10pt}
\global\figwidth=1\wd\figbox
\global\figheight=1\ht\figbox
\global\textwidth=\hsize
\global\advance\textwidth by - \figwidth }
\def\figtoksappend{\edef\temp##1{\global\figtoks=%
{\the\figtoks ##1}}\temp}
\def\figparmsa#1{\loop \global\advance\figcounta by 1
\ifnum \figcounta < #1
\figtoksappend{ 0pt \the\hsize }
\global\advance\figlines by 1
\repeat }
\def\figparmsb#1{\loop \global\advance\figcountb by 1
\ifnum \figcountb < #1
\figtoksappend{ \the\figwidth \the\textwidth}
\global\advance\figlines by 1
\repeat }
\def\figtext#1:#2:#3{\figreset%
\figparmsa{#1}%
\figparmsb{#2}%
\multiply\figmove by #1%
\global\setbox\figbox=\vbox to 0pt{\vskip \figmove  \hbox{\box\figbox}
\vss }
\parshape=\the\figlines\the\figtoks\the\zero\the\hsize
\noindent
\rlap{\box\figbox} #3}
\def\Buildrel#1\under#2{\mathrel{\mathop{#2}\limits_{#1}}}
\def\llongrarrow{\hbox to 40pt{\rightarrowfill}}



\def\boxit#1{\vbox{\hrule\hbox{\vrule\kern3pt
\vbox{\kern3pt#1\kern3pt}\kern3pt\vrule}\hrule}}
\newdimen\str
\def\fboxit#1#2{\vbox{\hrule height #1 \hbox{\vrule width #1
\kern3pt \vbox{\kern3pt#2\kern3pt}\kern3pt \vrule width #1 }
\hrule height #1 }}
\def\tran#1#2{\transpoint \hfuzz 5pt \gdef\label{#1}
\vbox to \the\vsize{\hsize \the\hsize #2} \par \eject }
\newdimen\baseskip
\newdimen\lskip
\lskip=\lineskip
\newdimen\transize
\newdimen\tall
\def\transpoint{\gdef\rm{\fam0\eighteenrm}
\font\twentyfourit = amti10 scaled \magstep5
\font\twentyfourrm = cmr10 scaled \magstep5
\font\twentyfourbf = ambx10 scaled \magstep5
\font\twentyeightsy = amsy10 scaled \magstep5
\font\eighteenrm = cmr10 scaled \magstep3
\font\eighteenb = ambx10 scaled \magstep3
\font\eighteeni = ammi10 scaled \magstep3
\font\eighteenit = amti10 scaled \magstep3
\font\eighteensl = amsl10 scaled \magstep3
\font\eighteensy = amsy10 scaled \magstep3
\font\eighteencaps = cmr10 scaled \magstep3
\font\eighteenmathex = amex10 scaled \magstep3
\font\fourteenrm=cmr10 scaled \magstep2
\font\fourteeni=ammi10 scaled \magstep2
\font\fourteenit = amti10 scaled \magstep2
\font\fourteensy=amsy10 scaled \magstep2
\font\fourteenmathex = amex10 scaled \magstep2
\parindent 20pt
\global\hsize = 7.0in
\global\vsize = 8.9in
\dimen\transize = \the\hsize
\dimen\tall = \the\vsize
\def\sy{\eighteensy }
\def\sl{\eighteens }
\def\bf{\eighteenb }
\def\it{\eighteenit }
\def\caps{\eighteencaps }
\textfont0=\eighteenrm \scriptfont0=\fourteenrm
\scriptscriptfont0=\twelverm
\textfont1=\eighteeni \scriptfont1=\fourteeni \scriptscriptfont1=\twelvei
\textfont2=\eighteensy \scriptfont2=\fourteensy
\scriptscriptfont2=\twelvesy
\textfont3=\eighteenmathex \scriptfont3=\eighteenmathex
\scriptscriptfont3=\eighteenmathex
\global\baselineskip 35pt
\global\lineskip 15pt
\global\parskip 5pt  plus 1pt minus 1pt
\global\abovedisplayskip  3pt plus 10pt minus 10pt
\global\belowdisplayskip 3pt plus 10pt minus 10pt
\def\rtitle##1{\centerline{\undertext{\twentyfourrm ##1}}}
\def\ititle##1{\centerline{\undertext{\twentyfourit ##1}}}
\def\ctitle##1{\centerline{\undertext{\caps ##1}}}
\def\vstrut{\hbox{\vrule width 0pt height .35in depth .15in }}
\def\cline##1{\centerline{\vstrut ##1}}
\output{\shipout\vbox{\vskip .5in
\pagecontents \vfill
\hbox to \the\hsize{\hfill{\tenbf \label} } }
\global\advance\count0 by 1 }
\rm }


%
%
%

%

%

%

%
{\obeyspaces\global\let =\ }
%
%
%
\widowpenalty 1000
\thickmuskip 4mu plus 4mu
\unlock
%
\Pubnum={${\twelverm IU/NTC}\  \the\pubnum $}
\pubnum={0000}
\def\p@nnlock{\begingroup \tabskip=\hsize minus \hsize
   \baselineskip=1.5\ht\strutbox \topspace-2\baselineskip
   \noindent\strut\the\Pubnum \hfill \the\date   \endgroup}
\def\titlepage{\FRONTPAGE\paperstyle\p@nnlock}
\def\displaylines#1{\displ@y
  \halign{\hbox to\displaywidth{$\hfil\displaystyle##\hfil$}\crcr
    #1\crcr}}
\def\addressee#1{\null
   \bigskip\medskip\rightline{\the\date\hskip 30pt}
   \vskip\lettertopfil
   \ialign to\hsize{\strut ##\hfil\tabskip 0pt plus \hsize \cr #1\crcr}
   \medskip\vskip 3pt\noindent}
\def\tmsaddressee#1#2{
   \vskip\lettertopfil
  \setbox0=\vbox{\singlespace \halign{\tabskip 0pt \strut ##\hfil\cr
   \noalign{\global\dt@ptrue}#1\crcr}}
  \line{\hskip 0.7\hsize minus 0.7\hsize \box0\hfil}
   \bigskip
   \vskip .2in
   \ialign to\hsize{\strut ##\hfil\tabskip 0pt plus \hsize \cr #2\crcr}
   \medskip\vskip 3pt\noindent}
\def\makeheadline{\vbox to 0pt{ \skip@=\topskip
      \advance\skip@ by -12pt \advance\skip@ by -2\normalbaselineskip
      \vskip\skip@  \vss
      }\nointerlineskip}
\def\signed#1{\par \penalty 9000 \bigskip \vskip .06in\dt@pfalse
  \everycr={\noalign{\ifdt@p\vskip\signatureskip\global\dt@pfalse\fi}}
  \setbox0=\vbox{\singlespace \halign{\tabskip 0pt \strut ##\hfil\cr
   \noalign{\global\dt@ptrue}#1\crcr}}
  \line{\hskip 0.5\hsize minus 0.5\hsize \box0\hfil} \medskip }
\def\lettersize{\hsize=6.25in\vsize=8.5in\hoffset=0in\voffset=1in
   \skip\footins=\smallskipamount \multiply\skip\footins by 3 }
%
%
%
%
%
\outer\def\newnewlist#1=#2&#3&#4&#5;{\toks0={#2}\toks1={#3}%
   \dimen1=\hsize  \advance\dimen1 by -#4
   \dimen2=\hsize  \advance\dimen2 by -#5
   \count255=\escapechar \escapechar=-1
   \alloc@0\list\countdef\insc@unt\listcount     \listcount=0
   \edef#1{\par
      \countdef\listcount=\the\allocationnumber
      \advance\listcount by 1
      \parshape=2 #4 \dimen1 #5 \dimen2
      \Textindent{\the\toks0{\listcount}\the\toks1}}
   \expandafter\expandafter\expandafter
    \edef\c@t#1{begin}{\par
      \countdef\listcount=\the\allocationnumber \listcount=1
      \parshape=2 #4 \dimen1 #5 \dimen2
      \Textindent{\the\toks0{\listcount}\the\toks1}}
   \expandafter\expandafter\expandafter
    \edef\c@t#1{con}{\par \parshape=2 #4 \dimen1 #5 \dimen2 \noindent}
   \escapechar=\count255}
\def\c@t#1#2{\csname\string#1#2\endcsname}
%
%
%
%
%
%
%
\def\noparGENITEM#1;{\hangafter=0 \hangindent=#1
    \ignorespaces\noindent}
\outer\def\noparnewitem#1=#2;{\gdef#1{\noparGENITEM #2;}}
\noparnewitem\spoint=1.5\itemsize;
%
%
%
\def\MEMO{\letterstyle\FRONTPAGE \letterfrontheadline={\hfil}
      \hoffset=1in \voffset=1.21in
    \line{\hskip .8in  \special{overlay ntcmemo.dat}
          \quad\fourteenrm NTC MEMORANDUM\hfil\twelverm\the\date\quad}
    \medskip\medskip \memod@f}

\def\memit@m#1{\smallskip \hangafter=0 \hangindent=1in
      \Textindent{\caps #1}}
\def\memod@f{\xdef\to{\memit@m{To:}}\xdef\from{\memit@m{From:}}%
     \xdef\topic{\memit@m{Topic:}}\xdef\subject{\memit@m{Subject:}}%
     \xdef\rule{\bigskip\hrule height 1pt\bigskip}}
\memod@f
\lock
%
\def\APrefmark#1{[#1]}

\singlespace
\newif\ifNuclPhys
\newif\ifAnnPhys
\PhysRevfalse
\NuclPhysfalse
\AnnPhystrue
\def\IUCFREF#1|#2|#3|#4|#5|#6|{\relax
    \ifPhysRev\REF#1{{\frenchspacing #2, #3 {\bf #4}, #6 (#5).}} \fi
    \ifNuclPhys\REF#1{{\frenchspacing #2, #3 {\bf #4} (#5) #6}}\fi
    \ifAnnPhys\REF#1{{\frenchspacing #2, {\it #3} {\bf #4} (#5), #6.}}\fi}
\def\IUCFBOOK#1|#2|#3|#4|{\relax
    \ifPhysRev\REF#1{{\frenchspacing #2, {\it #3} (#4).}}\fi
    \ifNuclPhys\REF#1{{\frenchspacing #2, {\sl #3} (#4)}} \fi
    \ifAnnPhys\REF#1{{\frenchspacing #2, ``#3'', #4.}} \fi}

\IUCFREF\rSTUECK|E. C. G. Stueckelberg and A. Peterman|Helv. Phys. Acta|
26|1953|499|
\IUCFREF\rGELL|M. Gell-Mann and F. E. Low|Phys. Rev.|95|1954|1300|
\IUCFBOOK\rBOGOL|N. N. Bogoliubov and D.V. Shirkov|Introduction to the
Theory of Quantized Fields|Interscience, New York, 1959|
\IUCFREF\rWILONE|K. G. Wilson|Phys. Rev.|140|1965|B445|
\IUCFREF\rWILTWO|K. G. Wilson|Phys. Rev.|D2|1970|1438|
\IUCFREF\rWILTHR|K. G. Wilson|Phys. Rev.|D3|1971|1818|
\IUCFREF\rWILFOUR|K. G. Wilson|Phys. Rev.|D6|1972|419|
\IUCFREF\rWILFIVE|K. G. Wilson|Phys. Rev.|B4|1971|3174|
\IUCFREF\rWILSIX|K. G. Wilson|Phys. Rev.|B4|1971|3184|
\IUCFREF\rWILSEVEN|K. G. Wilson and M. E. Fisher|Phys. Rev.
Lett.|28|1972|240|
\IUCFREF\rWILEIGHT|K. G. Wilson|Phys. Rev. Lett.|28|1972|548|
\IUCFREF\rWILNINE|K. G. Wilson and J. B. Kogut|Phys. Rep.|12C|1974|75|
\IUCFREF\rWILTEN|K. G. Wilson|Rev. Mod. Phys.|47|1975|773|
\IUCFREF\rWILELEVEN|K. G. Wilson|Rev. Mod. Phys.|55|1983|583|
\IUCFREF\rWILTWELVE|K. G. Wilson|Adv. Math.|16|1975|170|
\IUCFREF\rWILTHIRT|K. G. Wilson|Scientific American|241|1979|158|
\IUCFREF\rWEGONE|F. J. Wegner|Phys. Rev.|B5|1972|4529|
\IUCFREF\rWEGTWO|F. J. Wegner|Phys. Rev.|B6|1972|1891|
\REF\rWEGTHR{F. J. Wegner, {\it in} ``Phase Transitions and Critical
Phenomena''  (C. Domb and M. S. Green, Eds.), Vol. 6,
Academic Press, London, 1976.}
\IUCFREF\rKADANOFF|L. P. Kadanoff|Physica|2|1965|263|
\IUCFBOOK\rREBBI|C. Rebbi|Lattice Gauge Theories and Monte Carlo
Simulations|World Scientific, Singapore, 1983|
\IUCFREF\rCALLAN|C. G. Callan|Phys. Rev.|D2|1970|1541|
\IUCFREF\rSYMONE|K. Symanzik|Comm. Math. Phys.|18|1970|227|
\IUCFREF\rFEYN|R. P. Feynman|Rev. Mod. Phys.|20|1948|367|
\IUCFBOOK\rNEWTON|I. Newton|Philosophiae Naturalis Principia
Mathematica|S. Pepys, London, 1686|
\IUCFREF\rMAONE|S. K.  Ma|Rev. Mod. Phys.|45|1973|589|
\IUCFBOOK\rPFEUTY|G. Toulouse and P. Pfeuty|Introduction to the
Renormalization Group and to Critical Phenomena|Wiley, Chichester,
1977|
\IUCFBOOK\rMATWO|S. K. Ma|Modern Theory of Critical Phenomena|Benjamin,
New York, 1976|
\IUCFBOOK\rAMIT|D. Amit|Field Theory, the Renormalization Group, and
Critical Phenomena|Mc-Graw-Hill, New York, 1978|
\IUCFBOOK\rZINN|J. Zinn-Justin|Quantum Field Theory and Critical
Phenomena|Oxford, Oxford, 1989|
\IUCFBOOK\rGOLDEN|N. Goldenfeld|Lectures on Phase Transitions and the
Renormalization Group|Add-ison-Wesley, Reading Mass., 1992|
\IUCFREF\rDIRONE|P. A. M. Dirac|Rev. Mod. Phys.|21|1949|392|
\IUCFBOOK\rDIRTWO|P. A. M. Dirac|Lectures on Quantum Field Theory|
Academic Press, New York, 1966|
\REF\rLFREF{An extensive list of references on light-front physics
({\it light.tex}) is available via anonymous ftp from
public.mps.ohio-state.edu in the subdirectory pub/infolight.}
\IUCFREF\rWEI|S. Weinberg|Phys. Rev.|150|1966|1313|
\IUCFREF\rHVONE|A. Harindranath and J. P. Vary|Phys. Rev.|D36|1987|1141|
\IUCFREF\rHILLER|J. R. Hiller|Phys. Rev.|D44|1991|2504|
\IUCFREF\rSWENSON|J. B. Swenson and J. R. Hiller|Phys.
Rev.|D48|1993|1774|
\IUCFREF\rPERWIL|R. J. Perry and K. G. Wilson|Nucl. Phys.|B403|1993|587|
\IUCFREF\rFUB|S. Fubini and G. Furlan|Physics|1|1964|229|
\IUCFREF\rDASH|R. Dashen and M. Gell-Mann|Phys. Rev. Lett|
17|1966|340|
\IUCFREF\rBJORK|J. D. Bjorken|Phys. Rev.|179|1969|1547|
\IUCFBOOK\rFEY|R. P. Feynman|Photon-Hadron Interactions|Benjamin,
Reading, Massachusetts, 1972|
\IUCFREF\rKOGTWO|J. B. Kogut and L. Susskind|Phys. Rep.|C8|1973|75|
\IUCFREF\rCHANG|S.-J. Chang and S.-K. Ma|Phys. Rev.|180|1969|1506|
\IUCFREF\rKOGONE|J. B. Kogut and D. E. Soper|Phys. Rev.|D1|1970|2901|
\IUCFREF\rBJORKTWO|J. D. Bjorken, J. B. Kogut, and D. E. Soper|Phys.
Rev.|D3|1971|1382|
\IUCFREF\rCHAONE|S.-J. Chang, R. G. Root and T.-M. Yan|Phys. Rev.
|D7|1973|1133|
\IUCFREF\rCHATWO|S.-J. Chang and T.-M. Yan|Phys. Rev.|D7|1973|1147|
\IUCFREF\rYANONE|T.-M. Yan|Phys. Rev.|D7|1973|1760|
\IUCFREF\rYANTWO|T.-M. Yan|Phys. Rev.|D7|1973|1780|
\IUCFREF\rBRS|S. J. Brodsky, R. Roskies and R. Suaya|Phys.
Rev.|D8|1973|4574|
\IUCFREF\rBOUCH|C. Bouchiat, P. Fayet, and N. Sourlas|Lett. Nuovo Cim.|
4|1972|9|
\IUCFREF\rHARIONE|A. Harindranath and R. J. Perry|Phys.
Rev.|D43|1991|492|
\IUCFREF\rMUS|D. Mustaki, S. Pinsky, J. Shigemitsu, and K. Wilson|Phys.
Rev.|D43|1991|3411|
\IUCFREF\rBURKONE|M. Burkhardt and A. Langnau|Phys. Rev.|D44|1991|1187|
\IUCFREF\rBURKTWO|M. Burkhardt and A. Langnau|Phys. Rev.|D44|1991|3857|
\IUCFREF\rROBERT|D. G. Robertson and G. McCartor|Z. Phys.|C53|1992|661|
\IUCFREF\rPERQCD|R. J. Perry|Phys. Lett.|B300|1993|8|
\IUCFREF\rTAM|I. Tamm|J. Phys. (USSR)|9|1945|449|
\IUCFREF\rDAN|S. M. Dancoff|Phys. Rev.|78|1950|382|
\IUCFBOOK\rBET|H. A. Bethe and F. de Hoffmann|Mesons and Fields, Vol. II|
Row, Peterson and Company, Evanston, Illinois, 1955|
\IUCFREF\rPERONE|R. J. Perry, A. Harindranath and K. G. Wilson|
Phys. Rev. Lett.|65|1990|2959|
\REF\rTAN{A. C. Tang, Ph.D thesis, Stanford University, SLAC-Report-351,
June (1990).}
\IUCFREF\rPERTWO|R. J. Perry and A. Harindranath|Phys.
Rev.|D43|1991|4051|
\IUCFREF\rTANG|A. C. Tang, S. J. Brodsky, and H. C. Pauli|Phys.
Rev.|D44|1991|1842|
\IUCFREF\rKALUZA|M. Kaluza and H. C. Pauli|Phys. Rev.|D45|1992|2968|
\IUCFREF\rGLAZTHR|St. D. G{\l}azek and R.J. Perry|Phys.
Rev.|D45|1992|3740|
\IUCFREF\rHARITWO|A. Harindranath, R. J. Perry, and J. Shigemitsu|Phys.
Rev.|D46|1992|4580|
\IUCFREF\rWORT|P. M. Wort|Phys. Rev.|D47|1993|608|
\IUCFREF\rGHPSW|S. G{\l}azek, A. Harindranath, S. Pinsky, J. Shigemitsu,
and K. G. Wilson|Phys. Rev.|D47|1993|1599|
\IUCFREF\rLIU|H. H. Liu and D. E. Soper|Phys. Rev.|D48|1993|1841|
\IUCFREF\rGLAZEK|St. D. G{\l}azek and R.J. Perry|Phys.
Rev.|D45|1992|3734|
\IUCFREF\rSANDE|B. van de Sande and S. S. Pinsky|Phys.
Rev.|D46|1992|5479|
\REF\rGLAZWIL{S. D. G{\l}azek and K. G. Wilson, ``Renormalization of
Overlapping Transverse Divergences in a Model Light-Front Hamiltonian,''
Ohio State preprint (1992).}
\IUCFREF\rBERG|H. Bergknoff|Nucl. Phys.|B122|1977|215|
\IUCFREF\rELLER|T. Eller, H. C. Pauli, and S. J. Brodsky|Phys.
Rev.|D35|1987|1493|
\IUCFREF\rMA|Y. Ma and J. R. Hiller|J. Comp. Phys.|82|1989|229|
\IUCFREF\rBURK|M. Burkhardt|Nucl. Phys.|A504|1989|762|
\IUCFREF\rHORN|K. Hornbostel, S. J. Brodsky, and H. C. Pauli|Phys.
Rev.|D41|1990|3814|
\IUCFREF\rMCCART|G. McCartor|Zeit. Phys.|C52|1991|611|
\REF\rMO{Y. Mo and R. J. Perry, ``Basis Function Calculations for the
Massive Schwinger Model in the Light-Front Tamm-Dancoff Approximation,''
to appear in J. Comp. Phys. (1993).}
\REF\rLEPAGE{G. P. Lepage, S. J. Brodsky, T. Huang and P. B. Mackenzie,
{\it in} ``Particles and Fields 2'' (A. Z. Capri and A. N. Kamal,
Eds.), Plenum Press, New York, 1983.}
\IUCFREF\rNAM|J. M. Namyslowski|Prog. Part. Nuc. Phys.|74|1984|1|
\REF\rBROONE{S. J. Brodsky and G. P. Lepage, {\it in} ``Perturbative
quantum
chromodynamics'' (A. H. Mueller, Ed.), World Scientific,
Singapore, 1989.}
\REF\rBPMP{S. Brodsky, H. C. Pauli, G. McCartor, and S. Pinsky, ``The
Challenge of Light-Cone Quantization of Gauge Field Theory,''
SLAC preprint no. SLAC-PUB-5811 and Ohio State preprint no.
OHSTPY-HEP-T-92-005 (1992).}
\IUCFREF\rPAULI|W. Pauli and F. Villars|Rev. Mod. Phys.|21|1949|434|
\IUCFREF\rTHOONE|G. 't Hooft and M. Veltman|Nucl. Phys.|B44|1972|189|
\REF\rTHOTWO{G. 't Hooft and M. Veltman, ``Diagrammar,'' CERN preprint
73-9 (1973).}
\IUCFREF\rBLOONE|C. Bloch and J. Horowitz|Nucl. Phys.|8|1958|91|
\IUCFREF\rBLOTWO|C. Bloch|Nucl. Phys.|6|1958|329|
\IUCFREF\rBRANDOW|B. H. Brandow|Rev. Mod. Phys.|39|1967|771|
\IUCFREF\rLEU|H. Leutwyler and J. Stern|Ann. Phys. (New
York)|112|1978|94|
\IUCFREF\rOEHONE|R. Oehme, K. Sibold, and W. Zimmerman|
Phys. Lett.|B147|1984|115|
\IUCFREF\rOEHTWO|R. Oehme and W. Zimmerman|Commun. Math. Phys.|97|1985|569|
\IUCFREF\rZIMONE|W. Zimmerman|Commun. Math. Phys.|95|1985|211|
\IUCFREF\rKUBO|J. Kubo, K. Sibold, and W. Zimmerman|Nuc.
Phys.|B259|1985|331|
\IUCFREF\rOEHTHR|R. Oehme|Prog. Theor. Phys. Supp.|86|1986|215|
\IUCFREF\rLUCC|C. Lucchesi, O. Piguet, and K. Sibold|Phys.
Lett.|B201|1988|241|
\IUCFREF\rKRAUS|E. Kraus|Nucl. Phys.|B349|1991|563|
\IUCFREF\rCASH|A. Casher|Phys. Rev.|D14|1976|452|
\IUCFREF\rBARONE|W.A. Bardeen and R.B. Pearson|Phys. Rev.|D14|1976|547|
\IUCFREF\rBARTWO|W.A. Bardeen, R.B. Pearson and E. Rabinovici|Phys.
Rev.|D21|1980|1037|
\IUCFREF\rLEP|G.P. Lepage and S.J. Brodsky|Phys. Rev.|D22|1980|2157|
\REF\rDIRTHR{See P. A. M. Dirac,
in `Perturbative Quantum Chromodynamics'' (D. W. Duke and J. F. Owens,
Eds.), Am. Inst. Phys., New York, 1981.}
\REF\rWILQCD{K. G. Wilson, ``Light Front QCD,'' Ohio State internal
report, unpublished (1990).}
\IUCFREF\rTOM|E. Tomboulis|Phys. Rev.|D8|1973|2736|
\IUCFREF\rGRO|D. J. Gross and F. Wilczek|Phys. Rev.
Lett.|30|1973|1343|
\IUCFREF\rPOL|H. D. Politzer|Phys. Rev. Lett.|30|1973|1346|
\IUCFREF\rGOLD|J. Goldstone|Proc. Roy. Soc. (London)|A239|1957|267|
\IUCFREF\rGRIFFIN|P. A. Griffin|Nucl. Phys.|B372|1992|270|
\IUCFBOOK\rSCH|J. Schwinger|Quantum Electrodynamics|Dover, New York,
1958|
\IUCFBOOK\rBROWN|L. M. Brown|Renormalization|Springer-Verlag, New York,
1993|
\IUCFREF\rWILFORT|K. G. Wilson|Phys. Rev.|D10|1974|2445|
\IUCFBOOK\rQUARK|W. Buchm{\"u}ller|Quarkonia|North Holland, Amsterdam,
1992|
\IUCFREF\rTHORN| C. B. Thorn|Phys. Rev.|D20|1979|1934|
\IUCFREF\rGLAZFOUR|St. G{\l}azek|Phys. Rev.|D38|1988|3277|
\IUCFBOOK\rKOKK|J. J. J. Kokkedee|The Quark Model|Benjamin, New York,
1969|
%
%
%

\PhysRevfalse
\NuclPhysfalse
\AnnPhystrue

\def\calp{{\mit P}}
\def\calN{{\cal N}}
\def\ham{{\cal H}}
\def\order{{\cal O}}
\def\dkk{{dk^+ d^2k^\perp \over 16 \pi^3 k^+}\;}

\def\dqq{{dq^+ d^2q^\perp \over 16 \pi^3 q^+}\;}
\def\dsx{{d^2s^\perp dx \over 16 \pi^3 x(1-x)}\;}
\def\dq{{d^2q^\perp dx \over 16 \pi^3 x}\;}
\def\dr{{d^2r^\perp dy \over 16 \pi^3 y}\;}
\def\dss{{d^2s^\perp dz \over 16 \pi^3 z}\;}
\def\dqp{{d^2{q^\perp}' dx' \over 16 \pi^3 x'}\;}
\def\drp{{d^2{r^\perp}' dy' \over 16 \pi^3 y'}\;}
\def\dssp{{d^2{s^\perp}' dz' \over 16 \pi^3 z'}\;}
\def\ds{{d^2s^\perp dx \over 16 \pi^3 }\;}

\def\dkt{d\tilde{k}}
\def\dqt{d\tilde{q}}
\def\qp{{\bf q^\perp}}
\def\kp{{\bf k^\perp}}
\def\vp{{\bf v^\perp}}
\def\pp{{\bf p^\perp}}
\def\rp{{\bf r^\perp}}
\def\sp{{\bf s^\perp}}
\def\tp{{\bf t^\perp}}
\def\br{{\bf r}}
\def\bs{{\bf s}}
\def\bt{{\bf t}}
\def\qproj{{\cal Q}}
\def\pproj{{\cal P}}
\def\Q{{\cal Q}}
\def\P{{\cal P}}
\def\wave{|\Psi\rangle}
\def\wwave{|\Phi\rangle}
\def\awave{|a\rangle}
\def\bwave{|b\rangle}
\def\cwave{|c\rangle}
\def\iwave{|i\rangle}
\def\jwave{|j\rangle}

\def\lawave{\langle a|}
\def\lbwave{\langle b|}
\def\lcwave{\langle c|}
\def\liwave{\langle i|}
\def\ljwave{\langle j|}

\def\rb{{\cal R}}
\def\Hpp{H_{\P\P}}
\def\Hqq{H_{\Q\Q}}
\def\Hpq{H_{\P\Q}}
\def\Hqp{H_{\Q\P}}
\def\hpp{h_{\P\P}}
\def\hqq{h_{\Q\Q}}

\def\vpp{v_{\P\P}}
\def\vqq{v_{\Q\Q}}
\def\vpq{v_{\P\Q}}
\def\vqp{v_{\Q\P}}
\def\rnone{{\uppercase\expandafter{\romannumeral1 }}}
\def\rntwo{{\uppercase\expandafter{\romannumeral2 }}}
\def\rnthree{{\uppercase\expandafter{\romannumeral3 }}}
\def\rnfour{{\uppercase\expandafter{\romannumeral4 }}}
\def\rnfive{{\uppercase\expandafter{\romannumeral5 }}}
\def\rnsix{{\uppercase\expandafter{\romannumeral6 }}}
\def\rnseven{{\uppercase\expandafter{\romannumeral7 }}}
\def\rneight{{\uppercase\expandafter{\romannumeral8 }}}
\def\sqa{\sqrt{1-4 m^2/\Lambda_0^2}}
\def\sqb{\sqrt{1-4 m^2/\Lambda_1^2}}
\def\sqc{\sqrt{1-4 m^2/E}}
\def\sqd{\sqrt{1-4 y(1-y)}}
\def\tg{{\mathaccent "7E g}}
\def\eg{{\it e.g.}}
\def\ie{{\it i.e.}}
\def\apost{{\it a posteriori}}
\def\adhoc{{\it ad hoc~}}
\def\abinit{{\it ab initio~}}

\vsize=8.25truein
\hsize=6.5truein
\hoffset=.25truein
\voffset=.25truein
\hfill
OSU-NT-93-117
\bigskip
\bigskip
\bigskip
\centerline{\bf A renormalization group approach to}
\medskip
\centerline{\bf Hamiltonian light-front field theory}
\bigskip
\centerline{Robert J. Perry}
\centerline{Department of Physics}
\centerline{The Ohio State University}
\centerline{Columbus, OH 43210}
\bigskip
\bigskip

\centerline{ABSTRACT}
\medskip

A perturbative renormalization group is formulated for the study of
Hamiltonian light-front field theory near a critical
Gaussian fixed point.
The only light-front renormalization group transformations found here
that can be approximated by dropping irrelevant operators and using
perturbation theory near
Gaussian fixed volumes, employ invariant-mass cutoffs.  These cutoffs
violate covariance and cluster decomposition, and allow
functions of longitudinal momenta to appear in all relevant, marginal,
and irrelevant operators.  These functions can be determined by insisting
that the Hamiltonian display a coupling constant coherence,
with the number of
couplings that explicitly run with the cutoff scale being limited and
all other couplings depending on this scale only through their
dependence on the running couplings.
Examples are given that
show how coupling coherence restores Lorentz
covariance and cluster decomposition, as recently speculated
by Wilson and the author.
The ultimate goal of this work is
a practical Lorentz metric version of the renormalization
group, and practical
renormalization techniques for light-front quantum chromodynamics.

\bigskip
\bigskip

\centerline{April, 1993}

\bigskip
\vfill
\noindent Accepted for publication in Annals of Physics.
\eject


\bigskip
\noindent {\bf \rnone. Introduction}
\medskip

In a series of remarkable papers
Wilson reformulated the original renormalization group approach to
relativistic field theory \APrefmark{\rSTUECK-\rBOGOL}, initially developing
the modern renormalization group as
a tool for the study of the strong interaction in Minkowski
space \APrefmark{\rWILONE-\rWILTHR}.
He was later diverted
to the study of Euclidean field theory \APrefmark{\rWILFOUR} and
statistical field theory \APrefmark{\rWILFIVE-\rWILEIGHT},
where it was possible to implement perturbative and numerical
renormalization group transformations for theories of physical
interest.  He has written a number of reviews of this
work \APrefmark{\rWILNINE-\rWILELEVEN} and two simple
introductions \APrefmark{\rWILTWELVE,\rWILTHIRT}.  This paper relies
heavily on Ref. \rWILTEN, and also on Wegner's formulation of the
perturbative renormalization group \APrefmark{\rWEGONE-\rWEGTHR}.  Of
course, all of this work rests on the early development of the renormalization
group, especially on the work of Gell-Mann and Low \APrefmark{\rGELL};
as well as on the ideas of Kadanoff that inspired the modern
renormalization group \APrefmark{\rKADANOFF}.

The most promising area for
application of the renormalization group in the study of the strong
interaction is currently
provided by the Euclidean lattice formulation of quantum
chromodynamics (QCD) \APrefmark{\rREBBI},
but here the nonperturbative renormalization
group has found limited application and one is still forced to directly
include short distance scales
in large numerical calculations.  The lattice itself introduces
numerical complications that are not easily overcome, and alternative
nonperturbative tools should be developed.  At this point there is
no serious challenger to lattice field theory, as all other
nonperturbative algorithms rely on uncontrolled `approximations.'

The most significant barrier to the application of the renormalization
group is
algebraic complexity.  This complexity stems partially from the general
nature of the renormalization group as formulated by Wilson.  The
original renormalization group formalism developed by
Stueckelberg and Petermann \APrefmark{\rSTUECK}, and
Gell-Mann and Low \APrefmark{\rGELL}, as well as direct
derivatives such as that of Callan and
Symanzik \APrefmark{\rCALLAN,\rSYMONE}, are tailored to the
problem of renormalizing canonical field theories, and take
advantage of tools that
have been developed for Feynman perturbation theory \APrefmark{\rFEYN}.
These versions are of limited utility for some problems,
particularly those that cannot be
adequately solved with Feynman perturbation theory and those in
which one needs to remove degrees of freedom with explicit cutoffs.
In light-front field theory we
encounter a problem that requires the more general renormalization
group, because we need to use cutoffs that reduce the size
of Fock space to attack nonperturbative problems; and
because all cutoffs at our disposal
violate symmetries such as
Lorentz covariance and gauge invariance. Furthermore, we do not want to
include a complicated vacuum in our state vectors, so we need the more
general formulation of the renormalization group to allow
interactions induced by the vacuum to directly enter our Hamiltonians.

The modern
renormalization group is a pragmatic approach to any problem
that involves very many degrees of freedom that can be
profitably divided according
to a distance or
energy scale.  Instead of trying to solve such problems by considering
all scales at once, which usually fails even in perturbation theory,
one breaks the problem into pieces,
`solving' each scale in sequence.  There are many numerical advantages
to this approach.

As currently formulated the nonperturbative
renormalization group \APrefmark{\rWILTEN}
is reminiscent of the calculus as applied by
Newton \APrefmark{\rNEWTON}.
Most physicists would find it impossible to read the Principia,
and few would recognize the calculus in the form that Newton and his
peers employed it.
Fortunately, most of us can avoid the Principia;  but while there are
many good introductions to the perturbative renormalization
group \APrefmark{\rMAONE-\rGOLDEN}, the
nonperturbative renormalization group has not yet been developed to the
point where simple introductions exist.
This paper follows Wilson's remarkable
review article on the renormalization group and its application to the
Kondo problem \APrefmark{\rWILTEN}.
I concentrate only on the development of a perturbative
light-front renormalization group, and do not discuss possible
nonperturbative renormalization groups.  One of the chief purposes of
this paper, however, is to outline some of the problems one must face when
developing a nonperturbative light-front renormalization group.

Dirac formulated light-front field theory \APrefmark{\rDIRONE}
during his unsuccessful
search for a reasonable Hamiltonian formulation of relativistic field
theory \APrefmark{\rDIRTWO}.  For the most complete set of references
available on light-front physics, see Ref. \rLFREF.
Unfortunately, Dirac did not follow through with his initial
development of light-front field theory and
it was largely ignored until Weinberg developed the
closely related infinite momentum frame formalism \APrefmark{\rWEI}.
The principal advantages of the light-front formalism
are that boost invariance is kinematic,
and that the bare vacuum mixes only with modes that
have identically zero longitudinal momentum.  The first advantage allows
one to factor center-of-mass momenta from the equations of motion, which may be
extremely important if it proves possible to
formulate any relativistic problem so
that it becomes a few body problem; a fact that can be appreciated after
a study of the few-body problem in nonrelativistic quantum mechanics.
The second advantage is sometimes misrepresented to mean that the vacuum
is trivial in light-front field theory.  What is actually implied is
that one can isolate all modes that mix with the trivial vacuum to form
the physical vacuum, and then
note that all of these
have infinite energy in light-front field theory
(ignoring a set of measure
zero in theories with massless particles; \ie, states with identically
zero transverse momentum).
This observation may indicate that it is possible
to replace the problem of building the physical vacuum with the problem
of renormalizing the light-front Hamiltonian.
Using a boost-invariant renormalization
group one may be able to embed the vacuum problem into the larger
problem of using the light-front
renormalization group to remove high energy degrees
of freedom.

The light-front formalism does not automatically solve the physical
problems that force one to actually build the vacuum in Euclidean field
theory.  It merely allows us to reformulate these problems in what will
hopefully prove to be a more tractable form.
If a symmetry is broken by
the vacuum, we are not allowed to assume that symmetry when using the
renormalization group
to construct the renormalized Hamiltonian.  In other words, we are not
allowed to use a broken symmetry to restrict the space of
Hamiltonians, even when the symmetry is broken by the vacuum.
For example, if we want to study
$\phi^4$ theory in 1+1 dimensions beyond the critical coupling (\ie, in
the symmetry broken phase), we must allow
the space of Hamiltonians to include symmetry breaking interactions
(\eg, a $\phi^3$ interaction).
If such interactions are not allowed,
states with imaginary mass will typically appear in the
spectrum \APrefmark{\rHVONE-\rSWENSON}.  The
renormalization group allows one to isolate relevant and marginal
symmetry breaking interactions that might be tuned to reproduce the
vacuum effects.  A simple example involving spontaneous symmetry
breaking has recently been provided by Wilson and
the author \APrefmark{\rPERWIL}.

While the infinite momentum frame formalism
has been widely used,
especially in the study of current algebra \APrefmark{\rFUB,\rDASH,\rLFREF}
and the parton model \APrefmark{\rBJORK-\rKOGTWO,\rLFREF},
little formal work has been completed on
renormalization in light-front field theory, and
almost all early work concentrates primarily on developing a
map between light-front field theory
and equal-time field theory in perturbation
theory \APrefmark{\rCHANG-\rBRS}.  More recently a number of theorists have
begun to study light-front perturbation theory
directly \APrefmark{\rBOUCH-\rPERQCD},
and especially the renormalization of light-front field theory after a
Tamm-Dancoff \APrefmark{\rTAM-\rBET} truncation is
made \APrefmark{\rPERONE-\rGHPSW}.
The only formalism
currently available for nonperturbative renormalization is the
renormalization group, and
work on developing the renormalization group
for light-front field theory is in its
infancy \APrefmark{\rGLAZEK-\rGLAZWIL,\rPERQCD,\rPERWIL}.  In my
opinion, without a light-front version of the renormalization group,
light-front field theory may be
relegated to being a tool of last choice for doing perturbative
calculations in 3+1 dimensions.  Of course, in superrenormalizable
theories in 1+1 dimensions light-front field
theory has already proven to be very successful \APrefmark{\rBERG-\rMO};
and it can be
argued that light-front field theory is a much more
powerful tool for many nonperturbative
calculations in 1+1 dimensions than equal-time field theory.

In 3+1 dimensions we are in a situation where, because of serious
renormalization difficulties, we neither know
the correct light-front
Hamiltonians to study, nor can we compute
the physical ground
states to which these unknown Hamiltonians lead.  This is exactly
the type of problem for which the modern renormalization group is suited.

The
primary purpose of this paper is to develop
a perturbative light-front renormalization group as a tool
for the study of light-front Hamiltonian field theory,
with the hope that this work may aid the
development of a practical nonperturbative light-front
renormalization group.  The most interesting theory that one
may be able to
study with a perturbative light-front renormalization group is QCD;
however, the algebraic complexity of QCD makes it a poor development
ground.  Following tradition, I use scalar field theory for all of my
examples, as it is straightforward (\ie, difficult
but not impossible) to generalize
the formalism to other theories.

In the remainder of this Introduction I outline the rest of the
paper.  In the process I use both light-front and renormalization
group jargon, often without offering any definition.  I have tried to
carefully define most of the renormalization group
jargon in the text, so the reader who is
unfamiliar with the modern renormalization
group may want to read this Introduction again after reading the rest of
the text.  There are a number of articles that introduce or review most
of the
basic aspects of light-front field theory required in this work, and the
reader unfamiliar with this formalism may want to consult one or more of
these \APrefmark{\rKOGONE-\rYANTWO,\rLEPAGE-\rBPMP,\rPERTWO}.

To implement a renormalization group calculation one must
first delineate a space of Hamiltonians, and then define a
renormalization group transformation \APrefmark{\rKADANOFF}
that maps this space into itself.
The renormalization group
transformation must be carefully formulated because it is central to the
whole approach.  Several renormalization group
transformations are given for light-front field theory, including some
boost-invariant transformations.  These latter transformations allow one
to impose the constraint of boost invariance directly on the space of
Hamiltonians.
The next step in a renormalization group study
is to explore the topology of the Hamiltonian
space, first searching for fixed points of the transformation (\ie,
Hamiltonians that remain fixed under the action of the transformation)
and then studying the
trajectories of Hamiltonians near these fixed points.  The Hamiltonian
itself was originally formulated as a powerful tool for the study of
physical trajectories found in nature.  The renormalization group
can be considered to be a generalization of this idea, in which the
renormalization group transformation is used to derive physical
Hamiltonians found in nature.  As such, it is an alternative to the
canonical procedure that starts with classical equations of motion.

Almost all
analytic work concentrates on Gaussian fixed points (\ie, Hamiltonians
with no
interactions) and near-Gaussian fixed points where perturbation
theory can be used to approximate the renormalization group
transformation itself.  All of the examples in this paper are of this type,
and the entire investigation is directed toward the development of
transformations near Gaussian fixed points.  Actually, there are
fixed volumes rather than isolated fixed points, but I usually
refer to fixed points rather than fixed volumes.
Such examples may be of immediate relevance for QCD; and they illustrate
the basic light-front renormalization group
machinery.

In Section \rntwo~ I provide a brief summary of the modern renormalization
group formalism and the generalizations required when there are an
infinite number of relevant and marginal operators.  The generalizations
are not easily appreciated until one has studied the entire paper and
understood why they are required.
This Section is a poor substitute for Wilson's review
article \APrefmark{\rWILTEN},
but I have attempted to introduce the most important
concepts required by a perturbative renormalization group.  I have also
attempted to make the differences between a perturbative renormalization
group and a nonperturbative renormalization group clear, focusing on
the former.  The modern renormalization group formalism may be unfamiliar
to many students of relativistic field theory, being employed primarily
in the study of critical phenomena.  With a few notable
exceptions \APrefmark{\rAMIT,\rZINN}, most
field theory textbooks deal exclusively with the original
renormalization group formalism and its modern descendant, the
Callan-Symanzik formalism \APrefmark{\rCALLAN,\rSYMONE}.  Perhaps the most
important difference between these two types of renormalization group is
that the original renormalization group does not actually remove any
degrees of freedom, being concerned primarily with the problem of
divergences in perturbation theory
and techniques for allowing all cutoffs to be removed.
Here one typically uses either Pauli-Villars
regularization \APrefmark{\rPAULI},
which actually increases the number of degrees of freedom, or
dimensional regularization \APrefmark{\rTHOONE,\rTHOTWO},
which retains all degrees of freedom while
analytically continuing in the number of dimensions.  Neither of these
regulators is well-suited to many nonperturbative calculations.

In order to reduce the number of degrees of freedom, one must introduce
a real cutoff, such as a momentum cutoff or a lattice cutoff.
The cutoff is an artifice, so
results should not depend on its particular value and it should be
possible to change the cutoff without changing any physical result; \ie,
without changing the matrix elements of observables between cutoff
physical states.  (Later I often refer to such matrix elements as
observables, since the only operator that is discussed in any detail is
the Hamiltonian.)  This can
be accomplished by making the observables, and in particular the
Hamiltonian, depend on the cutoff in precisely the manner required to
yield cutoff independent matrix elements.  The modern renormalization
group is designed to achieve this goal.

In Section \rnthree~ I illustrate what is meant by a
space of Hamiltonians, and provide several light-front renormalization group
transformations.
I begin by discussing transformations that resemble those
developed by Wilson for Euclidean field theory;
however, these transformations later lead to pathologies because they
remove states of lower energy than some that are retained.
Almost all light-front transformations suffer from
these pathologies in perturbation theory,
and in the end I am forced to consider renormalization groups with some
unusual properties
to obtain a transformation that may be approximated by discarding at
least some irrelevant operators in perturbation
theory.  These transformations employ invariant-mass cutoffs, so I
refer to them as {\it invariant-mass transformations}.  The restriction
to transformations that may be approximated perturbatively
is an extreme limitation, and it is not even clear that the
invariant-mass transformations can be approximated in
perturbation theory with controlled errors.  We will find that couplings
depend on longitudinal momentum fractions when one uses a boost-invariant
cutoff, and that corrections to the Hamiltonian diverge
logarithmically for states containing particles with arbitrarily small
longitudinal momentum fractions, and for interactions involving
arbitrarily small longitudinal momentum transfer.
Thus, ultimately it is not clear that the
invariant-mass transformations are `better' than other transformations
one may use; and one should certainly consider other transformations
when developing a nonperturbative renormalization group.  This article
does not attack these nonperturbative problems, even though they may be
of more interest than the perturbative results obtained here; however, I try
to clarify some of the nonperturbative renormalization problems
light-front field theorists should be attacking.

An essential restriction in all Euclidean renormalization group
calculations is that long range interactions are excluded.  As Wilson
notes \APrefmark{\rWILTEN}, this is one of the most tenuous assumptions of the
renormalization group approach.
This locality
assumption must be altered in light-front field theory
where inverse powers of longitudinal derivatives are required already
at the Gaussian fixed points of interest; \ie, in free field theory.
Allowing inverse powers of longitudinal derivatives may seem to be
a rather minor conceptual modification of the Euclidean version of the
renormalization group; but the recognition that there are separate
renormalization group transformations that act on the longitudinal and
transverse directions may be profound.  This generalizes the fact that
there are separate power counting analyses for longitudinal and
transverse dimensions.
The analysis of
relevant, marginal and irrelevant longitudinal operators indicates
that it is inverse powers of longitudinal derivatives that arise as
irrelevant operators when one of the light-front renormalization group
transformations is applied to a
Hamiltonian.  Having allowed inverse longitudinal derivatives, one
naturally
considers the possibility that inverse powers of transverse
derivatives occur, a possibility that is especially
intriguing for the study of QCD; however, it is difficult to introduce
such operators in a controlled manner, and I avoid their
introduction in this article.

I assume that transverse interactions are local, or at least short range,
and refer to this
assumption as {\it transverse locality}.
This restriction is not merely a
technical convenience, because inverse transverse derivatives typically
lead to an infinite number of relevant operators near critical Gaussian
fixed points, including products of arbitrarily large numbers of field
operators.  Near fixed points that contain interactions, such
operators may appear without causing trouble, but I discuss only
Gaussian fixed points.
Once the space of Hamiltonians is restricted, the
renormalization group transformation may produce Hamiltonians that lie
outside the space.  We are only interested in
trajectories of Hamiltonians generated by repeated application of the
transformation that remain inside the restricted space.
Strict transverse locality is violated by the step
function
cutoffs I employ; however, these violations appear to be controllable.
Moreover, inverse transverse derivatives
are generated by the transformations; but they are accompanied by
cutoffs and I show that the resultant distributions do not produce long
range transverse interactions, and that they do not introduce relevant
operators.  In all cases that I have discovered where a product of inverse
transverse derivatives and cutoff functions arise, the product can be
shown to be an irrelevant operator with respect to transverse scaling;
although the operator may contain delta functions
or derivatives of delta functions of
longitudinal momentum fractions.

Each of the light-front renormalization group
transformations I consider consists of two steps.  In
the first step one alters a cutoff (\eg, lowers a cutoff on transverse
momenta) to remove degrees of freedom,  and computes a renormalized
Hamiltonian that produces the same eigenvalues and suitably
orthonormalized, projected eigenstates, in the remaining subspace.
All operators that correspond to
observables are renormalized,
not just the Hamiltonian, but in this paper I focus
only on the Hamiltonian.  Of particular interest for future work is the
renormalization of other Poincar{\'e} generators and various current
operators.  In the second step of each transformation the
variables (\eg, transverse
momenta) are rescaled to their original range, and the fields and
Hamiltonian are rescaled.  The most difficult part of this procedure is the
construction of an effective Hamiltonian, which is analogous to the
development of a block spin Hamiltonian in simple spin systems.  In
Section \rnfour~ I discuss two related procedures
for accomplishing this task; the first pioneered by Bloch
and Horowitz \APrefmark{\rBLOONE},
and the second
by Bloch \APrefmark{\rBLOTWO,\rBRANDOW}.
The second method was employed by Wilson in his first serious
numerical renormalization group study \APrefmark{\rWILTWO}.

In Section \rnfive~ I turn to the study of fixed points in perturbation
theory and linearized behavior of the transformations near these fixed
points \APrefmark{\rWEGONE-\rWEGTHR},
developing simple examples that hopefully clarify the
procedure.  The
study of linearized behavior near critical Gaussian fixed points in
light-front field theory is
dimensional analysis, as usual.
The fact that longitudinal boost
invariance corresponds to an exact scale
invariance \APrefmark{\rDIRONE,\rKOGTWO,\rLEU}
leads to the conclusion that all physical Hamiltonians are
fixed points with respect to a longitudinal light-front renormalization
group transformation.
This should
be true order-by-order in any perturbative expansion of the
Hamiltonian in powers of a parameter upon which it depends
analytically, and it
should be an extremely powerful tool for the analysis of fixed
points.  However, the
pathologies mentioned above make it difficult to
find applications.  Transformations that scale only the transverse
momenta produce an infinite number of relevant and marginal operators,
in addition to the familiar infinity of irrelevant operators.  This
happens because entire functions of longitudinal momentum fractions may
appear in any given operator without affecting the linear analysis that
determines this classification.  Transformations that scale only
longitudinal momenta also lead to an infinite number of relevant and
marginal operators because entire functions of transverse momenta are
allowed to appear.  The appearance of entire functions drastically
complicates the renormalization group analysis, but it may also
eventually lead to tremendous power in the application of the
renormalization group if one can learn how to accurately approximate
these functions.

In Section \rnsix~ I
study second-order perturbations about the critical Gaussian fixed point.  I
concentrate on Hamiltonians near the canonical massless
$\phi^4$ Hamiltonian, and
first show that several candidate transformations lead to divergences at
second-order.  I turn to a boost-invariant transformation and
concentrate primarily on a few marginal and relevant operators.  I show
that it is possible to find a closed set of two marginal operators and
one relevant operator in a second-order analysis,
despite the possible appearance of an infinite
number of such operators.  This simple analysis illustrates many of the
features of a full transformation.
I then study Lorentz covariance and
cluster decomposition, showing that both are violated by the
invariant-mass cutoff in second-order perturbation theory.  These
properties must be restored by counterterms, and the formalism should be
able to produce these counterterms without referring to covariant
results.

Since there are an infinite number of relevant and marginal operators in
the light-front renormalization group, and a simple perturbative
analysis indicates that Lorentz covariance and cluster decomposition
cannot be restored without actually employing an infinite number of such
operators, one must worry that the light-front renormalization group
analysis requires one to adjust an infinite number of independent
variables.  Wilson and I have recently proposed conditions under which a
finite number of variables are actually independent \APrefmark{\rPERWIL}.
These conditions turn out to be
a generalization of the coupling reduction conditions first
developed by Oehme, Sibold, and Zimmerman \APrefmark{\rOEHONE-\rKRAUS}.

Simply stated, we have proposed
that one should seek solutions to the renormalization group equations in
which a finite number of specified variables are allowed to be
independent functions of the cutoff.  While there are an infinite number
of relevant, marginal, and irrelevant variables, all but a finite number
of variables depend on the cutoff only through their dependence on these
independent variables.  This condition could merely be a
re-parameterization of the cutoff dependence,
but we also add the constraint that all dependent
variables must vanish when the independent variables are zero.  This
last constraint is motivated by considering the dependent variables to
be counterterms.  This means that as one changes the cutoffs, all
couplings (including masses) evolve coherently; whereas the general
solution to the renormalization group equations might allow much more
complicated behavior, at least for the relevant and marginal variables.
It is for this reason that we call our
conditions {\it coupling constant coherence conditions}, or more
briefly, coupling coherence.

In Section \rnsix~ I provide part of the demonstration
that coupling coherence
fixes the strengths of all
operators to second order in the canonical coupling, $\order(g^2)$,
and that the
resultant strengths are precisely the values required to restore
Lorentz covariance and cluster decomposition in second-order
perturbation theory.  I show that coupling coherence uniquely fixes the
relevant mass operator, and the dispersion relation associated with the
bare mass term is not that of a
free massive particle.  I determine the complete set of irrelevant
four-boson couplings, and show that a third-order (\ie, two loop)
calculation is required to fix the marginal four-boson couplings.  I do
not compute all couplings to $\order(g^2)$, but it is straightforward to
complete the analysis for the terms I do not evaluate.  I am primarily
interested in the marginal four-boson couplings, because these indicate
how one can expect couplings to run with longitudinal momenta in a
light-front renormalization group analysis.  In the scalar theory,
couplings decrease in strength as the longitudinal momentum fraction
carried by the bosons decreases, and as the longitudinal momentum
fraction transferred through the vertex decreases.  In QCD one expects
the opposite behavior.

There are two
essential expansions that are made in a perturbative renormalization
group analysis.  The first is the perturbative expansion of the
transformation itself, which may converge sufficiently near the
Gaussian fixed point.  The second is the expansion of the
transformed Hamiltonian in terms of relevant, marginal, and irrelevant
operators.  Each expansion must normally be
truncated at some finite order, and one should
try to show that each truncation leads
to controllable errors.
Much of the
discussion of the errors introduced by truncating the perturbative
expansion is delayed to Section \rnseven, although some of the most important
sources of errors are already evident
in a second-order analysis and are discussed in
Section \rnsix.  Many of the errors that arise when one truncates the
expansion of the Hamiltonian in terms of relevant, marginal and
irrelevant operators by discarding all or most of the irrelevant
operators can be studied using the second-order approximation
of the transformation.
Dropping some irrelevant
operators is essential to the program, because the transformations
become too complicated if one must follow the evolution of too many
operators.

Most transformations one might construct lead to uncontrollable
errors even in
a second-order analysis for a simple reason.  When a transformation is
applied once, irrelevant operators are produced; and when the
transformation is applied again these irrelevant operators produce
divergences.
The pathology is easily analyzed.  Simple transformations attempt
to remove degrees of freedom with much lower free energy than some of the
degrees of freedom retained.
As one step in making the expansion in terms of
relevant, marginal and irrelevant operators, one typically expands every
energy denominator encountered in the perturbative expansion of the
transformation.  This expansion is in powers of one of the energies of a
state that is retained after the transformation, and if this state has a
higher energy than one of the states removed, the expansion of the
energy denominator fails to converge.  Simply stated, $1/(E_{in}-E_{loop})$
cannot be expanded in powers of $E_{in}$ when $E_{in}$ is larger than
$E_{loop}$.  I believe that there are few solutions to this
problem.  One can abandon the perturbative expansion of the
transformation, one can abandon the renormalization group approach or
drastically modify it, or
one can design the transformation so that $E_{in}$ is always less than
$E_{loop}$.  In this paper, I choose the final option.
The simplest boost-invariant
transformations suffer from this same problem, and only the invariant-mass
transformation escapes.

In Section \rnseven~ I discuss third- and higher-order corrections to an
invariant-mass transformation near the critical Gaussian fixed point.  I first
show that the third-order analysis introduces new marginal operators
that contain logarithms of longitudinal momentum fractions.  I also show
that when one insists that these new marginal operators depend on the
cutoff only through their dependence on the original marginal $\phi^4$
interaction, their strength is precisely that required to restore
Lorentz covariance and cluster decomposition to the boson-boson
scattering amplitude computed in second-order perturbation theory.  I
then turn to a discussion of the errors introduced by various
approximations, showing that
these errors may be quite large.

Wilson has provided a thorough discussion of perturbative
renormalization group equations and the errors that result from various
approximations, so I focus on new approximations that must be made
in a light-front renormalization group analysis.  In a typical Euclidean
renormalization group there are a finite number of relevant and marginal
operators and one can accurately tune their boundary values at the
lowest cutoff.  In the
light-front renormalization group functions of longitudinal momenta
appear in each relevant and marginal operator, and these functions must
be approximated.  This means that errors are made in the relevant and
marginal operators themselves, and one must determine whether these
errors can be controlled.

These new approximations are
extremely problematic, because one expects that the strength of a
relevant operator will grow exponentially as the cutoff
is lowered;  which means that any error made in the approximation of a
relevant operator
will grow exponentially.  Of course, one partial solution to this
problem is to approximate the relevant operators at the lowest cutoff,
and solve the renormalization group equations for the relevant operators
in the direction of increasing cutoff, as is usually
done \APrefmark{\rWILTEN};
because the strengths of relevant operators
decrease exponentially when the cutoff is increased.  Marginal operators
present a more difficult problem, as usual; because errors tend to grow
linearly regardless of the direction in which one solves the equations
for marginal operators.  Moreover, we will see that operators develop
logarithmic singularities when one uses the invariant-mass
transformation.
These problems require careful
study, and this paper barely initiates such a study.
No convincing solution to these
problems is proposed in this paper, but I have tried to present the
problems in a clear manner; because any attempt to use Hamiltonian
light-front field theory to perform nonperturbative calculations must
address such problems.  I offer some speculation on how
one might approach this task, and argue that the same problems that
complicate the perturbative analysis may actually lead to a
simplification of the nonperturbative analysis.

The renormalization group analysis is drastically simplified
when a Tamm-Dancoff truncation \APrefmark{\rTAM,\rDAN} is made and
the light-front Tamm-Dancoff (LFTD)
approximation \APrefmark{\rPERONE-\rGHPSW} is used, so I
occasionally mention important
aspects of a renormalization group analysis of LFTD;
however, I do not consider such examples in this paper.
The Tamm-Dancoff
truncation can be simply included in the initial definition of the
space of Hamiltonians, after which the analysis proceeds exactly
as when no truncation is made.
The truncation preserves boost invariance, and thereby allows the
boost-invariant renormalization group transformations to be illustrated.
More importantly, it drastically simplifies the operators that are
included in the space of Hamiltonians, even allowing one to solve some
simple examples analytically.  Sector-dependent renormalization, in
which parameters appearing in the Hamiltonian and other observables are
allowed to depend on the Fock space sector(s) within or between which
they act, arises
in a natural manner; and the light-front renormalization group
drastically improves the
discussion of renormalization in LFTD.
However,  if one wants to use LFTD to study physical theories, it
is probably necessary to
remove some of the restrictions on the space of allowed Hamiltonians
that I assume, in particular restrictions associated with transverse
locality.  I do not yet know how one can introduce the required
nonlocalities and still control the
number of operators required; however,
simple perturbation theory arguments show that
nonlocal transverse
operators inevitably arise if one derives a LFTD Hamiltonian by
eliminating states with extra particles.  These issues are not discussed
in this paper.

In the conclusion I discuss the possible relevance of this work for the
study of Hamiltonian light-front
QCD \APrefmark{\rCASH-\rLEP,\rLEPAGE,\rBROONE,\rBPMP}, indicating some of the
difficult problems that I carefully avoid with my simple examples in
this paper.


\bigskip
\noindent {\bf \rntwo. The Renormalization Group}
\medskip

In classical mechanics the state of a system is completely specified by
a fixed number of coordinates and momenta.
The objective of classical mechanics is to compute the state as a
function of time, given initial conditions.  The state is
not regarded as fundamental; rather a
Hamiltonian that governs the
time evolution of the state is
regarded as fundamental.  In nonrelativistic quantum mechanics, one
must generalize the definition of the state, so that it is specified by a ket
in a state space, and one must drastically alter the theory of
measurement; but it remains possible to specify a
Hamiltonian that governs the time
evolution of the state.  In both cases the time
evolution of the state is a trajectory in a Hilbert space, and the
trajectory is determined by a Hamiltonian that must be discovered by
fitting data.
In principle, one would like to further
generalize this procedure for relativistic field theory;
however, any
straightforward generalization that maintains locality
leads to divergences that produce mathematical ambiguities.
To make mathematically meaningful
statements we must introduce an \adhoc regulator, to which I
refer as a cutoff; so that physical results can be derived as
limits of sequences of finite quantities.  The
renormalization group provides methods for constructing such limits
that are much more powerful than standard perturbation theory.

Even if divergences did not signal the need for a cutoff in field theory,
we would be forced to introduce a cutoff in some form to obtain finite
dimensional approximations for the state vector and Hamiltonian.
Fock space is an infinite dimensional sum of cross
products of infinite dimensional Hilbert spaces, and this is not a
convenient starting point for most numerical work.
If all interactions are weak and of nearly constant strength over the
entire range of scales that affect an observable, we can use standard
perturbation theory to compute the observable; however, if either of
these conditions is not met, we cannot directly compute
observables with realistic Hamiltonians.
This problem is easily appreciated by considering a simple
spin system in which 1024 spins are each allowed to take two values.  The
Hamiltonian for this system is a $2^{1024}\times2^{1024}$ matrix, and this
matrix cannot generally
be diagonalized directly.  This matrix is infinitely
smaller than the matrices we must consider when solving a
relativistic field theory.  If the interactions remain weak over all
scales of interest, but change in strength significantly,
we can use the perturbative renormalization group.
If the interactions become strong over a large number of scales of
interest, a nonperturbative renormalization group must be
developed.  A final possibility is that the interactions are weak over
almost all scales, becoming strong only over a few scales of
interest.  In this case, the perturbative renormalization group can be
used to eliminate the perturbative scales; after which one can use some
other method to solve the remaining Hamiltonian.

The introduction of an \adhoc cutoff in field theory complicates the
basic
algorithm for computing the time evolution of a state, because one must
somehow remove any dependence on the cutoff from physical matrix
elements.  This complication is so severe that
it has caused field theorists to essentially abandon many of the most
powerful tools employed in nonrelativistic quantum mechanics (\eg, the
Schr{\"o}dinger picture).  How can
we make reasonable estimates in relativistic quantum mechanics? How can
we guarantee that results are independent of the cutoff?  How
can we find a sequence of Hamiltonians
that depends on the cutoff in a manner that leads to correct results
as the cutoff approaches its limit?  These are the type of questions
that led Wilson to completely reformulate the original Gell-Mann--Low
renormalization group formalism.

Wilson adopted the same general strategy familiar from the study of the
time evolution of states, adding a layer of abstraction to the original
classical mechanics problem to compute `Hamiltonian trajectories'.
The strategy is universal
in physics, but the layer of
abstraction leads to a great deal of confusion.
In analogy to a formalism that yields the evolution of a state as
time changes,
he developed a formalism that yields the evolution of
a Hamiltonian as the cutoff changes.  In quantum mechanics a state is
represented by an infinite number of coordinates in a Hilbert space, and
the Hamiltonian is a linear operator that generates the time evolution
of these coordinates.   In the renormalization group formalism, the
existence of a space in which the Hamiltonian can be represented by an
infinite number of coordinates is assumed, and the cutoff evolution of
these coordinates is given by the renormalization group transformation.
The Hamiltonian is less fundamental than the renormalization group
transformation, which can be used to construct trajectories of
Hamiltonians.  Typically there are restrictions placed on the Hamiltonians
(\eg, no long-range interactions) that
make it possible for these trajectories to leave the space.
Trajectories of renormalized Hamiltonians remain in the space of
Hamiltonians, and are roughly
analogous to physical trajectories in classical mechanics
that meet some additional requirement
(\eg, do not leave a specified finite volume).
Transformations that change the cutoff by different amounts
are members of a renormalization
`group', which is actually
a semi-group since the transformations cannot be inverted.  The fact that
inverses do not exist is obvious because the transformations reduce the
number of degrees of freedom.

A more thorough, although by no means complete, discussion of the space
of Hamiltonians is given in Section \rnthree.  Here I will simply state
that each term in a
Hamiltonian can typically be written as a spatial integral whose
integrand is a product of derivatives and field operators.  The
definition of the Hamiltonian
space might be a set of rules that show how to
construct all allowed operators.
These operators should be thought of as unit vectors, and the coefficients in
front of these operators as coordinates.  This type of operator is not
usually
bounded, and this is a source of divergences in field theory.  To
regulate these
divergences the cutoff is included directly
in the definition of the space
of Hamiltonians.  The
cutoff one chooses has drastic effects on the renormalization group.  While
we will see several examples in Section \rnthree, one familiar example
of a cutoff is the lattice, which replaces spatial integrals by sums
over discrete points.  The facts that the Hamiltonian can be represented
by coordinates that correspond to the strengths of specific operators,
and that these operators are all regulated by a cutoff that is part of
the definition of the space, is all that one needs to appreciate at this
point.

Given a space of cutoff Hamiltonians, the next step is to construct
a suitable transformation.  This is slightly subtle and is usually the
most difficult conceptual step in a renormalization group analysis,
because one must
find a transformation that manages to alter the cutoff
without changing the space in which the Hamiltonian lies.  These two
requirements seem mutually contradictory at first.  An additional
problem for relativistic field theory is that all transformations one
can construct change the cutoff in the wrong direction.

To see how these
difficulties are averted, let me again use the lattice
as an example.  A
typical
lattice transformation consists of two steps.  In the first step one reduces
the number of lattice points, typically by grouping them into blocks
\APrefmark{\rKADANOFF}
and thereby doubling the lattice spacing; and
one computes a new effective Hamiltonian on the new lattice.  I do not
discuss how this effective Hamiltonian is constructed for a lattice, but
this issue is carefully discussed for light-front Hamiltonians in later
Sections.  At this point the lattice has changed, so the space in which
the Hamiltonian lies
has changed.  The second step in the transformation is to
rescale distances using a change of variables, so that the
lattice spacing is returned to its initial value, while one or more distance
units are changed.  After both steps are completed the lattice itself
remains unchanged, if it has an infinite volume,
but the Hamiltonian changes.  This shows how one can
alter the cutoff without leaving the initial space of Hamiltonians.
Numerically the cutoff does not change, but the units in which the
cutoff is measured change.

We want to study the limit in which the lattice spacing is taken to
zero, but the transformation increases the lattice spacing as measured
in the original distance units.
While no inverse transformation that directly decreases the lattice
spacing
exists, we can obtain the limit in which the lattice spacing
goes to zero by
considering a sequence of increasingly long trajectories.
Instead of fixing the initial lattice spacing, we fix the
lattice spacing at the end of the trajectory and we construct a sequence
of trajectories in which the initial lattice spacing is steadily
decreased.
We then directly
seek a limit for the last part of the trajectory
as it becomes infinitely long, by studying the sequence of increasingly
long trajectories.

This
procedure is illustrated by Wilson's triangle of renormalization, which
is briefly discussed below.
One must employ Wilson's algorithm
to perform a nonperturbative
renormalization group analysis; however, it is possible to study the
cutoff limit more directly when a reasonable
perturbative approximation exists.
In this case, the renormalization group
transformation can be approximated by an infinite number of coupled
equations for the evolution of a subset of
coordinates that are asymptotically complete,
and these equations can be inverted to allow direct study of
the Hamiltonian trajectory as the cutoff increases or
decreases \APrefmark{\rWEGONE-\rWEGTHR}.
If it can be shown that all but a finite number
of coordinates remain smaller than some chosen magnitude, it may be
possible to approximate the trajectory by simply ignoring the
small coordinates, retaining an increasing number of coordinates only as
one increases the accuracy of the approximation.  In this case the task
of approximating a trajectory of renormalized Hamiltonians is reduced to
the task of solving a finite number of coupled nonlinear difference
equations.  The
primary goal of this paper is the development of a perturbative
light-front renormalization group.

Given a transformation $T$ that maps a subspace of Hamiltonians into
the space of Hamiltonians, with the possibility that some Hamiltonians
are mapped to Hamiltonians outside the original space, we study
$T[H]$.  We can apply the transformation repeatedly, and construct a
trajectory of Hamiltonians, with the $l$-th point on the trajectory
being

$$H_l = T^l[H_0] \;. \eqno(2.1)$$

\noindent Any infinitely long trajectory that remains inside the space
is called a trajectory of renormalized Hamiltonians.  The motivation for
this definition of renormalization
is clarified further below.  It is assumed that the trajectory is
completely determined
by the initial Hamiltonian, $H_0$, and $T$; however,
the dependence on $H_0$
is usually not explicitly indicated. Moreover, we will see later that
boundary conditions may be specified in a much more general fashion.

Any renormalization group
analysis begins with the identification of at least one fixed point,
$H^*$.  A {\it fixed point} is defined to be any Hamiltonian
that satisfies the condition

$$H^*=T[H^*] \;. \eqno(2.2)$$

\noindent For perturbative renormalization groups the search for such
fixed points is relatively easy, as we will see in Section \rnfive;
however, in nonperturbative studies such a search typically involves
a difficult numerical trial and error
calculation \APrefmark{\rWILTWO,\rWILFOUR,\rWILTEN}.
If $H^*$ contains no
interactions (\ie, no terms with a product of more than two field
operators), it is called {\it Gaussian}.
If $H^*$ has a massless eigenstate,
it is called {\it critical}.
If a Gaussian fixed point has no mass term, it is
a {\it critical Gaussian} fixed point.
If it has a mass term, this mass must
typically be infinite, in which case it is a {\it trivial Gaussian} fixed
point.  In lattice QCD the trajectory of renormalized Hamiltonians stays
near a critical Gaussian fixed point until the lattice spacing becomes
sufficiently large that a transition to strong-coupling behavior
occurs.  If $H^*$ contains only weak interactions, it is called
{\it near-Gaussian}, and one may be able to
use perturbation theory both
to identify $H^*$ and to accurately approximate trajectories of
Hamiltonians near $H^*$ \APrefmark{\rWILNINE}.
Of course, once the trajectory leaves the
region of $H^*$ it is generally necessary to switch to a nonperturbative
calculation of subsequent evolution.  If $H^*$ contains a strong
interaction, one must use nonperturbative techniques to find $H^*$, but
it may still be possible to produce trajectories near the fixed point
using perturbation theory.  The perturbative analysis in this case
includes the interactions in $H^*$ to all orders, treating only the
deviations from these interactions in perturbation theory.

Consider the immediate neighborhood of the fixed point, and
assume that the trajectory remains in this neighborhood.  This assumption
must be justified \apost, but if it is true we should write

$$H_l=H^*+\delta H_l \;, \eqno(2.3)$$

\noindent and consider the trajectory of small deviations
$\delta H_l$.

As long as $\delta H_l$ is `sufficiently small', we can use a perturbative
expansion in powers of $\delta H_l$, which leads us to consider

$$\delta H_{l+1}= L \cdot \delta H_l + N[\delta H_l] \;. \eqno(2.4)$$

\noindent Here $L$ is the linear approximation of the full
transformation in the
neighborhood of the fixed point, and $N[\delta H_l]$ contains all
contributions to $\delta H_{l+1}$ of $\order(\delta H_l^2)$ and higher.

The object of the renormalization group calculation is to compute
trajectories and this requires
a representation for $\delta H_l$.  The problem of computing
trajectories is
one of the most common in physics, and a convenient basis for the
representation of $\delta H_l$ is provided by
the eigenoperators of $L$, since $L$ dominates the transformation near
the fixed point.  These eigenoperators and their eigenvalues are found
by solving

$$L \cdot O_m=\lambda_m O_m \;. \eqno(2.5)$$

\noindent If $H^*$ is Gaussian or near-Gaussian it is usually
straightforward to find $L$, and its eigenoperators and eigenvalues.
This is not typically true if $H^*$ contains strong interactions, and in
much of the remaining discussion I focus on formalism that is
primarily useful for the study of trajectories in the neighborhood of
Gaussian and near-Gaussian fixed points.
In Section \rnfive~ the linear approximations of several
simple light-front renormalization group transformations about a critical
Gaussian fixed point are derived,
and their eigenoperators and eigenvalues are computed.

Using the eigenoperators of $L$ as a basis we can represent
$\delta H_l$,

$$\delta H_l = \sum_{m\in R} \mu_{m_l}O_m +\sum_{m\in M} g_{m_l}O_m+
\sum_{m\in I} w_{m_l}O_m \;.\eqno(2.6)$$

\noindent Here the operators $O_m$ with $m\in R$ are {\it relevant} (\ie,
$\lambda_m>1$), the operators $O_m$ with $m\in M$ are {\it marginal} (\ie,
$\lambda_m=1$), and the operators with $m\in I$ are
either {\it irrelevant} (\ie, $\lambda_m<1$)
or become irrelevant after many applications of the
transformation.  The motivation behind this nomenclature is made clear
by considering repeated application of $L$, which causes the relevant
operators to grow exponentially, the marginal operators to remain
unchanged in strength, and the irrelevant operators to decrease in
magnitude
exponentially.  There are technical difficulties associated with the
symmetry of $L$ and the completeness of the eigenoperators that I
ignore \APrefmark{\rWEGONE-\rWEGTHR}.
I occasionally refer to the relevant variables as
masses, and the marginal and irrelevant variables as couplings; but I
also occasionally refer to all variables, including relevant
variables, as couplings.  What is meant should be clear from context.

$L$ depends both on the transformation and the fixed point, but there
are always an infinite number of irrelevant operators.  On the other
hand, transformations of interest for Euclidean lattice field theory
typically lead to a finite number of relevant and marginal operators.
One of the most serious problems for a perturbative light-front
renormalization group is that {\it an infinite number of relevant and
marginal operators are required.}  This result is derived in Section
\rnfive, and I discuss some of the consequences below.
In the case of scalar field theory, an infinite number of
relevant and marginal operators arise because the light-front cutoffs
violate Lorentz covariance and cluster decomposition.
These are continuous symmetries, and their
violation leads to an infinite number of constraints on the
Hamiltonian.  The key to showing that the light-front renormalization
group may not be rendered useless by an infinite number of relevant and
marginal operators is the observation that both the strength and the
evolution of all but a finite number of
relevant and marginal
operators are fixed by Lorentz covariance and cluster
decomposition.  However, one does not want to employ either of
these properties directly in the construction of Hamiltonians,
because they are never explicit in the
renormalization group calculation of a Hamiltonian trajectory.  Lorentz
covariance and cluster decomposition are properties of observations that
are obtained using the full Hamiltonian, and one does not want to solve
problems that require the entire Hamiltonian to compute the
Hamiltonian itself.  The alternative that Wilson and I have
proposed \APrefmark{\rPERWIL,\rOEHONE-\rKRAUS} is
to insist that the new relevant and marginal variables are not
independent functions of the cutoff, but depend only on the cutoff
through their dependence on canonical variables.  While this requirement
obviously fixes the manner in which the new variables evolve with the
cutoff, it also fixes their value at all cutoffs once the values of the
canonical variables are chosen.  The remarkable feature of this
procedure is that the value it gives to the new variables is precisely
the value required to restore Lorentz covariance and cluster
decomposition.  This conclusion is not proven to all
orders in perturbation theory, but it is illustrated by a nontrivial
second order example in this paper.

To simplify subsequent discussion, the
statement that $\delta H_l$ is small is assumed to mean that all
masses and couplings in the expansion of $\delta H_l$ are small.  The
analysis itself should signal when this assumption is naive.
A rigorous discussion would require consideration of the
spectra of the eigenoperators.  In Section \rnsix~ I show that
several candidate light-front renormalization group transformations lead
to unbounded operators, including unbounded irrelevant operators, even
though there are cutoffs.  In some of
these cases, corrections that are $\order(\delta H^2)$ are shown to be
infinite; \ie, not small.
If the coefficient of a single operator
(\eg, a single mass) becomes large, it may be straightforward to alter
the analysis so that this coefficient is included to all orders in an
approximation of the transformation, so that one perturbs only in the small
coefficients; but this possibility is not pursued in this paper.

For the purpose of illustration, let me assume that $\lambda_m=4$ for
all relevant operators, and $\lambda_m=1/4$ for all irrelevant
operators.
The transformation can be represented
by an infinite number of coupled, nonlinear difference equations:

$$\mu_{m_{l+1}}=4 \mu_{m_l} + N_{\mu_m}[\mu_{m_l}, g_{m_l}, w_{m_l}] \;,
\eqno(2.7)$$

$$g_{m_{l+1}}=g_{m_l} + N_{g_m}[\mu_{m_l}, g_{m_l}, w_{m_l}] \;,
\eqno(2.8)$$

$$w_{m_{l+1}}={1 \over 4} w_{m_l} + N_{w_m}[\mu_{m_l}, g_{m_l},
w_{m_l}] \;. \eqno(2.9)$$

\noindent
Sufficiently near a critical Gaussian fixed point,
the functions $N_{\mu_m}$,
$N_{g_m}$, and $N_{w_m}$ should be
adequately approximated by an expansion in
powers of $\mu_{m_l}$, $g_{m_l}$, and $w_{m_l}$.
The assumption that the Hamiltonian remains in the neighborhood of
the fixed point, so that all $\mu_{m_l}$, $g_{m_l}$, and $w_{m_{l}}$
remain
small must be justified {\it a posteriori}.  Any precise definition of
the neighborhood of the fixed point within which all approximations are
valid must also be provided {\it a posteriori}.

Wilson has given a general discussion of how these
equations are solved \APrefmark{\rWILTEN},
and I repeat only the most important points.
In perturbation theory these equations are
equivalent to an infinite number of coupled,
first-order, nonlinear differential equations.
To solve them
we must specify `boundary' values for
every variable,
possibly at different $l$, and then employ a {\it stable}
numerical algorithm to find the variables
at all other values of $l$ for which
the trajectory remains near the fixed point.
We want to apply the transformation $\calN$ times,
letting $\calN \rightarrow \infty$, and adjusting the initial Hamiltonian
so that this limit exists.
Eq. (2.7) must be solved by
`integrating' in the exponentially stable direction of decreasing $l$
(\ie, typically toward larger cutoffs), while Eq. (2.9) must be solved in the
direction of increasing $l$.  Eq. (2.8) is linearly unstable in either
direction.  The
coupled equations must be solved using an iterative algorithm.  Such
systems of coupled difference equations and the algorithms required for
their solution are familiar in numerical analysis.  In this context the
need for renormalization can be understood by considering the fact that
the renormalization group difference equations need to be solved over an
infinite number of scales in principle.

The final output of the renormalization group analysis is the cutoff
Hamiltonian $H_\calN$.  If this Hamiltonian is the final point in an
infinitely long trajectory of Hamiltonians, it will yield the
same observables below the final cutoff as $H_0$; but for an infinitely
long trajectory $H_0$ contains no cutoff, so $H_\calN$ {\it will yield
results that do not depend on the cutoff}.  It is for this reason that
$H_\calN$ and all other Hamiltonians on any infinitely long
trajectory are referred to
as {\it renormalized Hamiltonians}.  How one solves the final cutoff
Hamiltonian problem using $H_\calN$ depends on the theory.  For the
scalar theories used as examples in this paper I assume that
perturbation theory can be used to predict observables.  For QCD, even
if $H_\calN$ can be derived by purely perturbative techniques, it
will have to be solved nonperturbatively because of
confinement.  In either case, we must have an accurate
approximation for $H_\calN$; however, we do not necessarily need to
explicitly construct accurate approximations for all $H_l$.

The boundary values for the
irrelevant variables should be set at $l=0$, because we need to solve
Eq. (2.9) in the direction of increasing $l$.
At large $l$
all variables are exponentially insensitive to the irrelevant boundary
values.  Therefore, they
can be chosen arbitrarily (universality); and the values
of the irrelevant variables at $l=\calN$ are output by the modern
renormalization group.  This is one of the crucial differences between
the modern renormalization group and the Gell-Mann--Low renormalization
group in which irrelevant variables are not treated.  Irrelevant
operators are important in $H_\calN$ unless the final cutoff is much
larger than the scale of physical interest.  The fact that they are
irrelevant implies that their final values
are exponentially insensitive to their initial values; and it
implies that they are driven at an
exponential rate
toward a function of the relevant and marginal variables, as discussed
below.  The fact that they are irrelevant does not necessarily imply
that they are unimportant.  This depends on how sensitive the physical
observables in which one is interested are to physics near the scale
of the cutoff.

The boundary values required by Eqs. (2.7) and (2.8)
can be given at $l=\calN$.
Sufficiently far from $l=0$ the irrelevant variables are
exponentially driven to
maintain polynomial dependence on relevant and marginal variables, and
sufficiently far from $l=\calN$ the relevant variables are exponentially
driven toward similar polynomial dependence on the
marginal variables.  While the
calculation of transient behavior near $l=0$ and $l=\calN$ usually
requires a numerical computation, the relevant and irrelevant variables
are readily approximated by polynomials that involve only marginal
variables in the intermediate region.  These polynomials are
determined by the expansions of $N_{\mu_m}$, $N_{g_m}$,
and $N_{w_m}$; and they can
be fed back into Eq. (2.8) to find an
approximate equation for the marginal variables that requires direct
knowledge only of the marginal variables themselves.

These points may be confusing, so let me consider a simple example.
Consider coupled differential equations for a relevant variable $m$, a
marginal variable $g$, and an irrelevant variable $w$,

$${\partial m \over \partial t} = -2 m + c_1 g^2 + c_2 g w
\;,\eqno(2.10)$$

$${\partial g \over \partial t} = -c_3 g^3 + c_4 g^2 m + c_5 w^2
\;,\eqno(2.11)$$

$${\partial w \over \partial t} = 2 w + c_6 g^2 \;.\eqno(2.12)$$

\noindent  Here the cutoff increases as $t$ increases.  We want to fix
the boundary condition for $w$ at $t \rightarrow \infty$, and the
boundary condition for $m$ at $t=0$.  We are only interested in the
solution near $t=0$, and this should not depend on the boundary
condition for $w$; as long as $w(t)$ remains finite
as $t \rightarrow \infty$.  Let
$m(0)=m_0$, and $g(0)=g_0$.

To satisfy the boundary condition for $w$
we must have

$$w(t)= -{c_6 \over 2} g^2(t)+\order(cubic) \;,\eqno(2.13)$$

\noindent for all finite values of $t$, where all cubic and higher order
terms in $g$ and $m$ are readily computed.
We can substitute this result
in Eqs. (2.10) and (2.11) and we see that the irrelevant variable has no
effect to leading order in an expansion in powers of $g$.
For very large, but finite values of $t$, we
find

$$m(t)={c_1 \over 2} g^2(t)+\order(g^3) \;.\eqno(2.14)$$

\noindent Therefore, the marginal variable must satisfy

$${\partial g \over \partial t} = -c_3 g^3 + \order(g^4)
\;,\eqno(2.15)$$

\noindent for large but finite values of $t$.  Eq. (2.15) can now be
solved easily
to obtain an accurate approximation for $g(t)$ for large $t$,

$$g^2(t)={g^2(0) \over 1+2 c_3 g^2(0) t} \;. \eqno(2.16)$$

To
obtain an accurate approximation for small $t$, we can continue to use
Eq. (2.13), but we need to use an iterative algorithm to improve our
estimate of $g(t)$ and $m(t)$.  This is done by integrating Eqs.
(2.10)-(2.11) near $t=0$ repeatedly, using the estimates from one iteration
on the right-hand-sides of these equations to generate a subsequent
estimate.  This process is repeated until a convergence criterion is
met.  The initial seed is given by Eq. (2.13),
the solution of Eq. (2.15) for $g(t)$,
and

$$m(t)=m_0 e^{-2t}+{c_1 \over 2}
\Bigl(g^2(t)-g_0^2 e^{-2t}\Bigr)
\;,\eqno(2.17)$$

\noindent for example.  The transient behavior near $t=0$ in this
approximation of $m(t)$ is wrong, but any guess that is sufficiently
near the solution should lead to convergence.
After this iterative process converges, the
desired result is obtained.  In this case, the only output is $w(0)$,
because $m(0)$ and $g(0)$ are input.

In the simplest case there is only one marginal variable and a finite
number of relevant variables.  It is assumed that all variables are
small near the critical
Gaussian fixed point, and in particular it is assumed
that the marginal variable is small.  There are an infinite number of
irrelevant variables, but one can classify these variables according to
the eigenvalues in Eq. (2.5) and according to
their magnitude in terms of the marginal variable.  One can first
replace the irrelevant variables with an appropriate polynomial
involving marginal and relevant variables.  To leading order these are
given by the zeroes of the polynomials on the right-hand-side of Eq.
(2.9).
Substituting these results in Eq.
(2.7) one can next
determine the strength of each relevant variable in terms of the single
marginal variable.  The appropriate polynomials are given to leading
order by the zeroes of the polynomials on the right-hand-side of Eq.
(2.7).  After this one also has expansions for the irrelevant variables
in terms of the single marginal variable.
Every irrelevant variable has a leading term of
$\order(g^{p_m})$, where $g$ is the single marginal variable.  There are
typically a finite number of irrelevant variables with an eigenvalue
$\lambda_m$ from Eq. (2.5) greater than any given value and
with any given value
of $p_m$, and one can construct a first approximation by keeping only
the `leading' irrelevant operators.
In other words, one constructs a perturbative
approximation for the trajectory in which the order of perturbation
theory is determined by the single marginal variable $g$.  Generalization
of this procedure to the case where there is any finite number of
marginal and relevant variables is straightforward.

Next consider the case where there are an infinite number of relevant
variables, in addition to an infinite number of irrelevant variables; but
there are only a finite number of marginal variables.  The irrelevant
variables are treated exactly as above, except now the evolution of the
irrelevant variables may be sensitive to an infinite number of relevant
variables.
There are an infinite number of boundary values that must be specified
at $l=\calN$ in principle,
and an infinite number of polynomials that must be
considered in Eqs. (2.7) and (2.9).  One can simultaneously
classify all of the irrelevant and relevant operators
in terms of the
leading powers of the marginal variables that appear in the
zeroes of these polynomials, just as the irrelevant operators were
classified above.  To leading order,
one can replace the irrelevant and relevant variables
in the equations for the marginal variables using these zeroes, and one
can construct a perturbative approximation for the marginal variables
for $0 \ll  l \ll \calN$ by dropping sub-leading
irrelevant and relevant variables.  This part of the procedure is a
straightforward generalization of
the procedure used to handle an infinite number of
irrelevant variables; however, there is a crucial difference between
relevant and irrelevant variables.  As discussed above, the boundary
values chosen for the irrelevant variables are arbitrary.  The boundary
values chosen for the relevant variables do not affect the trajectory
for $l \ll \calN$, because their effects are exponentially suppressed as
$l$ decreases in Eq. (2.7); however, we need to construct $H_\calN$, and
the boundary values for the relevant variables may have important effects
on all irrelevant, marginal, and relevant variables
near $l=\calN$.

Remember that the boundary values of the relevant and marginal variables
are input, while the values of the irrelevant variables at $l=\calN$ are
output by the renormalization group equations.  If there are an infinite
number of relevant variables, we are forced to fix an infinite number of
boundary conditions.  Moreover, even if we want to compute a finite
number of `leading' irrelevant variables at $l=\calN$, in principle
we must
approximate the evolution of an infinite number of relevant variables
near $l=\calN$, because all of the relevant variables
affect the evolution of each irrelevant operator; and near $l=\calN$
transient behavior may prevent us from replacing the relevant variables
with functions of the marginal variables.
Does this render the
perturbative renormalization group useless?  Hopefully not, for two reasons.
First, there are many problems in
physics where an infinite number of boundary conditions must be fixed
(\eg, the value of a field on a surface).  There are also many problems
in which the evolution of an infinite number of variables must be
computed (\eg, the value of a field in a volume).  The key to the
solution to such problems is to show that it is possible to approximate
an infinite number of variables with a few well-chosen variables (\eg,
parameters in functions of the original variables). In the case of
light-front field theory we encounter an infinite number of relevant
operators because functions of longitudinal momenta appear in these
operators.  However, a
finite number of functions appear, and specifying the
infinite number of boundary conditions is accomplished by specifying
these functions.

Further consideration of the renormalization group equations reveals
a second reason that the perturbative renormalization group may survive,
even though there are an infinite number of relevant operators.
It may be natural for only a few of the relevant
variables to be `independent'.  Suppose that we independently fix the
boundary values of an infinite number of relevant variables and then
solve Eqs. (2.7)-(2.9).  Since we assume that $H_\calN$ is near the
Gaussian fixed point, all relevant variables are small at $l=\calN$, and
we expect all relevant variables to approach
polynomials of the marginal variables at an exponential rate.
It is only possible for these variables to deviate significantly from
these polynomials near the end of the trajectory, so on
any trajectory of renormalized Hamiltonians near a Gaussian fixed point
all
relevant variables are exponentially near these polynomials over an infinite
number of cutoff scales.  Thus, it is natural to speculate that almost all
of these variables track these polynomials exactly, never
deviating from this behavior.  If only a finite number of relevant
variables depart from these polynomials,
we can approximate the trajectory by dropping all sub-leading relevant
variables and numerically computing the behavior
only of those relevant variables that depart from these polynomials.
I call a relevant variable whose evolution is exactly given by
a polynomial a dependent relevant variable.  A
relevant variable whose value departs from such a polynomial is called an
independent relevant variable.
I simply postulate that there are only a finite number of
independent relevant variables in theories of physical interest (\eg,
Lorentz covariant theories).  The
boundary values of dependent relevant variables cannot
be adjusted independently because they are determined by the
polynomials; and this
places severe constraints on the theories that satisfy this postulate.
These are the coupling constant coherence
conditions, and Wilson and I have shown
that they arise
naturally when there is a hidden
symmetry \APrefmark{\rPERWIL}.

If there are an infinite number of marginal variables, in addition to an
infinite number of relevant and irrelevant variables, their behavior is
unpredictable without further assumptions.  I
postulate that there are a
finite number of independent marginal variables, and that the
infinite number of remaining
dependent marginal variables can each be replaced by
a polynomial expansion in powers of the independent marginal variables.

To see that these postulates are reasonable, one must try to understand
why light-front renormalization group transformations lead to an
infinite number of relevant and marginal variables.  We will see in
Section \rnthree~ that every cutoff that is run by a light-front
renormalization group transformation breaks Lorentz covariance, and
most violate cluster decomposition.
Moreover, in gauge theories these cutoffs violate gauge invariance.  The
price one pays for violating a continuous symmetry in this case
is the appearance of
an infinite number of relevant and marginal variables.  This price
should not be surprising, because these symmetries must be restored to
all physical observables and this requirement imposes an infinite number
of conditions on the renormalized Hamiltonian.  These conditions
relate the strengths of operators in the Hamiltonian, providing an
infinite number of relationships between the relevant and marginal
operators at $l=\calN$.  Thus, it is expected that the appearance of an
infinite number of relevant and marginal variables does not imply that
there are an infinite number of free parameters in the theory.  There
should be exactly as many independent relevant and marginal variables in
a light-front renormalization group analysis as there are in a Euclidean
renormalization group analysis, and the strength of all dependent
relevant and marginal variables should be fixed by the requirement that
the broken symmetries are restored.  These symmetries can be restored by
the adjustment of the strength of the dependent variables at the
boundary, and the requirement that the symmetries be maintained for all
other cutoffs places an infinite number of conditions on the
renormalization group.  These points are illustrated by examples
in Sections \rnsix~ and \rnseven.

In order to summarize the most important aspects of a perturbative
renormalization group analysis and clarify the difference between
perturbative and nonperturbative analyses,
I introduce Wilson's triangle of
renormalization, shown in figure 1.  The triangle displays a sequence of
renormalization group trajectories of increasing length, $\calN$.
We can label the Hamiltonian using a superscript to denote the
absolute cutoff and a subscript to denote the effective cutoff,
$H^{\Lambda_0}_{\Lambda_n}$.  Assume that the cutoff is a cutoff on
energy.  The object of the renormalization group is
to make it possible
to let $\Lambda_0 \rightarrow \infty$ while keeping
$\Lambda_\calN$ at some fixed value, say 2 GeV.  The
subscript $\calN$ indicates how many times the transformation must be
applied to the original Hamiltonian to lower the effective cutoff to its
final value, so one has $\Lambda_\calN = \Lambda_0/(2^\calN)$
for example.  To
fix $\Lambda_\calN$ and let $\Lambda_0 \rightarrow \infty$, we must also let
$\calN \rightarrow \infty$.

The renormalization group enables one to
compute renormalized Hamiltonians, shown as the
right-most column in figure 1,
by providing an operational definition of
renormalization.  In the perturbative renormalization group this task is
reduced to solving a finite number of difference or differential
equations, as shown above.
At each stage of a nonperturbative
renormalization group calculation
one selects a cutoff Hamiltonian $H^{\Lambda_0}_{\Lambda_0}$,
and applies the transformation, $T$,
$\calN$ times to generate the Hamiltonian
$H^{\Lambda_0}_{\Lambda_\calN}$.  In a
successive stage one selects a new Hamiltonian
$H^{\Lambda_0}_{\Lambda_0}$ and increments $\calN$ by one.
The sequence of initial Hamiltonians are related in a manner that
must be determined as part of an algorithm tailored to the specific
theory.  In a nonperturbative calculation one probably must construct
the triangle of Hamiltonians directly, being satisfied with numerical
evidence that the limiting trajectory of renormalized Hamiltonians
exists.

If $H^{\Lambda_0}_{\Lambda_0}$ lies near a Gaussian
fixed point we have seen that the irrelevant
variables in $H^{\Lambda_0}_{\Lambda_0}$ should have an exponentially
small effect on most of the trajectory.  We have also seen that we want
to fix the strength of some operators at the end of the trajectory, not
at the beginning.  Similar features should appear in a nonperturbative
analysis, but there is no general procedure to identify and order
operators in this case.
We may have to simply search on initial
values of parameters that are identified as important,
to find $H^{\Lambda_0}_{\Lambda_0}$ that yields
desired operators in $H^{\Lambda_0}_{\Lambda_\calN}$.
After
convincing oneself that an arbitrarily long trajectory can be constructed if
$H^{\Lambda_0}_{\Lambda_0}$ is adjusted to sufficient precision, in
practice one hopefully
needs to explicitly construct only the last part of a finite trajectory to
obtain an accurate approximation of a renormalized Hamiltonian,
$H^{\Lambda_0}_{\Lambda_\calN}$.

In light-front renormalization groups
there are an infinite number of relevant and marginal operators,
because undetermined functions of longitudinal momenta appear in a
finite number of operators that are relevant or marginal according to
transverse power counting.
In a perturbative analysis one can use
continuous symmetries such as Lorentz covariance to fix these functions
in each order of an expansion in terms of a single coupling; however, in
a nonperturbative analysis one may have to parameterize each
function and seek an approximation of the renormalization group
transformation that is represented by a set of coupled equations for the
evolution of these parameters.

Wilson has given an excellent discussion of the relationship between
divergences in standard perturbation theory (\eg, Feynman perturbation
theory) and the perturbative renormalization
group \APrefmark{\rWILTEN}, and I close this Section by repeating
the most salient points.  There are usually no
divergences encountered when one applies the renormalization group
transformation once; however, divergences can arise in
the form of powers of $l$ and exponents containing $l$,
when $T$ is applied a large number of times; and these divergences
are directly related to the divergences in Feynman perturbation theory.
There are no divergences apparent when one solves the perturbative
renormalization group equations using a stable numerical algorithm;
however, if one attempts to expand a coupling $g_l$ in powers of $g_0$,
for example, powers of $l$ appear.  As $l \rightarrow \infty$ these
powers of $l$ lead to the divergences familiar in Feynman perturbation
theory.  One can see an example of this by studying Eq. (2.16), where
$t$ represents the logarithm of a ratio of cutoffs.  If the right-hand
side of Eq. (2.16) is expanded in powers of $g(0)$, each term
diverges like a power of $t$; so this expansion is numerically useless.
On the other hand, $g(t)$ is perfectly well-behaved if $g(0)$ is small,
and the divergences result from the fact that $g(t)$ and $g(0)$ differ
by orders-of-magnitude for sufficiently large $t$.

Eq. (2.16) illustrates the significant improvement over standard
perturbation theory offered by the perturbative renormalization group.
In standard
perturbation theory one deals only with bare and renormalized
parameters.  This is analogous to dealing only with the
parameters in $H^{\Lambda_0}_{\Lambda_0}$ and
$H^{\Lambda_0}_{\Lambda_\calN}$, without ever encountering separate
parameters for every Hamiltonian in the trajectory.  Except in
super-renormalizable theories, the
ratio of bare and renormalized parameters goes to
infinity (or zero). If a perturbative expansion of an observable in
powers of the renormalized parameters converges, the expansion for the
same observable in terms of the bare parameters cannot converge.  This
leads to some interesting departures from logic in standard perturbation
theory \APrefmark{\rDIRTHR}.
A small contribution of the renormalization group is that logic
may sometimes be restored to perturbation theory.


\bigskip
\noindent {\bf \rnthree. Light-Front Hamiltonians and Renormalization Group
Transformations}
\medskip

The first step in defining a renormalization group
transformation is to define the space of Hamiltonians upon which
this operator acts.  I give no precise definition of this
space, partially because it must usually be defined after studying a
transformation, not before.
I restrict myself to scalar field theory, as it is
straightforward but tedious to generalize the
discussion to theories that include more complicated fields.  I
indicate what kind of operators are allowed in the Hamiltonians by
example, and I display these operators in several forms that prove
useful.  Ultimately the Hamiltonians must be expressed in terms of
Fock space eigenstates of the Gaussian fixed point Hamiltonian (\ie,
in terms of projection operators) if one
wants to use an invariant-mass transformation, so much of the early
discussion is schematic.
A brief summary of canonical light-front scalar field theory is
given in Appendix A.

The operators and constants with which Hamiltonians can be formed in a 3+1
dimensional scalar field theory, and their naive engineering dimensions, are

$$ \partial^+ = \biggl[{1 \over x^-}\biggr] \;,\;
\partial^\perp = \biggl[{1 \over x^\perp}\biggr] \;,\;
\Lambda = \biggl[{1 \over x^\perp}\biggr] \;,\;
\epsilon= \biggl[{1 \over x^-}\biggr] \;,\;
\phi(x) = \biggl[{1 \over x^\perp}\biggr] \;.  \eqno(3.1)
$$

\noindent  I should also note that the Fourier transform of the field
operator, $\phi(q)$, has the dimension $\bigl[x^\perp\bigr]$.
I work with a metric in which $x^\pm=x^0 \pm x^3$.  In addition to
derivative operators and the scalar field operator, I indicate
that there may be a
cutoff with the dimension of transverse momentum ($\Lambda$)
that can be
used, and there may be a cutoff with the dimension of
longitudinal momentum ($\epsilon$)
that can be
used.  All masses are expressed as dimensionless constants
multiplying $\Lambda$.
In general there may be
many cutoffs (\eg, different cutoffs in different sectors of Fock
space), but all of them can be expressed in terms of $\Lambda$ and
$\epsilon$.

Perhaps the most important feature of Eq. (3.1) is that
transverse and longitudinal dimensions are treated
separately \APrefmark{\rWILQCD}, just as
one treats time and space differently in nonrelativistic physics.
There is no analog of physical mass with the dimensions of
longitudinal momentum instead of transverse momentum, because
longitudinal boost invariance is a scale
invariance \APrefmark{\rKOGTWO,\rLEU}, and
physical masses (not necessarily bare masses) violate
scale invariance.  The cutoff $\epsilon$ is the only
constant with the dimensions of longitudinal momentum that can enter
the definition of the Hamiltonian, and it must enter in
a manner that restores boost invariance to observables
despite the violation of explicit boost invariance caused by the cutoff
itself.  In general one cannot be
sure that naive engineering dimensions are significant in an
interacting theory; however, near a Gaussian fixed point naive power
counting is appropriate for the same reasons it is appropriate in
standard perturbation theory.  This is explicitly
shown in Section \rnfive~ for light-front
transformations.

The assumption of transverse locality naively
means that no inverse powers of
$\partial^\perp$ are allowed.  Restrictions on inverse powers of
$\partial^\perp$ are clarified in Section \rnsix~ where they first appear
in the second-order behavior of the light-front transformations.

Masses appear in the Hamiltonian as dimensionless constants multiplying
$\Lambda$.
I always assume that physical masses are
much smaller than $\Lambda$, and I make
an operational
distinction between physical masses and mass counterterms.  Mass
counterterms are present
even when the physical mass is zero unless a symmetry protects the mass
operator, and I know of no examples in cutoff
light-front field theory where
this occurs.  Mass counterterms can have a very different
dependence on longitudinal momenta than physical mass terms, as shown
in Section \rnsix.  Except in Appendix C I usually assume that the
physical mass is zero and focus on the critical theory.

I make no initial restriction on the manner in which longitudinal
derivatives appear.  In canonical scalar field theory longitudinal
derivatives appear only as inverse powers (see Appendix A); however,
in canonical light-front QCD in light-cone gauge \APrefmark{\rTOM,\rCASH} one
finds both inverse powers and powers
of the longitudinal derivative in the three-gluon vertex
and in the exchange
of an instantaneous gluon between a quark and gluon or between two
gluons.

The Hamiltonian, $H$, is the integral of the Hamiltonian
density, $H = \int dx^- d^2x^\perp \ham$.  Their dimensions are
easily derived in canonical free field theory, and here I simply
take them as given to be

$$ H = \biggl[{x^- \over x^{\perp 2}}\biggr] \;,\;\;\;
\ham = \biggl[{1 \over x^{\perp 4}}\biggr] \;. \eqno(3.2) $$

\noindent
Given the catalog of operators from which the Hamiltonian can be formed,
the space of Hamiltonians that I initially
consider consists of all operators
that can be formed from the basic catalog and that have the appropriate
engineering dimension.
Inverse powers of the transverse derivative operator are excluded
initially
and inverse powers of the field
operator are always forbidden.
Furthermore, cutoffs must be
imposed to complete the definition.

I work in momentum space rather than position space,
and the Hamiltonian can be written schematically as

$$\eqalign{ H = &\;\;\;\; {1 \over 2} \int \dqt_1 \; \dqt_2 \; (16 \pi^3)
\delta^3(q_1+q_2) \;
u_2(q_1,q_2) \;
\phi(q_1) \phi(q_2) \cr
&+{1 \over 4!}
\int \dqt_1\; \dqt_2\; \dqt_3\; \dqt_4 \; (16 \pi^3)
\delta^3(q_1+q_2+q_3+q_4) \cr
&\qquad\qquad\qquad\qquad u_4(q_1,q_2,q_3,q_4)\; \phi(q_1)
\phi(q_2) \phi(q_3) \phi(q_4)  \cr
&+{1 \over 6!}
\int \dqt_1 \cdot \cdot \cdot \dqt_6\; (16 \pi^3)
\delta^3(q_1+\cdot\cdot\cdot+q_6) \; u_6(q_1,...,q_6)
\; \phi(q_1) \cdot
\cdot \cdot \phi(q_6) \cr
&+\qquad \cdot \cdot \cdot
\;,}\eqno(3.3)
$$

\noindent where,

$$\dqt = {dq^+ d^2q^\perp \over 16 \pi^3 q^+} \;. \eqno(3.4) $$

\noindent
We will see that this leads to
a free energy $(\qp^2+m^2)/q^+$ when $u_2=\qp^2+m^2$.
In terms of plane wave creation and
annihilation operators,

$$\phi(q)=a(q) \;\;\;(q^+>0)\;; \;\;\;\;\;\;\;\;\;\phi(q)=-a^\dagger(-q)
\;\;\;(q^+<0) \;.  \eqno(3.5)$$

The above restrictions on transverse derivatives
become restrictions on the
functions $u_2, u_4$, etc. At this point the integrals include both
positive and negative longitudinal momenta.  The next step toward an
expression for the Hamiltonian that can be directly manipulated involves
replacing the field operators in Eq. (3.3)
with creation and annihilation operators, normal-ordering the
Hamiltonian and changing variables so that only positive longitudinal
momenta appear.  There are no modes with zero longitudinal momentum.
This complicates the Hamiltonian algebraically,
but the advantages far outweigh this complication.  I
should mention that there is no need to define the normal-ordering
operation until after the cutoffs required by the transformation are
implemented, and that after this no divergences are encountered in the
normal-ordering procedure.  The initial Hamiltonian is simply assumed to
be normal-ordered.
The schematic Hamiltonian becomes

$$\eqalign{ H = &\qquad  \int \dqt_1 \; \dqt_2 \; (16 \pi^3)
\delta^3(q_1-q_2) \; u_2(-q_1,q_2)
\;a^\dagger(q_1) a(q_2) \cr
&+{1 \over 6} \int \dqt_1\; \dqt_2\; \dqt_3\; \dqt_4 \; (16 \pi^3)
\delta^3(q_1+q_2+q_3-q_4) \cr
&\qquad\qquad\qquad\qquad\qquad
u_4(-q_1,-q_2,-q_3,q_4)\; a^\dagger(q_1) a^\dagger(q_2)
a^\dagger(q_3)
a(q_4)  \cr
&+{1 \over 4} \int \dqt_1\; \dqt_2\; \dqt_3\; \dqt_4 \; (16 \pi^3)
\delta^3(q_1+q_2-q_3-q_4) \cr
&\qquad\qquad\qquad\qquad\qquad u_4(-q_1,-q_2,q_3,q_4)\; a^\dagger(q_1)
a^\dagger(q_2) a(q_3) a(q_4)  \cr
&+{1 \over 6} \int \dqt_1\; \dqt_2\; \dqt_3\; \dqt_4 \; (16 \pi^3)
\delta^3(q_1-q_2-q_3-q_4) \cr
&\qquad\qquad\qquad\qquad\qquad u_4(-q_1,q_2,q_3,q_4)\; a^\dagger(q_1)
a(q_2) a(q_3) a(q_4)  \cr
&+ \qquad \cdot \cdot \cdot
\;.}\eqno(3.6)
$$

Since modes with identically zero longitudinal momentum are not
allowed, there are no operators in this light-front Hamiltonian that
contain only creation operators or only annihilation operators.  This is
only natural in light-front field theory.
{}From this point forward it is assumed that all
longitudinal momenta are positive.  The Hamiltonian is
simply assumed to be normal-ordered, and it is assumed as part of the
definition of each transformation that the transformed Hamiltonian is
normal-ordered.  This is readily insured in perturbation theory, where
one only encounters products of operators that are easily
normal-ordered, as shown below.

When one studies a boost-invariant renormalization group transformation,
the Hamiltonians must be written in terms of projection and transition
operators that are constructed from the free many-body states generated
by products of the above
creation and annihilation operators acting on the vacuum.  This
introduces severe notational complications,  but such representations
are trivially constructed and manipulated.  All
boost-invariant cutoffs involve the total longitudinal
and transverse momenta of a {\it state}, and not simply a few momenta
carried by individual particles.  When the cutoff depends on extensive
quantities such as the total momentum,
spectator-dependence inevitably appears
in the operators, as we will find
in Section \rnsix.  Such spectator-dependence has been studied recently
in LFTD \APrefmark{\rPERONE-\rGHPSW},
but I want to emphasize that it is required even if one does
not make a Tamm-Dancoff truncation.
Since interactions depend on
spectators, they also depend on the Fock space sector(s) in or between
which they act \APrefmark{\rPERONE,\rPERTWO,\rGLAZTHR}.
In this case, one might worry that the distinction
between the functions $u_2$, $u_4$, etc. might become altered or
blurred.  However, these distinctions are easily maintained by
considering how many individual particle momenta are altered by the
operator.  Examples below should clarify these points.

In order to develop simple examples of the
light-front renormalization group
one can use a Tamm-Dancoff truncation \APrefmark{\rTAM,\rDAN},
which simply limits the number of particles and introduces an additional
source of sector-dependence in the operators \APrefmark{\rPERONE}.
In LFTD the Hamiltonian again must
be written in terms of projection operators \APrefmark{\rGLAZTHR}.
Let me emphasize that some
of the most interesting results one can find in a renormalization group
analysis are apparently lost when a Tamm-Dancoff truncation is made, and
I do not use LFTD in any examples in this paper.  However, to
illustrate what a Hamiltonian written in terms of projection
and transition operators looks like, let me truncate Fock space to allow
only the one-boson and two-boson
sectors.  In this case the complete Hamiltonian is

$$\eqalign{ H = &\;\; \int \dqt\; {u_2^{(1)}(q)
\over q^+}\;
|q\rangle \langle q|  \cr
&+\int \dkt_1\;\dkt_2\; \Biggl[
{u_2^{(2)}(k_1) \over k_1^+}\;+\;
{u_2^{(2)}(k_2) \over k_2^+} \Biggr]
\;|k_1,k_2\rangle \langle k_1,k_2|  \cr
&+\;{1 \over 3}\;\int \dqt \; \dkt \; \Biggl[
{u_3^{(1,2)}(q,-k,k-q) \over q^+-k^+}\; |q\rangle \langle k,q-k|
\;+\;{u_3^{(2,1)}(k,q-k,-q) \over q^+-k^+} |k,q-k \rangle \langle q|
\Biggr] \cr
&+\;{1 \over 4}\;\int \dkt_1\; \dkt_2\; \dkt_3\; \Biggl[
{u_4^{(2,2)}(k_1,k_2,-k_3,-k_1-k_2+k_3) \over k_1^+ + k_2^+ - k_3^+}
\Biggr] \; |k_1,k_2 \rangle \langle k_3,k_1+k_2-k_3|
\;.}\eqno(3.7)
$$

\noindent Note that $|k_1,k_2\rangle = a^\dagger(k_1) a^\dagger(k_2)|0
\rangle$, etc. More complicated examples are readily constructed.
A superscript is added to $u_2$, $u_3$, etc. to indicate the
Fock space sector(s) within which or between which
the operator acts.  In this example only the superscript on $u_2$ is
required.  The Hamiltonians that are actually manipulated in the
light-front renormalization group are similar to this last expression,
not the expressions in Eqs. (3.3) and (3.6).  However, to see the
connection with the more familiar canonical
formalism, it is necessary to start with
the latter expressions.



As mentioned above, I
have simply dropped all terms that explicitly involve zero
modes only; \ie, terms containing creation operators only or
annihilation operators only.  The motivation for dropping such terms was
briefly discussed in the Introduction, but here I simply assume this
restriction is placed on the space of Hamiltonians \abinit.  I do not
question whether this assumption is reasonable and I do not believe that
it is possible to answer such questions with a perturbative
renormalization group analysis.  The only issue at present is whether
this restriction can be maintained or whether the transformations
themselves regenerate pure creation or pure annihilation operators.
To insure that this does not occur I simply drop the zero modes in every
term in the Hamiltonian, and assume that this does not affect any
momentum integral because a set of measure zero is being subtracted.  The
zero modes are automatically removed by several of the cutoffs studied
below, and there is no need to be careful about how the zero modes are
removed in any of the examples considered in this paper.  This may not
be the case in QCD, where severe divergences associated with small
longitudinal momenta force one to be more careful.  I
have nothing more to say about this issue until the Conclusion.

Up to this point the operators act in an infinite dimensional Fock
space unless otherwise specified,
and momenta are left unconstrained.  The definition of the space of
Hamiltonians is not complete until the momentum cutoffs that enter the
renormalization group transformation are added.
Let me begin by introducing the two simplest
light-front renormalization group
transformations, $T^\perp$ and $T^+$.

In the renormalization group one
starts with degrees of freedom that are already
restricted by momentum cutoffs.  There is never any point at which one
explicitly considers Hamiltonians that contain no cutoffs, although
renormalized Hamiltonians are obtained by considering what happens when
the initial cutoff is taken to its limit.
After introducing a cutoff one removes additional degrees of
freedom and computes an effective Hamiltonian that is required, for
example, to reproduce all of the low-lying eigenvalues and `accurately'
approximate the low-lying eigenstates.  This step is radically different
in the light-front renormalization group
from the integration over large momentum components in a
path integral \APrefmark{\rWILNINE} employed
in a Euclidean renormalization group, and details are not
discussed until Section \rnfour.  The final step in a light-front
renormalization group
transformation is to rescale all momenta so that
their original
numerical range is restored, to rescale the field operators, and to
rescale the Hamiltonian itself.
At this point the Hamiltonian is
identical in form to the original Hamiltonian, but the functions $u_2,
u_4$, etc. may all change.  The aim of the analysis is to understand how
these functions change.  In general the analysis is most useful if one
can reduce the task of following these functions to the infinitely
simpler task of following a few constants.  Naturally, the most
difficult step is demonstrating that this simplification is a legitimate
approximation.

For the light-front renormalization group
transformation $T^\perp$ one begins by cutting off all
transverse momenta, so that

$$0 \le \pp^2 \le \Lambda_0^2 \;. \eqno(3.8)$$

\noindent One then removes any state in which one or more momenta lie in
the range

$$\Lambda_1^2 < \pp^2 \le \Lambda_0^2 \;. \eqno(3.9)$$

\noindent Typically $\Lambda_1=\Lambda_0/2$ or $\Lambda_1$ differs by an
infinitesimal amount from $\Lambda_0$.  In most examples in this paper I
use $\Lambda_1=\Lambda_0/2$.
A new Hamiltonian must be found
that is able to reproduce the eigenvalues and approximate the
eigenstates of the original Hamiltonian without the degrees of freedom
that have been removed.  Techniques for computing such `effective'
Hamiltonians are discussed in the next Section.
This new Hamiltonian must be written in the same form in which the
initial Hamiltonian is written, and in particular it cannot include any
explicit energy dependence.  Let the initial Hamiltonian be called
$H_{\Lambda_0}$, with the new effective Hamiltonian being $H_{\Lambda_1}$.

The renormalization group transformation $T^\perp$ is completed by
changing variables to

$$\pp' = {\Lambda_0 \over \Lambda_1} \pp \;,
\eqno(3.10)$$

\noindent scaling the field operator $\phi$ by a constant so that

$$ \phi(\pp,p^+) = \zeta^\perp \phi'(\pp',p^+) \;, \eqno(3.11)$$

\noindent and multiplying the entire Hamiltonian by a constant
$z^\perp$.  The constants $\zeta^\perp$ and $z^\perp$ are introduced so
that fixed points can exist.  They are part of the definition of the
transformation and can be chosen freely.  The only practical restriction
on the choice of $\zeta^\perp$ and $z^\perp$ is that the choice should
lead to a fixed point of physical interest.
$z^\perp$ is not essential, but it
simplifies the task of comparing Hamiltonians, because it allows the
range of eigenvalues of the transformed Hamiltonian to be the same as
the range of eigenvalues of the initial Hamiltonian.  The price paid for
introducing $z^\perp$ is that one must multiply the Hamiltonian by
$1/z^\perp$ to obtain eigenvalues in the original units.  If $\calN$
transformations are made, one must multiply the resultant Hamiltonian by
$(1/z^\perp)^\calN$.  This point is clarified by examples below.
The cutoffs in the original Hamiltonian introduce step
functions into the momentum integrals in Eq. (3.3), for example.  The
elimination of the degrees of freedom with momenta in Eq. (3.9) leads to
a new set of step functions, and the rescaling leads back to the
original step functions.  Each term in $H_{\Lambda_1}$ changes in a
simple manner as a result of these various rescaling operations, and the
final result is the transformed Hamiltonian, $T^\perp[H]$.  The only
difficult step is constructing $H_{\Lambda_1}$.

I do not discuss any examples in detail until Section \rnfive, where
Gaussian fixed points and linear approximations of the
transformations are constructed.
However, to orient the reader I consider the simple
Hamiltonian needed to find the Gaussian fixed point of $T^\perp$ in
Section \rnfive.  If there are no interactions the Hamiltonian is

$$\eqalign{ H_{\Lambda_0} = &\;\; \int \dqt \;
\theta(\Lambda_0^2-\qp^2)
u_2(q) \;
{a^\dagger(q) a(q) \over q^+}
\;,} \eqno(3.12)$$

\noindent and when the cutoff changes the effective Hamiltonian is
simply

$$\eqalign{ H_{\Lambda_1} = &\;\; \int \dqt \;
\theta(\Lambda_1^2-\qp^2)
u_2(q) \;
{a^\dagger(q) a(q) \over q^+}
\;.} \eqno(3.13)$$

\noindent  The complicated procedure for producing effective
Hamiltonians developed in Section \rnfour~ is not needed when there are no
interactions, because the original Hamiltonian projected onto the
subspace of allowed states exactly reproduces all of the
eigenvalues and eigenstates in this subspace.
Next we change variables according to Eqs. (3.10) and (3.11), and
multiply the Hamiltonian by a constant.  The final transformed
Hamiltonian is

$$\eqalign{ T^\perp[H_{\Lambda_0}] = \;\;  z^\perp \zeta^{\perp\;2}
\Bigl({\Lambda_1 \over \Lambda_0}\Bigr)^2 &\int \dqt \;
\theta(\Lambda_0^2-\qp^2)
\; u_2({\Lambda_1 \over \Lambda_0} \qp,q^+) \;
{a^\dagger(q) a(q) \over q^+}
\; .} \eqno(3.14)$$

\noindent Naturally, when interactions are included the transformation
is far more complicated.  In fact, we will see that perturbation theory
generally cannot be used to approximate $T^\perp[H]$.

For the light-front renormalization group
transformation $T^+$, one begins by cutting off all
longitudinal momenta, so that

$$\epsilon_0 \le p^+ < \infty \;. \eqno(3.15)$$

\noindent  It is immediately evident that this transformation is going
to be radically different from the transformation on transverse momenta
or any transformation considered in Euclidean field theory, because small
momenta are removed instead of large momenta.  This is an appropriate
cutoff because the energy is $(\pp^2+m^2)/p^+$ near the Gaussian fixed
point of interest.
States with
small longitudinal momenta correspond to high energy states unless
both $\pp$ and $m$ vanish, leading to Eq. (3.15).

The first step in the transformation is to remove all states in which
one or more particle momenta lie in the range

$$\epsilon_0 \le p^+ < \epsilon_1 \;. \eqno(3.16)$$

\noindent Typically $\epsilon_1=2 \epsilon_0$ or $\epsilon_1$ differs by
an infinitesimal amount from $\epsilon_0$.  Again a new Hamiltonian must
be found that reproduces the original low-lying eigenvalues and
approximates the low-lying eigenstates of the original Hamiltonian
without the degrees of freedom that have been removed.  The
procedures for computing such Hamiltonians are identical to those used
for $T^\perp$.  However,
because longitudinal boosts are scale transformations, as seen in Eq.
(3.19) below, it is possible to
say a great deal about the transformed Hamiltonian without going through
an explicit construction.  In Section \rnfive~ I prove that physical
Hamiltonians are fixed points of the transformation $T^+$.

The light-front renormalization group
transformation $T^+$ is completed by changing variables to

$$p^{+'} = {\epsilon_0 \over \epsilon_1} p^+ \;, \eqno(3.17)$$

\noindent scaling the field operator $\phi$ by a constant so that

$$\phi(\pp,p^+) = \zeta^+ \phi'(\pp,p^{+'}) \;, \eqno(3.18)$$

\noindent and multiplying the entire Hamiltonian by a constant $z^+$.
Again the constants $\zeta^+$ and $z^+$ are introduced so that a fixed
point can exist.  I do not discuss
$T^+$ further in this Section, but perturbation
theory generally cannot be used to approximate $T^+$ either.

At this point a general procedure for inventing light-front
renormalization group transformations
should be apparent, although many important details may be obscure.
I close this Section by discussing boost-invariant light-front
renormalization group
transformations.  Both $T^\perp$ and $T^+$ break manifest
boost invariance because
they employ cutoffs that are not boost-invariant.  Under a longitudinal
boost the Hamiltonian and all particle momenta are
transformed \APrefmark{\rKOGTWO,\rLEU},

$$p^+ \rightarrow e^\nu p^+ \;, \;\; \pp \rightarrow \pp \;, \;\; H
\rightarrow e^{-\nu} H \;,
\eqno(3.19)$$

\noindent and under a transverse boost,

$$p^+ \rightarrow p^+ \;, \;\; \pp \rightarrow \pp + p^+ \vp \;,
\;\; H \rightarrow H + 2 {\cal \bf P}^\perp \cdot \vp + {\cal P}^+
\vp^2 \;.
\eqno(3.20)$$

\noindent In Eq. (3.19) $\nu$ is an arbitrary real number, and in
Eq. (3.20) $\vp$ is an arbitrary velocity, while ${\cal P}^+$ is the total
longitudinal momentum and ${\cal \bf P}^\perp$ is the total
perpendicular momentum.  Since the cutoffs are not changed by these
transformations, these transformations place severe restrictions on the
possible form of physical Hamiltonians that are not easily satisfied.
Let me mention that this is not the chief problem one encounters when
studying $T^\perp$ and $T^+$.  The chief problem, as mentioned above, is
that no perturbative expansion exists for these transformations in
general.  When a
perturbative expansion exists for a transformation, it should be
possible to implement Lorentz covariance order-by-order in
perturbation theory.

Physical Hamiltonians must transform according to Eqs. (3.19) and
(3.20), and if possible one would like to build these constraints into the
space of Hamiltonians {\it ab initio} and construct transformations that
automatically maintain these constraints.  Boosts are part of the
kinematic subgroup of Poincar{\'e} transformations in light-front field
theory, and this should allow one to make the kinematic invariances
manifest by choosing the correct variables.  This kinematic subgroup is
isomorphic to the two-dimensional Galilean group,
and the use of appropriate variables
resembles the separation of center-of-mass motion in
nonrelativistic quantum mechanics \APrefmark{\rKOGTWO,\rLEU}.  The appropriate
variables are

$$x={p^+ \over {\cal P}^+} \;, \;\; {\bf \kappa}^\perp = \pp - x {\cal
\bf P}^\perp \;. \eqno(3.21)$$

\noindent One can easily verify that $x$ and ${\bf \kappa}^\perp$ are
invariant under the above boosts.  Therefore any cutoff constructed from
these variables does not interfere with our ability to maintain manifest
boost invariance.  In particular, if we use a cutoff formed from these
variables we may simply fix ${\cal P}^+$ and choose ${\cal \bf
P}^\perp=0$
as part of the definition of the space of Hamiltonians, without loss of
generality.

An obvious feature of such cutoffs is their use of the total momenta
${\cal P}^+$ and ${\cal \bf P}^\perp$, which are themselves extensive
quantities.  The momentum available to a system in one room, which
determines what free states can be superposed to form physical states,
may depend on
the state of a system in the next room; and this may introduce
nonlocalities into the Hamiltonian when it is constrained to produce
exact physical results despite such nonlocal constraints on phase space.
This could be a very severe price to pay for manifest boost invariance, but it
does not seem likely that one can avoid paying this price without
sacrificing the possibility of using perturbation theory if cutoffs that
remove degrees of freedom are used, as is shown in
Section \rnsix.  The important
point for a perturbative
renormalization group analysis is to show that such cutoff
nonlocalities introduce new irrelevant, marginal, and relevant
operators that are readily computed in perturbation theory.
It is essential to show that cutoff
nonlocalities do not lead to inverse powers of transverse momenta that
produce long-range interactions.  We do encounter inverse powers of
transverse momenta that do {\it not} produce long-range interactions because
of the cutoffs that accompany them.

There are many boost-invariant renormalization group transformations one
can introduce, and I introduce three.  All of these transformations
are identical in form to $T^\perp$ and $T^+$, so in each case it is
sufficient to specify the cutoffs.  There is no reason to consider a
cutoff on longitudinal momentum fractions alone,
because if we change such a cutoff we cannot rescale momenta to recover
their original range.  When individual longitudinal momenta are
rescaled, the total longitudinal momentum is rescaled by the same
amount, and the longitudinal momentum fractions are invariant.  In other
words, longitudinal boost invariance is a scale invariance.  The first
cutoff one might consider starts by cutting off the relative transverse
momenta defined in Eq. (3.21), so that

$$0 \le {\bf \kappa^\perp}^2 \le \Lambda_0^2 \;. \eqno(3.22)$$

\noindent
One then proceeds by lowering this cutoff, computing a new Hamiltonian,
and then by rescaling the transverse momenta exactly as in Eq. (3.10).
This cutoff should violate locality in a minimal fashion, but we shall
see that perturbation theory again cannot be used to approximate this
transformation.

The second boost-invariant transformation begins with the introduction
of the cutoff

$$0 \le \sum_i {\pp_i^2 \over p_i^+} \le {{\cal \bf
P^\perp}^2+\Lambda_0^2 \over {\cal P}^+} \;. \eqno(3.23)$$

\noindent
By expanding this sum one easily demonstrates that this cutoff acts only
on the relative transverse momenta defined in Eq. (3.21).  This cutoff
is the invariant-mass cutoff for massless theories, and again the remaining
steps in the transformation involve lowering the cutoff, computing a new
Hamiltonian and rescaling transverse momenta and fields.  We will find
that this transformation can be approximated perturbatively if there are
no physical
mass terms in $u_2$.  A variety of `mass'
terms arise in light-front field theory, because any function of
longitudinal momentum fractions
can accompany a mass.  {\it Physical mass terms} appear
in $u_2$ with no additional function of longitudinal momenta; \eg,
$u_2=\pp^2+m^2$ results
in a free energy of the form $(\pp^2+m^2)/p^+$.  We will find in Section
\rnsix~ that the invariant-mass transformation leads to mass
counterterms that appear in $u_2$ in the form $g^2 \Lambda_0^2\; p^+/\calp^+$,
where $g$ is a coupling constant and
$\calp^+$ is the total longitudinal momentum of a state.  This
type of mass counterterm can be treated perturbatively, whereas a
perturbative treatment of the physical mass terms leads to divergent
longitudinal momentum integrals.

The only transformation that I have been able to approximate
using perturbation
theory when there are physical mass terms in $u_2$ begins with the cutoff

$$0 \le \sum_i {\pp_i^2 + m_i^2 \over p_i^+} \le {{\cal \bf
P^\perp}^2+\Lambda_0^2 \over {\cal P}^+} \;. \eqno(3.24)$$

\noindent  Here the cutoff masses $m_i$ that appear must be specified, and the
appropriate values can only be chosen after an analysis of the
transformation.  In Wilson's triangle of renormalization discussed in
Section \rntwo, one must consider the limit in which the initial cutoff
$\Lambda_0 \rightarrow \infty$.  When taking this limit one should
consider the masses, $m_i$, to be fixed.  Clearly, if this limit can
actually be taken, the specific values of the cutoff
masses should enter
primarily in the justification of perturbation theory and
are adjusted for that purpose.

I go through each step of this transformation, because cutoff masses
introduce significant new features.  The first step in the
transformation is to remove all states in which

$$ {{\cal \bf P^\perp}^2+\Lambda_1^2 \over {\cal P}^+} < \sum_i
{\pp_i^2 + m_i^2 \over p_i^+} \le {{\cal \bf P^\perp}^2+\Lambda_0^2
\over {\cal P}^+} \;. \eqno(3.25)$$

\noindent  After computing the effective Hamiltonian one rescales all
transverse momenta by changing variables as in Eq. (3.10).  However,
after this change of variables the states satisfy the constraint

$$0 \le \sum_i {\pp_i^2 + (\Lambda_0/\Lambda_1)^2 m_i^2 \over p_i^+} \le
{{\cal \bf P^\perp}^2+\Lambda_0^2 \over {\cal P}^+} \;. \eqno(3.26)$$

The momenta do not have the same range after rescaling as they initially
had, because the cutoff masses change.  We will see that masses in the
Hamiltonian rescale in exactly this manner.  If one applies the
transformation a large number of times, eventually the factor
$(\Lambda_0/\Lambda_1)^2 m_i^2$ becomes large and all of phase space is
eliminated by the transformation.  Clearly the transformation must be
highly nonperturbative at this point; however, there is no reason to
believe that one can ever lower the cutoff on transverse momenta to a
scale comparable to physical mass scales in the problem without the
transformation becoming highly nonperturbative.  One should consider the
cutoff masses to be of the same scale as physical masses.  This
transformation still defines a semi-group, and I show in Appendix C that
all of the basic features of the renormalization group apparently
survive when one
allows such a transformation.  In particular, one can define the
functions $u_2$, $u_4$, etc. independently of the cutoffs, and study their
evolution as analytic functions of their arguments over the original range
of momenta.  This discussion is clarified by examples in Appendix C.


\bigskip
\noindent {\bf \rnfour. Perturbation Formulae for the Effective Hamiltonian}
\medskip

I discuss two related methods
for computing the effective Hamiltonian after the cutoff is changed,
one developed originally by Bloch and
Horowitz \APrefmark{\rBLOONE}
and a second developed by Bloch \APrefmark{\rBLOTWO}.  I call the first
Hamiltonian a Bloch-Horowitz Hamiltonian and the second simply a Bloch
Hamiltonian.
The Bloch Hamiltonian was used
by Wilson in early work on the renormalization group \APrefmark{\rWILTWO}.
More sophisticated
methods must be employed if one wants to work far from the
Gaussian fixed point, but I concentrate only on the development of
the effective Hamiltonian in perturbation theory.

The primary
requirement is that the effective Hamiltonian produce the same low-lying
eigenvalues in the second cutoff subspace that the initial Hamiltonian
produces in the initial cutoff subspace.
A second requirement is that the eigenstates of the
effective Hamiltonian be orthonormalized projections of the
original eigenstates.  The Bloch Hamiltonian is designed to satisfy these
requirements, while the Bloch-Horowitz Hamiltonian is not.  One
of the primary reasons that one wants to preserve the eigenstates in
addition to the eigenvalues is to compute
measurable quantities
in addition to the energy.  Each measurable quantity corresponds to
the absolute value of a
matrix element, and when one introduces cutoffs all operators must be
properly renormalized so that their matrix elements in the cutoff space
are independent of the cutoffs.  An intermediate step in the
construction of the Bloch Hamiltonian is the construction of an
operator ${\cal R}$.  ${\cal R}$ can be used to renormalize all
observables.

I add two additional requirements that must be met by the effective
Hamiltonian.  First, it must be manifestly Hermitian.  Second,
its representation can
only involve the unperturbed (\ie, free) energies in any denominators
that occur.  The second
property is a severe limitation that
limits the utility
of the Hamiltonian to studies of Gaussian and near-Gaussian fixed
points.  This property is not desirable {\it a priori} and is adopted
only because I do not
know of a general procedure for constructing Hamiltonians
outside of perturbation theory.  This
limitation does not imply that one is always limited to the use of
perturbation theory, after one uses a perturbative construction of the
light-front renormalization group
Hamiltonian.  The perturbative construction only needs to be valid
near the Gaussian fixed point.  After one has used the transformation
many times to reduce the cutoff to an acceptable value, the final
Hamiltonian may have been accurately computed using a perturbative
renormalization group transformation,
but its diagonalization is typically
nonperturbative.  I am particularly interested in QCD, where one can
hopefully use a light-front renormalization group
to remove high energy partons through sequential
application of a perturbative renormalization group transformation,
justified by asymptotic freedom \APrefmark{\rGRO,\rPOL};
after which one must employ suitable
nonperturbative techniques to diagonalize the final cutoff Hamiltonian
and obtain low energy observables.

I begin with the Bloch-Horowitz Hamiltonian because it
is most easily derived.  In fact, the primary reason I include a
discussion of the Bloch-Horowitz Hamiltonian is because its development
is particularly transparent and allows a reader unfamiliar with such
many-body techniques to readily grasp the main ideas.
Let the operator that projects onto all of the states removed when the
cutoff is lowered be $\qproj$, and let $\pproj=1-\qproj$.  Of course,
$\qproj^2=\qproj$ and $\pproj^2=\pproj$.
I occasionally refer to a subspace using the appropriate projector.
With these projectors,
Schr{\"o}dinger's equation can be divided into two parts,

$$\pproj H \pproj \wave + \pproj H \qproj \wave = E \pproj \wave \;,
\eqno(4.1)$$

$$\qproj H \pproj \wave + \qproj H \qproj \wave = E \qproj \wave \;.
\eqno(4.2) $$

\noindent
Solving Eq. (4.2) for $\qproj \wave$ and substituting the result into
Eq. (4.1),
one finds an operator, $H'$, that produces the same
eigenvalue $E$ in the subspace $\pproj$ as the original Hamiltonian,

$$H'(E)= \pproj H \pproj + \pproj H \qproj {1 \over E - \qproj H \qproj}
\qproj H \pproj \;. \eqno(4.3)$$

\noindent This is the Bloch-Horowitz effective Hamiltonian.  It
is not satisfactory, however,
because it contains the eigenvalue, $E$; and it is not Hermitian unless
$E$ is held fixed.

The development of the Bloch Hamiltonian begins with the definition of
the same projection operators $\qproj$ and $\pproj$ used above.
After defining these operators one looks for an operator $\rb$ that
satisfies

$$\qproj \wave = \rb \pproj \wave \;, \eqno(4.4)$$

\noindent
for all eigenstates of the Hamiltonian that have support in the subspace
$\pproj$.  To construct this operator, multiply Eq. (4.1) by $\rb$ and
replace $\qproj \wave$ in Eq. (4.2) with $\rb \pproj \wave$, and
subtract the resultant equations to obtain

$$\rb \Hpp-\Hqq \rb+\rb \Hpq \rb -\Hqp = 0 \;. \eqno(4.5)$$

\noindent  Here I have introduced the notation $\Hpp=\pproj H \pproj$,
etc.  This is the fundamental equation for $\rb$ and one
of the most difficult
steps in constructing the Bloch Hamiltonian is solving this equation.  I
am only interested in the perturbative solution, and one can
already see that $\rb$ should be proportional to $\Hqp$.

For notational convenience I let $H=h+v$ for the Bloch Hamiltonian,
where $h$ is a `free'
Hamiltonian and I assume $[h,\qproj]=0$.  Since $\pproj
\qproj=0$, Eq. (4.5) can also be written as

$$\rb \hpp - \hqq \rb - \vqp + \rb \vpp - \vqq \rb + \rb \vpq \rb = 0
\;. \eqno(4.6)$$

\noindent This equation is now in a form that can be solved using an
expansion in powers of $v$, and it is apparent that $\rb$ starts at
first order in $v$.  Before solving this equation, let me write the
effective Hamiltonian and the eigenstates in terms of $\rb$.

The states $\pproj \wave$ are not orthonormal.  However, if we assume
that no two eigenstates of $H$ have the same projection in the subspace
$\pproj$ (\ie, that $\rb$ is single-valued), then one can readily check that

$$H' \wwave = E \wwave \;, \eqno(4.7)$$

\noindent when,

$$\wwave = \sqrt{1+\rb^\dagger \rb}\;\;\pproj\; \wave \;, \eqno(4.8)$$

\noindent and,

$$H'= {1 \over \sqrt{1+\rb^\dagger \rb}} (\pproj+\rb^\dagger) H
(\pproj+\rb) {1 \over \sqrt{1+\rb^\dagger \rb}} \;. \eqno(4.9)$$

\noindent The states $\wwave$ are orthonormalized projections of the
original eigenstates $\wave$, and the manifestly Hermitian
effective Hamiltonian $H'$ yields the same eigenvalues as the original
Hamiltonian $H$.  $H'$ is the Bloch Hamiltonian.

To construct the Bloch Hamiltonian in perturbation theory, one first
solves Eq. (4.6) in perturbation theory.  Since we cannot assume that
$[\rb,\hpp]$ or $[\rb,\hqq]$ are zero, we must employ the eigenstates

$$\hpp \awave = \epsilon_a \awave \;,\;\;\;\hqq \iwave = \epsilon_i
\iwave \;, \eqno(4.10)$$

\noindent to develop algebraic equations for the matrix elements of
$\rb$.  I use $\awave$, $\bwave$ , ... to indicate free eigenstates
in $\pproj$; and $\iwave$, $\jwave$, ... to indicate free eigenstates in
$\qproj$.  In order to expand $H'$ through third order in the
interaction, we only need the first two terms in an expansion of $\rb$,
and these are

$$\liwave \rb_1 \awave = {\liwave v \awave \over \epsilon_a-\epsilon_i}
\;, \eqno(4.11)$$

$$\liwave \rb_2\awave = \sum_j {\liwave v \jwave \ljwave v \awave \over
(\epsilon_a-\epsilon_i) (\epsilon_a-\epsilon_j)} - \sum_b {\liwave v
\bwave \lbwave v \awave \over (\epsilon_a-\epsilon_i)
(\epsilon_b-\epsilon_i)} \;. \eqno(4.12)$$

\noindent There are a few general features of each term in this expansion
that I want to note.  First, note that $\rb$ only has nonzero matrix
elements between a bra in $\qproj$ and a ket in $\pproj$.
Second, every energy denominator involves a
difference between a free energy in $\pproj$ and a free energy in
$\qproj$.  Ultimately the convergence of this expansion is going to rest
not only on the weakness of $v$, but also on the fact that high energy
states are being eliminated so that these denominators are large
throughout the most important regions of phase space.  We are ultimately
interested only in the very low energy eigenstates and eigenvalues, and
almost all of the states we eliminate to obtain the effective
Hamiltonian for these states are extremely far off shell.  The energy
denominators may vanish, but this should only happen for a set of measure
zero if the expansion is to converge.  One can avoid all but accidental
degeneracies by putting the system in a box, but I do not need to do
this for the examples considered.  Serious problems should only arise
when this entire series needs to be re-summed because a particular
nonperturbative effect must be properly included.  Unfortunately, such
serious problems are common.
Higher order terms are easily constructed using the recursion relation

$$\rb_n \hpp-\hqq \rb_n+\rb_{n-1} \vpp - \vqq \rb_{n-1} +
\sum_{m=1}^{n-2} \rb_m \vpq \rb_{n-m-1} = 0 \;. \eqno(4.13)$$

\noindent Clearly the number of terms in each successive order grows
exponentially, and it is likely that the expansion is at best
asymptotic.

Given a perturbative expansion for $\rb$, one can use Eq. (4.9) to
derive a perturbative expansion for the Bloch Hamiltonian.  After some
simple algebra, one finds that through third order in $v$,

$$\eqalign{
\lawave H' \bwave = &\lawave h+v \bwave + {1 \over 2} \sum_i \Biggl(
{\lawave v \iwave \liwave v \bwave \over (\epsilon_a-\epsilon_i)} +
{\lawave v \iwave \liwave v \bwave \over (\epsilon_b-\epsilon_i)}
\Biggr) \cr
&+ {1 \over 2} \sum_{i,j} \Biggl( {\lawave v \iwave \liwave v \jwave
\ljwave v \bwave \over (\epsilon_a-\epsilon_i) (\epsilon_a-\epsilon_j)}
+ {\lawave v \iwave \liwave v \jwave \ljwave v \bwave \over
(\epsilon_b-\epsilon_i) (\epsilon_b-\epsilon_j)} \Biggr) \cr
&- {1 \over 2} \sum_{c,i} \Biggl( {\lawave v \iwave \liwave v \cwave
\lcwave v \bwave \over (\epsilon_b-\epsilon_i) (\epsilon_c-\epsilon_i)}
+ {\lawave v \cwave \lcwave v \iwave \liwave v \bwave \over
(\epsilon_a-\epsilon_i) (\epsilon_c-\epsilon_i)} \Biggr) \;\;+\;{\cal
O}(v^4)
\;.}\eqno(4.14)$$

This expression can be used to compute the renormalization of the
quark-gluon vertex through third order in the bare coupling, for
example \APrefmark{\rPERQCD}.
It is quite similar to expressions one encounters in
time-ordered perturbation theory for off-shell Green's functions, but
there are some distinct differences.  Given Eq.
(4.14), or any extension of higher order,
it is straightforward to derive a set of diagrammatic rules
that allow one to summarize the operator algebra in fairly simple
diagrams.  These diagrams are similar to Goldstone
diagrams \APrefmark{\rGOLD},
familiar from many-body quantum mechanics,
but they require
one to display the energy denominators.  I refer to them as
{\it Hamiltonian diagrams}.
In a standard time-ordered
diagram, there is a factor $1/(E-\epsilon_i)$ for every intermediate
state, $\iwave$,
and the energy $E$ is the same in every denominator.  In the
Hamiltonian diagrams there are a wide variety of denominators,
always involving differences of eigenvalues of $H_0$; however, these
eigenvalues may be associated with widely separated states in the
Hamiltonian diagram.  While
the above process of generating a perturbative expansion for the
effective Hamiltonian is easily mechanized, I have not found a simple
set of diagrammatic rules that summarize this process to all orders.  It
is quite possible that such rules already exist in the literature, but for the
purposes of this paper it is only necessary to understand Eq. (4.14) and
to appreciate the fact that higher order terms are readily generated.

In figure
2 I show a few typical Hamiltonian
diagrams that occur when $H$ contains a $\phi^4$
interaction.  Energy denominators
are denoted by lines connecting the relevant states, with
the arrow pointing toward the state whose eigenvalue occurs last in Eq.
(4.14).
One can infer from these lines which states lie in $\pproj$ and which
states lie in $\qproj$, because
energy denominators always involve differences of energies in $\pproj$
and energies in $\qproj$ and the arrow points toward a state in
$\qproj$.
External states always lie in $\pproj$ of
course.
In a realistic calculation there are many different
vertices, corresponding to the many different functions $u_4, u_6$, etc.
in the Hamiltonian.  Sometimes $u_2$ determines $h$; however, in
perturbation theory one usually needs to include part of $u_2$ in $v$.
This issue is clarified in the
next Section.

The division of the Hamiltonian into a `free' and
`interaction' part above is arbitrary, except for the requirement
that $[h,\qproj]=0$.  Let us suppose that this requirement is met, but
some of the eigenvalues of $h$ in $\pproj$ are larger than some of the
eigenvalues of $h$ in $\qproj$.
The energy denominators in Eq. (4.14) pass
through zero when this happens and one must question the
convergence of the expansion in Eq. (4.14).  The fact that the
denominators may vanish does not automatically imply that the expansion
does not converge, but at the minimum it forces one to carefully
consider the boundary conditions required to construct the Green's
functions in Eq. (4.14).  However, even if the expansion in Eq. (4.14)
does converge, it may prove useless for a renormalization group study.
In general, each term in Eq. (4.14) leads to an infinite number of
operators in the transformed Hamiltonian, because of the potentially
complicated dependence each term may have on the momenta of the
`incoming' and `outgoing' states.  We will see in the next Section that
when some of the eigenvalues of $h$ in $\pproj$ are larger than some
of the eigenvalues of $h$ in $\qproj$, operators that are classified
as irrelevant in the linearized renormalization group analysis occur
with large and sometimes infinite
coefficients that render this classification scheme
useless.


\bigskip
\noindent {\bf \rnfive. Critical Gaussian
Fixed Points and Linearized
Behavior Near Critical Gaussian Fixed Points}
\medskip

A fixed point is defined to be any Hamiltonian that satisfies

$$T[H^*]=H^* \;. \eqno(5.1)$$

\noindent The fixed point is central to the renormalization group
analysis, and unless one has found a fixed point it is unlikely that
anything can be accomplished with perturbation theory.
The simplest example is the Gaussian fixed point, one for
which $u_4=0, u_6=0$, etc.  In this case the transformation that
generates $H_{\Lambda_1}$ is trivial.  Let me begin by discussing the
Gaussian fixed point for the transformation $T^\perp$.
To find the Gaussian fixed point
we need to consider

$$\eqalign{ H_{\Lambda_0} = &\;\; \int \dqt \;
\theta(\Lambda_0^2-\qp^2)
\;u_2(q)\;
{a^\dagger(q) a(q) \over q^+}
\;,} \eqno(5.2)$$

\noindent and when the cutoff changes the effective Hamiltonian is
simply

$$\eqalign{ H_{\Lambda_1} = &\;\; \int \dqt \;
\theta(\Lambda_1^2-\qp^2)\;
u_2(q)\;
{a^\dagger(q) a(q) \over q^+}
\;.} \eqno(5.3)$$

\noindent
Next we change variables according to Eqs. (3.10) and (3.11), and
multiply the Hamiltonian by a constant.  The final transformed
Hamiltonian is

$$\eqalign{ T[H_{\Lambda_0}] = \;\; z^\perp \zeta^{\perp\;2}
\Bigl({\Lambda_1 \over \Lambda_0}\Bigr)^2 &\int \dqt \;
\theta(\Lambda_0^2-\qp^2)
\;u_2({\Lambda_1 \over \Lambda_0} \qp,q^+) \;
{a^\dagger(q) a(q) \over q^+}
\;.} \eqno(5.4)$$

The factor of $(\Lambda_1/\Lambda_0)^2$ in front of the integral arises
because the Hamiltonian itself has the dimension given in Eq. (3.2).  I
absorb this overall rescaling by setting

$$z^\perp=\Bigl({\Lambda_0 \over \Lambda_1}\Bigr)^2 \;. \eqno(5.5)$$

\noindent  $z^\perp$ is introduced so that the transformed Hamiltonian
can be directly compared with the initial Hamiltonian.  To obtain
eigenvalues in the initial units, one must multiply by $1/z^\perp$.
With $z^\perp$ determined, the Gaussian fixed point is found
by insisting that

$$ u_2^*(\qp,q^+)=\zeta^{\perp\;2}\;u_2^*({\Lambda_1 \over \Lambda_0}
\qp,q^+) \;. \eqno(5.6)$$

\noindent The general solution to this equation is familiar from
Euclidean field theory \APrefmark{\rWILNINE},
being a monomial in $\qp$.  The general
solution is

$$u_2^*(\qp,q^+)= f(q^+) (\qp)^n \;, \eqno(5.7)$$

$$\zeta^\perp=\Bigl({\Lambda_0 \over \Lambda_1}\Bigr)^{(n/2)} \;.
\eqno(5.8)$$

\noindent One must decide {\it ab initio} which Gaussian fixed point is to be
studied by fixing the dispersion relation in the free theory, and I am
only interested in the case $n=2$.  Since there is no mass term in
this fixed point, the correlation length is infinite and it is called a
critical fixed point.

The fixed point of $T^\perp$
contains an undetermined function of longitudinal momentum, $f(q^+)$.  This
has important implications for the light-front renormalization group,
which will become clearer
below.

One can look for fixed points that contain interactions, but no such
fixed point has been found for a scalar field
theory in $3+1$ dimensions.  It is possible to change the number of
transverse dimensions \APrefmark{\rWILSEVEN}
to $2-\epsilon$ and construct an analog
of the analysis found, for
example, in the review article of Wilson and Kogut \APrefmark{\rWILNINE}.
I do not pursue
this idea, and turn to
the next step in the
analysis of $T^\perp$, the construction of the linearized
transformation about the Gaussian fixed point.  To construct the
linearized transformation, consider Hamiltonians that are almost
Gaussian,

$$H_l=H^* + \delta H_l \;. \eqno(5.9)
$$

\noindent Here the subscript $l$ labels the number of times the
transformation has been applied, so that a sequence of Hamiltonians can
be studied.  I assume that $\delta H_l$ is
`small'.  Applying the full transformation we find that

$$\eqalign{
\delta H_{l+1} &= T[H^*+\delta H_l]-H^* \cr
&= L_{H^*} \cdot \delta H_l + {\cal O}(\delta H_l^2) \;.} \eqno(5.10)$$

\noindent This equation defines the linear operator $L_{H^*}$, which
depends explicitly on the fixed point.  I typically drop the
subscript on $L_{H^*}$ and simply refer to $L$.

In general the construction of the operator $L$ is complicated;
however, when $H^*$ is a Gaussian fixed point, $L$ is easily
constructed.  In the first step of a renormalization group
transformation a cutoff is changed and degrees of freedom are removed.
Effective interactions result when an incoming state experiences an
interaction, so that some momenta are altered and fall into the range
being removed.  However, both the incoming and outgoing states must fall
in the sector retained by the transformation, so a second interaction is
always required to return the state to the allowed sector.  The only
exception to this occurs when zero modes are allowed and one can form a
`loop' with a single interaction.  I drop zero modes, so
I can ignore this possibility and conclude
that near the Gaussian fixed point all new interactions and changes to
initial
interactions in the effective Hamiltonian are ${\cal O}(\delta H^2)$.
This means that the only part of the full transformation that affects
the linear operator is the rescaling.

We immediately conclude that $L$ for the Gaussian fixed point of
$T^\perp$ is given by the rescaling operations in Eqs. (3.10) and
(3.11), along with the overall multiplicative transformation of the
Hamiltonian.  $z^\perp$ is given by Eq. (5.5), and $\zeta^\perp$ is
given by Eq. (5.8), so one readily finds that

$$\eqalign{
u_n(\qp_1,q^+_1,...) & \rightarrow z^\perp \zeta^{\perp \; n}
\Bigl({\Lambda_1 \over \Lambda_0}\Bigr)^{(2n-2)} \; u_n({\Lambda_1 \over
\Lambda_0} \qp_1,q^+_1,...) \cr
& \rightarrow \Bigl({\Lambda_1 \over \Lambda_0}\Bigr)^{(n-4)} \;
u_n({\Lambda_1 \over
\Lambda_0} \qp_1,q^+_1,...)
\;.} \eqno(5.11) $$

Subsequent analysis of the full transformation should be simplified by
identifying the eigenoperators and eigenvalues of this linearized
transformation.  We seek solutions to the equation

$$L\cdot O = \lambda O \;, \eqno(5.12)$$

\noindent and the solutions are readily found to be homogeneous
polynomials of transverse momenta.  We can label $O$ with a subscript
that displays the number of field operators in $O$ and a
superscript that displays the number of powers of transverse momenta.
For example,

$$O_2^4 =
\int \dqt \;
\theta(\Lambda_0^2-\qp^2)
\bigl[\qp \bigr]^4 \;
{a^\dagger(q) a(q) \over q^+}
\;. \eqno(5.13)$$

\noindent
In general more labels are required because there may be many possible
polynomials of the same degree; however, the eigenvalue is determined by
these two labels.  Applying $L$ to any operator $O_n^m$ one finds

$$L\cdot O_n^m = \Bigl( {\Lambda_1 \over \Lambda_0}\Bigr)^{(m+n-4)}
O_n^m \;. \eqno(5.14) $$

\noindent Inspection of this result reveals that the eigenvalue is
determined by the naive transverse engineering dimension of the operator.
This is not surprising since the linear approximation of the
transformation is given by the rescaling operations alone.

{\it Relevant operators} are defined to be operators for which $\lambda > 1$;
and since $\Lambda_0 > \Lambda_1$, relevant operators have $m+n-4 < 0$.
Transverse locality implies that $m \ge 0$, and we always require $n \ge
2$, so the only relevant operator is one satisfying $m=0$, $n=2$, if we
assume that a $\phi \rightarrow -\phi$ symmetry is maintained.  This is
a mass term.
There is no {\it a priori} reason to assume that such a symmetry
can be maintained, and if it is violated a $n=3$ relevant operator will
appear.
The $m=1$, $n=2$
operator is ruled out by rotational invariance about the z-axis.
If one includes a power of a transverse momentum, the Lorentz indices on
this momentum must be contracted with a second Lorentz index.  In the
absence of spin the only transverse Lorentz indices are carried by
transverse momenta, and therefore single powers of transverse momenta
cannot occur.
Note
that the dependence of $O_n^m$ on longitudinal momenta does not affect
this analysis; and in this sense there are an infinite number of
relevant operators if there is one.

{\it Marginal operators} are defined to be operators for which $\lambda = 1$.
Allowing $\phi \rightarrow -\phi$ symmetry to be broken, there appear
to be
$(n=2,m=2)$, $(n=3,m=1)$, and $(n=4,m=0)$ marginal operators.  The
$(n=3,m=1)$ operator is excluded by rotational symmetry about
the z-axis.  Again, any dependence on longitudinal momenta does not
affect this analysis.

{\it Irrelevant operators} are defined to be operators for which
$\lambda < 1$, and one sees that almost all operators are irrelevant by
this definition.  If inverse powers of transverse momenta are allowed,
this conclusion is destroyed, and there are an infinite number of
relevant and marginal operators in addition to those discussed above.

This analysis is itself relevant only if there is a region near the
fixed point in which the linearized transformation is a reasonable
approximation of the full transformation.  In this case there may be
a convergent or asymptotic perturbative expansion for the
transformation in a region near the fixed point.
A perturbative analysis is probably useful only for theories
that display asymptotic freedom.  In this paper I simply assume
that all couplings are small, even though we know that
this condition cannot be satisfied for an interacting
symmetric scalar theory
if the
initial cutoff approaches infinity, because it is not asymptotically
free.

Before discussing $T^+$, I would like to explicitly show what happens when a
physical mass is added to $u_2^*$.
In this case we have

$$\eqalign{ H_{\Lambda_0} = &\;\; \int \dqt \;
\theta(\Lambda_0^2-\qp^2)\;
\bigl[\qp^2+m^2\bigr]\;
{a^\dagger(q) a(q) \over q^+}
\;.} \eqno(5.15)$$

\noindent One can readily verify that after $T^\perp$ is applied $\calN$
times the result is

$$\eqalign{{T^\perp}^\calN\bigl[H_{\Lambda_0}\bigr]= \;\; \int \dqt \;
\theta(\Lambda_0^2-\qp^2) \; \bigl[\qp^2+4^\calN m^2\bigr]\; {a^\dagger(q)
a(q) \over q^+}
\;,} \eqno(5.16)$$

\noindent where I assume $\Lambda_1=\Lambda_0/2$.
As $\calN \rightarrow \infty$ the Hamiltonian approaches another
Gaussian fixed point, the so-called trivial fixed point at which the mass is
infinitely large in comparison to the kinetic energy.  Note that the
original energy spectrum can always be obtained by multiplying the
Hamiltonian by $4^{-\calN}$.

The analysis of the Gaussian fixed point and linearized
approximation for $T^+$ closely parallels that of $T^\perp$.
To find the Gaussian fixed point we must
analyze

$$\eqalign{ H_{\epsilon_0} = &\;\; \int \dqt \;
\theta(q^+-\epsilon_0)\;
u_2(q)\;
{a^\dagger(q) a(q) \over q^+}
\;,} \eqno(5.17)$$

\noindent and when the cutoff changes the effective Hamiltonian is

$$\eqalign{ H_{\epsilon_1} = &\;\; \int \dqt \;
\theta(q^+-\epsilon_1)\;
u_2(q) \;
{a^\dagger(q) a(q) \over q^+}
\;.} \eqno(5.18)$$

\noindent Next we change variables according to Eqs. (3.17) and (3.18),
and multiply by an overall constant to obtain

$$\eqalign{ T[H_{\epsilon_0}] = \;\; z^+ \zeta^{+2}
\Bigl({\epsilon_0 \over \epsilon_1}\Bigr) &\int \dqt \;
\theta(q^+-\epsilon_0) \;
u_2(\qp,{\epsilon_1 \over \epsilon_0} q^+)\;
{a^\dagger(q) a(q) \over q^+}
\;.} \eqno(5.19)$$

Here the factor of $(\epsilon_0 / \epsilon_1)$ in front of the integral
arises because the Hamiltonian has the dimension given in Eq. (3.2), and
I absorb this overall rescaling by setting

$$z^+=\Bigl({\epsilon_1 \over \epsilon_0}\Bigr) \;. \eqno(5.20)$$

\noindent Given $z^+$, the Gaussian fixed point satisfies

$$ u_2^*(\qp,q^+)=\zeta^{+2}\;u_2^*(\qp,{\epsilon_1 \over \epsilon_0}
q^+)\;. \eqno(5.21)$$

\noindent The general solution is a monomial in $q^+$,

$$u_2^*(\qp,q^+)= f(\qp) \Bigl({1 \over q^+}\Bigr)^n \;, \eqno(5.22)$$

$$\zeta^+=\Bigl({\epsilon_1 \over \epsilon_0}\Bigr)^n \;. \eqno(5.23)$$

\noindent Again, one must decide {\it ab initio} which Gaussian fixed
point to investigate, and I am interested in the case $n=0$.

$T^\perp$ and $T^+$ both have Gaussian fixed volumes rather than fixed
points, because neither scales all momentum components simultaneously.
The single fixed point that both share, with the definitions of the
scaling constants
above, is

$$u_2^*(\qp,q^+)= \qp^2 \;. \eqno(5.24)$$

The linearized approximation for $T^+$ is determined
by the rescaling operations in Eqs. (3.17) and
(3.18).  Combined with Eq. (5.23) these yield

$$\eqalign{
u_n(\qp_1,q^+_1,...) & \rightarrow z^+ \zeta^{+ \; n} \Bigl({\epsilon_0
\over \epsilon_1}\Bigr)
\; u_n(\qp_1,{\epsilon_1 \over \epsilon_0} q^+_1,...) \cr
& \rightarrow
u_n(\qp_1,{\epsilon_1 \over \epsilon_0} q^+_1,...)
\;.} \eqno(5.25) $$

\noindent The eigenoperators of this transformation are homogeneous
polynomials of inverse longitudinal momenta.  I choose to employ inverse
derivatives for reasons that will become clear.
For $T^+$ the solutions need not
be labelled by the number of field operators, because
the scalar field is longitudinally dimensionless, as in Eq.(3.1).  However,
for convenience I use the same notation found in Eq. (5.13), but
with the superscript now indicating the power of inverse longitudinal
momenta.  In
this case one finds that

$$O_2^4 = \int \dqt \; \theta(q^+-\epsilon_0) \Bigl[{1 \over q^+} \Bigr]^4 \;
{a^\dagger(q) a(q) \over q^+} \;, \eqno(5.26)$$

\noindent and

$$L\cdot O_n^m = \Bigl( {\epsilon_0 \over \epsilon_1} \Bigr)^m  O_n^m \;.
\eqno(5.27)$$

\noindent I have not included the inverse power of $q^+$ found in the
measure $\dqt$, nor the inverse power of $q^+$ that accompanies any
product of creation and annihilation operators found in the Hamiltonian,
in the index $m$.  This is clear in Eq. (5.26).  The eigenvalue in this
case is determined by the naive longitudinal
engineering dimension of the
operator.

$\epsilon_0 < \epsilon_1$, so longitudinal relevant operators must have
$m<0$; longitudinal marginal operators have $m=0$ and irrelevant
operators have $m>0$.  This explains the choice of inverse derivatives,
because powers of inverse derivatives produce irrelevant operators.  The
same type of simplification that occurs in a Euclidean renormalization
group when locality is assumed, might
occur when one formulates a principle
of {\it longitudinal nonlocality} for the longitudinal transformation.
When we study
second-order corrections in Section \rnsix, we will find that $T^+$
leads to problems that prevent us from exploiting any simplifications
from longitudinal nonlocality, unfortunately.

This completes the analysis of $T^+$ to leading order near the fixed
point.  The analysis of the next order shows that $T^+$ cannot be
approximated in perturbation theory unless one allows arbitrary functions
of transverse momenta, including functions that violate transverse locality.
Worse, one encounters arbitrarily large coefficients of irrelevant
operators in perturbation theory, so these operators cannot be dropped.

It is fairly easy to see that Hamiltonians of physical interest
are fixed points of $T^+$, using Eq. (3.19).  I demonstrate
this and discuss some implications before closing this Section.  In the
first step of the transformation $T^+$, we must remove degrees of freedom
following Eq. (3.16), and find a new Hamiltonian that yields the same
eigenvalues and properly renormalized eigenstate projections.  A
Hamiltonian that meets these criteria is readily found by applying a
longitudinal boost, so that

$$p^+ \rightarrow {\epsilon_1 \over \epsilon_0} p^+ \;. \eqno(5.28)$$

\noindent  After this boost all longitudinal momenta satisfy the
constraint $\epsilon_1 < p^+ < \infty$.  Under this same boost we know
that

$$H \rightarrow {\epsilon_0 \over \epsilon_1} H \;, \eqno(5.29)$$

\noindent which reveals that when we multiply the boosted Hamiltonian by
the constant $(\epsilon_1 / \epsilon_0)$, we retrieve the
original Hamiltonian, but with momenta restricted to a new range.
$T^+$ is completed by rescaling
the variables according to Eqs. (3.17) and (3.18), and with
$z^+=(\epsilon_1 / \epsilon_0)$ and $\zeta^+=1$, this merely returns the
longitudinal momenta to their original values before the boost and has
no other effect on the Hamiltonian.  In other words, as long as Eq.
(5.29) holds, we have

$$T^+[H]=H \;, \eqno(5.30)$$

\noindent so that $H$ is a fixed point of the full transformation $T^+$.

This may be an important result.
However, fixed points of $T^+$ that contain interactions
invariably contain operators that violate transverse locality and leave
one with no perturbative
means to study the dependence of the Hamiltonian on the
transverse scale.  At this point I have discovered no benefit
to implementing longitudinal boost-invariance by seeking fixed points of
$T^+$.

All of the boost-invariant renormalization group transformations
have the same Gaussian fixed
points as $T^\perp$, and the same linear analysis as $T^\perp$.  Therefore,
for each of these transformations, operators are classified according to
their transverse dimensionality.  Longitudinal dimensionality plays
no role whatsoever, and the linear analysis cannot be used to control the
longitudinal functions in the Hamiltonian.  There is no need to repeat
the discussion leading up to Eq. (5.14).
The only caveat is that
strictly speaking the
invariant-mass transformation that employs the cutoff in Eq. (3.24) has
no fixed points.  As discussed in Section \rnthree,
after a finite number of transformations all of phase
space is eliminated and the transformed Hamiltonian approaches zero,
because of the masses in this cutoff;
unless the initial cutoff is allowed to approach infinity.
However, using analytic continuation one can extend the functions $u_2$,
$u_4$, etc. outside the range of the invariant mass cutoff and define
the fixed point using the analytically continued Hamiltonian.  In this
case, the Gaussian fixed points are again identical to the
Gaussian fixed points of
$T^\perp$.


\bigskip
\noindent {\bf \rnsix.
Second-Order Behavior Near Critical Gaussian Fixed Points}
\medskip

The perturbative renormalization group analysis becomes much more
interesting when second-order corrections to the Hamiltonian are
included.  In fact, almost all features of a complete
perturbative analysis are displayed in a second-order analysis.
In this Section I begin by
adding one simple interaction to the Hamiltonian and studying the
interactions it generates to second-order.  This analysis demonstrates
that most candidate transformations produce irrelevant interactions with
divergent coefficients.  I next concentrate on the invariant-mass
transformation that runs the cutoff in Eq. (3.23).  This
transformation is used throughout the rest of the paper, except in
Appendix C, where the effects of physical masses are briefly studied and
a transformation that runs the cutoff in Eq. (3.24) must be used.  It is
shown that if one drops all irrelevant operators, a complete sequence of
second-order transformations may be summarized by a few coupled
difference equations.  These equations are solved, and the physics
implied by their solution is discussed.

The final subject in this Section is the most important.  I
first show that when one uses the canonical $\phi^4$ interaction and the
invariant-mass cutoff, the boson mass shift and
the boson-boson scattering amplitude violate
both Lorentz covariance and cluster decomposition.  The calculation of
these expectation values
reveals the `counterterms' required to restore Lorentz
covariance and cluster decomposition; and these counterterms include an
infinite number of irrelevant operators, as well as a marginal operator
that contains logarithms of longitudinal momentum fractions.  This
result is not surprising, and the important question is how to find a
Hamiltonian that restores these properties to the mass shift
and the scattering amplitude.
Of course, one method is to simply compute amplitudes in perturbation
theory, and add to the Hamiltonian the difference between the result
desired and the result obtained.
This is
not very satisfactory, because it requires reference to a separate
calculation, so I ask a question that Wilson and I have
recently posed \APrefmark{\rPERWIL}.
What happens if one allows the Hamiltonian to become
arbitrarily complicated, but {\it one insists that a single coupling is
allowed to explicitly depend on the cutoff}, with all other interactions
depending on the cutoff only through their dependence on this
coupling?  The answer may depend on the coupling one allows to run, and
I select the $\phi^4$ canonical coupling, which a manifestly covariant
analysis would show is the only coupling that must be allowed to run if
$\phi \rightarrow -\phi$ symmetry is maintained.
In this Section I show that coupling coherence
uniquely fixes all
relevant and irrelevant operators to second-order in the $\phi^4$
coupling constant; and that these operators restore Lorentz covariance and
cluster decomposition to the boson mass shift, which is a relevant
operator, and to the
irrelevant part of the boson-boson scattering amplitude.
I then show that a third-order analysis is required
to determine the marginal operators to second-order in the
canonical coupling.
The third-order analysis is completed in Section \rnseven.

The second term on the right-hand side of Eq. (4.14) can be used to
determine the second-order behavior of any transformation near
a critical Gaussian fixed point. The fixed point Hamiltonian is $h$ in Eq.
(4.14), and all deviations from the fixed point are part of $v$.  I
should note that deviations of $u_2$ from $u_2^*$ are produced by the
second-order part of the transformation, but they do not directly affect
the subsequent second-order behavior of the
transformations.  This is because we need two
interactions to first produce an intermediate state with
an energy above the new cutoff and then produce a final state with an energy
below the new cutoff, and
$u_2$ does not change the state.

If we consider the second-order
behavior of a transformation when acting on a Hamiltonian in which $u_4$
is the only nonzero interaction, for example, we encounter the
Hamiltonian diagrams
in figure 3.  A detailed derivation of the expression for the
Hamiltonian diagram in
figure 3a is given in Appendix B as an illustrative example,
but throughout the text I simply
give the appropriate expressions without explicit derivation.  The
diagrammatic rules for Hamiltonian diagrams are almost identical to the
diagrammatic rules for time-ordered Green's functions given in Appendix
A.  The principal differences are that an infinite number of vertex
rules are required, each of which is easily determined from the related
interaction; and the rules for energy denominators differ, as discussed
in Section \rnfour.

I have not
bothered to indicate the energy denominators in figure 3, because as one
sees in Eq. (4.14), there are only two choices in the second-order term.
The denominator contains either a difference of the incoming state
energy and the intermediate state energy or the outgoing state energy
and the intermediate state energy.
A transformation is completed by evaluating these
Hamiltonian diagrams with the
appropriate cutoffs, and then rescaling the external variables.

Given the transformations in Section \rnthree, and the techniques for
constructing effective Hamiltonians in Section \rnfour, we can readily
compute
the second-order behavior of the transformations `near' the
critical Gaussian fixed
point in Eq. (5.24).
The first step in this process is to choose an initial
Hamiltonian, $H^{\Lambda_0}_{\Lambda_0}$.  If we choose a Hamiltonian
for which only $u_2$ is nonzero, the linear analysis is exact, so we need
to add at least one interaction to study second-order behavior.
The simplest examples should result
from adding only one interaction, and the strength of this interaction
must be sufficiently weak that $H^{\Lambda_0}_{\Lambda_0}$ is near the
Gaussian fixed point.  Regardless of what interaction we add to the
fixed point Hamiltonian to create $H^{\Lambda_0}_{\Lambda_0}$, we will
find that $H^{\Lambda_0}_{\Lambda_1}$ contains an infinite number of
interactions.  Any interaction we add leads to an infinite number of
irrelevant interactions at least, so I first
consider the
relevant or marginal interactions that can be added to the Gaussian
fixed point.

Let me specialize to the discussion of $T^\perp$ and the
boost-invariant transformations.
For these transformations, there are relevant and
marginal interactions with $u_3$ nonzero, and there are marginal
operators with $u_4$ nonzero.  I assume that the
symmetry $\phi \rightarrow -\phi$ is maintained manifestly,
so that only $u_4$ is
nonzero.  If spontaneous symmetry breaking occurs, so that the $\phi
\rightarrow -\phi$ symmetry is hidden, $u_3$ must be allowed
to appear in the Hamiltonian; however, in this situation it is a
function of the variables appearing in the manifestly symmetric theory,
as Wilson and I have discussed \APrefmark{\rPERWIL}.
There can be no powers of transverse momenta in the marginal part of $u_4$,
because these lead only to irrelevant operators according to the
analysis of Section \rnfive.  On the other hand, functions of longitudinal
momenta have no effect on the linear analysis, so we should consider

$$H^{\Lambda_0}_{\Lambda_l} = H^* + \delta H_l \;, \eqno(6.1)$$

\noindent with,

$$\eqalign{\delta H_0 &=
{1 \over 6} \int \dqt_1\; \dqt_2\; \dqt_3\; \dqt_4 \; (16 \pi^3)
\delta^3(q_1+q_2+q_3-q_4) \cr
&\qquad\qquad\qquad\qquad\qquad
\tg(-q_1^+,-q_2^+,-q_3^+,q_4^+)\; a^\dagger(q_1) a^\dagger(q_2)
a^\dagger(q_3)
a(q_4)  \cr
&+{1 \over 4} \int \dqt_1\; \dqt_2\; \dqt_3\; \dqt_4 \; (16 \pi^3)
\delta^3(q_1+q_2-q_3-q_4) \cr
&\qquad\qquad\qquad\qquad\qquad \tg(-q_1^+,-q_2^+,q_3^+,q_4^+)\; a^\dagger(q_1)
a^\dagger(q_2) a(q_3) a(q_4)  \cr
&+{1 \over 6} \int \dqt_1\; \dqt_2\; \dqt_3\; \dqt_4 \; (16 \pi^3)
\delta^3(q_1-q_2-q_3-q_4) \cr
&\qquad\qquad\qquad\qquad\qquad \tg(-q_1^+,q_2^+,q_3^+,q_4^+)\; a^\dagger(q_1)
a(q_2) a(q_3) a(q_4)
\;,}\eqno(6.2)$$

\noindent  where $\tg(q_i^+)$ is the marginal part of $u_4$.

This is still too general for an initial discussion,
because of the function
$\tg$.  The only constraints on this function are
that it be symmetric under the interchange of any two arguments with the
same sign and that it be
dimensionless, so that ratios of momenta must be employed
unless there is a cutoff on longitudinal momenta that can be used
to form $\tg$.  I assume that such a cutoff does not
exist, because one must typically employ $T^+$ to control a cutoff on
longitudinal momenta, and I show below that perturbation theory cannot
be used to construct $T^+$.  We will see below that spectator momenta
also enter $\tg$ for some transformations.
Ultimately one would like to use Lorentz
covariance to place restrictions on $u_4$, but boost invariance alone
places no further restrictions.  Let me first
concentrate on the simplest choice,

$$\tg(q_1^+,q_2^+,q_3^+,q_4^+) = g \;. \eqno(6.3)$$

\noindent  This leads to the familiar canonical
$\phi^4$ Hamiltonian, with no mass
term.  We will see that a mass term is always generated by the
transformation.

Figure 3 shows the second-order corrections to the effective
Hamiltonian that arise when we use $T^\perp$ or any other
transformation.  Let me begin by studying
one of the corrections to $u_4$, the first
Hamiltonian diagram in figure 3b.
The analytic expression corresponding to this diagram is
determined by the second-order term in Eq. (4.14), and I further
simplify the calculation by considering only the part of the expression
that involves the energy of the incoming state.
The relevant momenta are shown in the
figure, but the analytic expression is drastically simplified if one
chooses the Jacobi variables,

$$\eqalign{ &(p_1^+,\pp_1) = (y \calp^+, y \calp^\perp+\rp)
\;,\;\;\; (p_2^+,\pp_2) = ((1-y) \calp^+, (1-y) \calp^\perp-\rp)\;, \cr
&(k_1^+,\kp_1) = (x \calp^+, x \calp^\perp+\sp)
\;,\;\;\; (k_2^+,\kp_2) = ((1-x) \calp^+, (1-x) \calp^\perp-\sp)\;.}
\eqno(6.4)$$

\noindent  Note that all longitudinal momentum fractions, such as $x$
and $y$ in Eq. (6.4), range from 0 to 1.  The total momentum is
$\calp$.
Using these variables one finds that the correction is

$$\eqalign{ \delta v_4 = {1 \over 2}\cdot {1 \over 2} \cdot g^2\; \int
& \dsx  \Bigl[{\rp^2 \over y(1-y)}
- {\sp^2 \over x(1-x)}\Bigr]^{-1} \cr
&\theta\bigl(\Lambda_0^2-(x
\calp^\perp+\sp)^2\bigr)
\theta\bigl((x \calp^\perp+\sp)^2 -
\Lambda_1^2\bigr) \cr
&\theta\bigl(\Lambda_0^2-((1-x)\calp^\perp-\sp)^2\bigr)
\theta\bigl(((1-x) \calp^\perp-\sp)^2 - \Lambda_1^2\bigr) \cr
&\theta\bigl(\Lambda_1^2-(y\calp^\perp+\rp)^2\bigr)
\theta\bigl(\Lambda_1^2-((1-y)\calp^\perp-\rp)^2\bigr)
\;.}\eqno(6.5)$$

\noindent  I use the notation $\delta v_4$ to indicate that the
rescaling operations in $T^\perp$ have not been completed.
The first factor of $1/2$ is seen in Eq. (4.14),
while the second factor of $1/2$
is a combinatoric factor that can be seen in Eq. (A.16) in Appendix A.
The final step functions are associated with the incoming and outgoing
particles, whose momenta must satisfy $0 \le \pp^2 \le \Lambda_1^2$
before the rescaling operations.  Usually I do not display the step
function cutoffs associated with the incoming and outgoing states,
because they are always understood to be present.
After completing the
rescaling operations of Eqs. (3.10) and (3.11), and multiplying by
$z^\perp$, one obtains one of the second-order contributions to $\delta
u_4$.

There are three important features of Eq. (6.5) upon which I want to focus.
First, the cutoffs employed in $T^\perp$ lead to a somewhat complicated
analytical analysis, because of their dependence on
the total transverse momentum of the incoming
particles.  However, one might normally be willing to live with this
difficulty because it affects only irrelevant operators in $u_4$.  Such
difficulties also occur for marginal operators in $u_2$.
The second feature
of Eq. (6.5) is its dependence on $\rp$ and $y$.  We started with a very
simple expression for $u_4$ given in Eq. (6.3), and after one
application of $T^\perp$ we generate a complicated $u_4$.  The
renormalization group analysis is useful if we can show that most
of this complexity is associated with irrelevant operators.  Remembering
that any powers of transverse momenta that appear in $u_4$ are
irrelevant operators, we should expand the integrand in powers of $\rp$.
This leads to

$$\eqalign{ \delta v_4 = -{g^2 \over 4}\;  \int
& \ds  {1 \over \sp^2}  \sum_{n=0}^\infty \Bigl[{x(1-x) \over y(1-y)}
{\rp^2 \over \sp^2}\Bigr]^n \cr
&\theta\bigl(\Lambda_0^2-(x
\calp^\perp+\sp)^2\bigr)
\theta\bigl((x \calp^\perp+\sp)^2 -
\Lambda_1^2\bigr) \cr
&\theta\bigl(\Lambda_0^2-((1-x)\calp^\perp-\sp)^2\bigr)
\theta\bigl(((1-x) \calp^\perp-\sp)^2 - \Lambda_1^2\bigr)
\;.}\eqno(6.6)$$

\noindent I do not display the cutoffs associated with the incoming and
outgoing particles.
There may be little
problem with the convergence of this expansion as
far as the ratio $\rp^2/\sp^2$ is concerned, because the cutoffs insure
that this ratio is usually less than one.  It is not guaranteed that
this ratio is always less than one; however, we will find a far more
serious problem.
I choose
$\Lambda_1=\Lambda_0/2$ in remaining calculations, unless specified
otherwise.
Consider the simplest case in which
$\calp^\perp=0$, so that

$$\delta v_4 = -{g^2 \over 32 \pi^2}
\Biggl\{ln(2)+\sum_{n=1}^\infty {\sqrt{\pi} \over 2^{2n+1}} {\Gamma(n+1)
\over \Gamma(n+3/2)} {1-(1/2)^{2n} \over 2n} \Bigl[{\rp^2 \over
y(1-y)\Lambda_1^2}\Bigr]^n \Biggr\} \;. \eqno(6.7)$$

\noindent  After rescaling the momenta according to Eq. (3.10) we
obtain a Hamiltonian that contains

$$\delta u_4 = -{ g^2 \over 32 \pi^2}
\Biggl\{ln(2)+\sum_{n=1}^\infty {\sqrt{\pi} \over 2^{2n+1}} {\Gamma(n+1)
\over \Gamma(n+3/2)} {1-(1/2)^{2n} \over 2n} \Bigl[{\rp^2 \over
y(1-y)\Lambda_0^2}\Bigr]^n \Biggr\} \;. \eqno(6.8)$$

The first term in this expansion is identical in form to the interaction
with which we began, and it causes the strength of this marginal
operator to decrease as the cutoff decreases.
If we add the correction involving the outgoing
energy, the first
Hamiltonian diagram in figure 3b alters the marginal part of
$u_4$ by a factor of $-ln(2)/(16 \pi^2)\;g^2$.
The remaining terms all correspond to irrelevant
operators.
Before we can drop such
irrelevant operators as a first approximation, however, we should show
that not only do they occur with small coefficients, they also continue
to lead to small corrections.  In order to see that this is {\it not} the
case, let me simplify the problem by considering a Hamiltonian that
contains only the irrelevant interaction

$$u_4(q_1,q_2,q_3,q_4)=h \Bigl[(q_1^+ + q_2^+) \bigl(
{\qp^2_1 \over q_1^+ \Lambda_0^2} + {\qp^2_2 \over q_2^+ \Lambda_0^2}\bigr)
+ \; permutations \; \Bigr] \;. \eqno(6.9)$$

\noindent While it may not be obvious, after a change of variables
this interaction leads to the
first irrelevant operator in Eq. (6.8).  When we apply the
transformation $T^\perp$ to the Hamiltonian containing this irrelevant
interaction we encounter
Hamiltonian diagrams identical in form to those shown in
figure 3; however, instead of the vertex in Eq. (6.3) we have the
vertex in Eq. (6.9).  Following steps analogous to those leading up to
Eq. (6.6) one set of resultant terms is

$$\eqalign{ \delta v_4 = -{h^2 \over 4} \; \int
& \dsx  {1 \over \Lambda_0^2}  \sum_{n=0}^\infty \Bigl[{x(1-x) \over y(1-y)}
{\rp^2 \over \sp^2}\Bigr]^n \cr
&\theta\bigl(\Lambda_0^2-(x
\calp^\perp+\sp)^2\bigr)
\theta\bigl((x \calp^\perp+\sp)^2 -
\Lambda_1^2\bigr) \cr
&\theta\bigl(\Lambda_0^2-((1-x)\calp^\perp-\sp)^2\bigr)
\theta\bigl(((1-x) \calp^\perp-\sp)^2 - \Lambda_1^2\bigr)
\;.}\eqno(6.10)$$

\noindent While this is almost identical to Eq. (6.6), the $n=0$ term is
infinite.  This problem inevitably results when one applies $T^\perp$;
but it is much worse than this of course, because one generates
arbitrarily high inverse powers of longitudinal momentum fractions, as
is apparent in Eq. (6.7).  The
cutoffs do not prevent these fractions from becoming arbitrarily small,
and divergences inevitably result.  Let me note that $T^\perp$ is the
type of transformation one must employ when transverse lattice
regularization is used and one wants to vary the transverse lattice
spacing \APrefmark{\rBARONE,\rBARTWO,\rGRIFFIN}.

Placing cutoffs on the longitudinal momentum, as required if we use
$T^+$, alters this problem slightly; but it does not cure the
problem.  Once we have placed a cutoff on longitudinal momentum,
$\epsilon_0$, we are
free to consider the effective four-point interaction for particles that
have much larger longitudinal momentum, $\calp^+$,
than this cutoff.  The longitudinal momentum
cutoff then shows up in Eq. (6.10) as a small cutoff on the $x$
integration, leading to a factor of $ln(\calp^+/\epsilon_0)$ instead of
$\infty$.  This is an improvement, but the analysis of higher-order
irrelevant operators in Eq. (6.17), for example, leads to
arbitrarily large powers of $\calp^+/\epsilon_0$,
and we have no way to prevent this ratio from
producing arbitrarily large coefficients.  The sceptical reader should
explore this problem much further, but I do not believe that there are
any simple solutions short of abandoning perturbation theory.

All of the above problems can be traced to the fact that some of the
energies of states retained by $T^\perp$, as determined by the fixed
point Hamiltonian, are larger than some of the
energies of states removed.  The energy denominators must be expanded in
powers of the energy of incoming and outgoing states to separate
relevant, marginal, and irrelevant operators; and this expansion does
not converge.  The problem is extremely severe for $T^\perp$ because
states of infinite energy are retained, leading to infinite errors when
irrelevant operators are dropped.
Similar problems are encountered when one studies $T^+$.
The cutoffs in $T^+$ require one to integrate over
transverse momenta from zero to the transverse momentum
cutoff.  For massless Hamiltonians
this leads to transverse infrared divergences, and even for massive
Hamiltonians one again finds that the
irrelevant longitudinal operators end up producing arbitrarily
large effects if
the external transverse
momenta are large.  I do not detail this problem further,
because the calculations are quite similar to those above, with the only
changes being in the cutoffs that occur inside the integrals.

If we consider a boost-invariant transformation that places any cutoff
on transverse momenta without limiting small longitudinal momenta, we
encounter exactly the same problems discussed above.  Thus, there
is no reason to consider the transformation that utilizes the cutoff in
Eq. (3.22).  The cutoffs in Eqs. (3.23) and (3.24) limit the
longitudinal momenta, and I show that the transformations
associated with them apparently avoid the above problems.
However, both of these cutoffs involve a sum over all particles present
in a given state.  As we shall see, this leads to second-order
corrections that are spectator-dependent.
Spectator
dependence may not drastically complicate a perturbative
renormalization group analysis.

The boost-invariant transformation that utilizes the cutoff in Eq.
(3.23)
avoids the problems found for $T^\perp$ in second-order.
Using this transformation, and starting with a
Hamiltonian that contains the interaction in Eq. (6.3), the correction
in figure 3b becomes,

$$\eqalign{ \delta v_4 = {g^2 \over 4}\;  \int
& \dsx  \Bigl[{\rp^2 \over y(1-y)}
- {\sp^2 \over x(1-x)}\Bigr]^{-1} \cr
&\theta\Bigl(\Lambda_0^2-{\sp^2 \over x(1-x)}\Bigr)
\theta\Bigl({\sp^2 \over x(1-x)} -
\Lambda_1^2\Bigr) \theta\Bigl(\Lambda_1^2-{\rp^2 \over y(1-y)}\Bigr)
\;.}\eqno(6.11)$$

\noindent This is identical to Eq. (6.5) except for the cutoffs.  There
is an important assumption that is made in writing Eq. (6.11).  I have
assumed that there are no spectators, so that the interactions displayed
in figure 3 occur without any disconnected lines present.
I return to
this issue later, and discuss spectator effects.  At this point we can
expand the denominator exactly as was done above, but now the cutoffs
guarantee that

$$ {\sp^2 \over x(1-x)} \ge {\rp^2 \over y(1-y)} \;. \eqno(6.12)$$

\noindent Following the same steps that led to Eq. (6.7) we find

$$\eqalign{ \delta v_4 &= -{g^2 \over 4}\;  \int
\ds  {1 \over \sp^2}  \sum_{n=0}^\infty \Bigl[{x(1-x) \over y(1-y)}
{\rp^2 \over \sp^2}\Bigr]^n \cr
&\qquad \qquad \qquad \theta\Bigl(\Lambda_0^2-{\sp^2 \over x(1-x)}\Bigr)
\theta\Bigl({\sp^2 \over x(1-x)} -
\Lambda_1^2\Bigr) \cr
&= -{ g^2 \over 32 \pi^2}
\Biggl\{ln(2)+\sum_{n=1}^\infty
{1-(1/2)^{2n} \over 2n} \Bigl[{\rp^2 \over
y(1-y)\Lambda_1^2}\Bigr]^n \Biggr\}
\;.}\eqno(6.13)$$

\noindent I have not displayed the step function cutoffs on the incoming
energy.
To complete the transformation we need to rescale according to
Eq. (3.10), and as we found above in Eq. (6.8), the only effect this has
on $\delta u_4$ is to change $\Lambda_1$ in Eq. (6.13) into $\Lambda_0$.

This result
is similar to the result for $T^\perp$, and to see
that problems do not arise when we add these irrelevant operators and
apply the transformation a second time we need to follow the steps
leading to Eq. (6.10).  Starting with a Hamiltonian that contains the
interaction given in Eq. (6.9), and applying an invariant-mass
transformation, we derive

$$\eqalign{ \delta v_4 &= -{h^2 \over 4} \; \int
\dsx  {1 \over \Lambda_0^2}  \sum_{n=0}^\infty \Bigl[{x(1-x) \over y(1-y)}
{\rp^2 \over \sp^2}\Bigr]^n \cr
&\qquad \qquad \qquad \theta\Bigl(\Lambda_0^2-{\sp^2 \over x(1-x)}\Bigr)
\theta\Bigl({\sp^2 \over x(1-x)} -
\Lambda_1^2\Bigr) \cr
&= -{h^2 \over 32 \pi^2}
\Biggl\{{3 \over 8}+ {ln(2) \over 4}
{\rp^2 \over y(1-y)\Lambda_1^2} +\sum_{n=2}^\infty
{1-(1/2)^{2n-2} \over 2n-2} \Bigl[{\rp^2 \over
y(1-y)\Lambda_1^2}\Bigr]^n \Biggr\}
\;.}\eqno(6.14)$$

\noindent Here again the transformation must be completed by rescaling
the momenta, which again replaces $\Lambda_1$ with $\Lambda_0$.  Not
only is every operator in the sum bounded by cutoffs on external
momenta, every coefficient in the expansion is small when $h$ is
small.  Since these irrelevant operators occur at ${\cal O}(g^2)$,
the corrections in Eq. (6.14) are ${\cal O}(g^4)$.
This should give us hope that irrelevant operators do indeed
lead to small corrections in an analysis that starts by simply
discarding such operators at each stage, at least
while the Hamiltonian remains
near the Gaussian fixed point.

Unfortunately, it is not sufficient to show that the coefficients of the
irrelevant operators are small,  and that each subsequent correction
produced by each irrelevant operator is small.  One should also worry
about the convergence of the corrections produced by the entire sum of
irrelevant operators.  For example,
the reader should be concerned with the convergence of the sum in Eq.
(6.13).  Completing the transformation by rescaling $\rp$, and
completing the sum leads to

$$\eqalign{ \delta u_4
&= -{ g^2 \over 32 \pi^2}
\Biggl\{ln(2)+ {1 \over 2}\;ln\Bigl( {4y(1-y)\Lambda_0^2-\rp^2 \over
4y(1-y)\Lambda_0^2-4\rp^2}\Bigr)
\Biggr\}\;\theta\Bigl(\Lambda_0^2 - {\rp^2 \over y(1-y)} \Bigr)
\;.}\eqno(6.15)$$

\noindent I have restored the cutoff on the incoming energy, and one can
clearly see a potential problem coming from the fact that the logarithm
diverges when the incoming energy reaches its maximum value.  This
divergence occurs because the energy denominator in Eq. (6.11) vanishes
on a surface when the incoming energy equals $\Lambda_1^2$.  However,
this divergence is simply
being buried in the irrelevant operators.  How can this
make sense?

There are two issues one must address.  First, one must determine
whether this divergence actually shows up in observables if the
Hamiltonian containing the interaction in Eq. (6.15) is solved.  While I
do not go through the details, if one studies two-particle scattering
in perturbation theory
with such a Hamiltonian, the logarithmic divergence in Eq. (6.15) enters
first-order perturbation theory with a strength of ${\cal O}(g^2)$.
In second-order perturbation theory there is an additional contribution
of ${\cal O}(g^2)$ coming from the interaction in Eq. (6.3).  This
contribution also has a logarithmic divergence that tends to cancel the
divergence from the interaction in Eq. (6.15).  If one drops the
irrelevant operators, one finds that this latter divergence is not
canceled and a large error is made if the invariant-mass of the
scattering state is near the cutoff.  Of course, one expects large
errors near the cutoff; but well below the cutoff one hopes that errors
are small even when irrelevant operators are dropped, and that the
results are systematically improved as the leading irrelevant operators
are retained.  At least to leading orders in the coupling it is easy to
verify that this happens.

The second issue is directly relevant to the renormalization group.  One
must determine whether the divergence in Eq. (6.15) causes the
irrelevant operators to have a significant effect on the next
Hamiltonian produced by a transformation.  To determine this, include
the entire logarithm in a vertex and study the second-order correction
in the first
Hamiltonian diagram of figure 3b, using this logarithmic vertex in
combination with the original vertex from Eq. (6.3).  Keeping only the
marginal part of $\delta v_4$, and not worrying about combinatoric
factors, we obtain

$$\eqalign{ \delta v_4 \approx  g^3\; \int
& \dsx  \;{x(1-x) \over \sp^2}
\Biggl\{ln(2)+\sum_{n=1}^\infty
{1-(1/2)^{2n} \over 2n} \Bigl[{\sp^2 \over
x(1-x)\Lambda_0^2}\Bigr]^n \Biggr\} \cr
& \qquad \qquad \theta\Bigl(\Lambda_0^2-{\sp^2 \over x(1-x)}\Bigr)
\theta\Bigl({\sp^2 \over x(1-x)} -
\Lambda_1^2\Bigr)
\;.}\eqno(6.16)$$

\noindent I have restored the sum to facilitate the discussion of
convergence.  The integrals are easily completed, leading to

$$\eqalign{ \delta v_4 \approx & {g^3 \over 8 \pi^2}\;
\Biggl\{\bigl[ln(2)\bigr]^2 +\sum_{n=1}^\infty \Bigl[{1-(1/2)^{2n} \over
2n}\Bigr]^2 \Biggr\} \cr
\approx & {g^3 \over 8 \pi^2}\;\Biggl\{0.480+0.141+0.055+0.027+\cdot
\cdot \cdot \Biggr\}
\;.}\eqno(6.17)$$

Clearly this sum converges.  It is also clear that it converges rather
slowly, with an error that falls off as the inverse power of the number
of terms included.
If one wants to include corrections of ${\cal
O}(g^3)$, such as the third-order corrections to the transformation
studied in the next Section,
without making a large (\eg, 25\%) error in the
coefficient of these corrections, it is also necessary to include
some irrelevant operators.  However, if $g$ is small, it is
apparent that these irrelevant operators produce small corrections to
the second-order analysis, despite the fact that the logarithm arising
in Eq. (6.13) diverges.  This is possible because the divergence is
integrable.

Next I want to show that the transformation generates a mass term in
$u_2$,
and that the mass counterterm required when we try to
keep the
physical mass zero is rather unusual.
In figure 3a I show the second-order correction to the Hamiltonian that
affects $u_2$, and in Appendix B I analyze this correction for arbitrary
$u_4$.  I use
the interaction in Eq. (6.3).
To proceed,
define the Jacobi variables

$$\eqalign{
&(k_1^+,\kp_1) = (x \calp^+, x \calp^\perp+\qp)
\;,\;\;\; (k_2^+,\kp_2) = (y \calp^+, y \calp^\perp+\rp)\;, \cr
& \qquad \qquad \qquad (k_3^+,\kp_3) = (z \calp^+, z \calp^\perp+\sp).}
\eqno(6.18)$$

\noindent  Using these variables the second-order correction to $u_2$
is

$$\eqalign{ \delta v_2 = - {g^2 \over 3!}\; \int & \dq \dr \dss
\Bigl[{\qp^2 \over x} + {\rp^2 \over y} + {\sp^2 \over z}
\Bigr]^{-1} \cr
&(16 \pi^3) \delta(1-x-y-z) \delta^2(\qp+\rp+\sp)\cr
&\theta\Bigl(\Lambda_0^2- {\qp^2 \over x} - {\rp^2 \over y} - {\sp^2
\over z} \Bigr)
\theta\Bigl({\qp^2 \over x} + {\rp^2 \over y} + {\sp^2 \over z} -
\Lambda_1^2\Bigr)
\;.}\eqno(6.19)$$

\noindent This expression does not depend on $\calp$, so it leads only
to a mass shift and does not change the marginal operators in $u_2$ or
produce any irrelevant operators in $u_2$.  This conclusion changes
when spectators are included.

It should be obvious that the
integral does not vanish,
and therefore $u_2$ develops a negative mass-squared term
if we start with a
Hamiltonian that has no mass term.

It is important to note
that the mass shift is not infinite, despite the fact that particles
with small transverse momenta are removed.  If transverse infrared
divergences appear, transverse nonlocalities may follow.  Consider
a Euclidean field theory and place a cutoff on the four-momentum
squared, $q^2$.  When states with a given large range of $q^2$ are
removed, it is not required that every component of the momentum be
small.  If $q_0^2$ is large, for example, states with small
values of $q_3^2$ are removed, and one might naively worry that this
produces long-range forces in the $x^3$ direction.  Of course this
does not happen, because of rotational invariance.  In light-front field
theory the invariant-mass cutoff violates strict rotational invariance;
however, it retains some features of this symmetry and apparently allows
one to remove states with small transverse momenta without producing
long-range transverse interactions, at least in low orders of
perturbation theory.  The most interesting examples of
this principle are discussed later in this Section.

As discussed in Section \rntwo, we want to choose
the Hamiltonian $H^{\Lambda_0}_{\Lambda_0}$
so that $H^{\Lambda_0}_{\Lambda_\calN}$ gives reasonable results when it is
diagonalized.
Since there is no inverse transformation, this
process is typically trial and error in a nonperturbative analysis.
One first tries a particular
$H^{\Lambda_0}_{\Lambda_0}$, and constructs $H^{\Lambda_0}_{\Lambda_\calN}$.
If $H^{\Lambda_0}_{\Lambda_\calN}$ does not produce reasonable observables,
$H^{\Lambda_0}_{\Lambda_0}$ must be altered.
While there are an
infinite number of operators that can be adjusted in
$H^{\Lambda_0}_{\Lambda_0}$, only the relevant and marginal operators
are expected to produce effects that survive after many applications of
$T$; therefore one hopes that it is necessary to control only a finite
number of readily identified terms in $H^{\Lambda_0}_{\Lambda_0}$
to produce desired results in $H^{\Lambda_0}_{\Lambda_\calN}$.  One
also hopes that the irrelevant operators in $H^{\Lambda_0}_{\Lambda_\calN}$
are less important than the relevant and marginal operators, when
$H^{\Lambda_0}_{\Lambda_\calN}$ is diagonalized, but there is no guarantee
of this and it is not essential to the renormalization group analysis.
If this happens, one is led to consider the relationship
between the coefficients of the relevant and marginal operators in
$H^{\Lambda_0}_{\Lambda_0}$ to the corresponding coefficients in
$H^{\Lambda_0}_{\Lambda_\calN}$.  There is no guarantee that this
relationship is simple.  It could even be chaotic, in which case one
may want to find a new problem.  However, when one evolves Hamiltonians near
a Gaussian fixed point, the relationship between coefficients at the
beginning
and end of a long trajectory should not be highly nonlinear, as each
transformation introduces only small nonlinearities.  Wilson has given a
general discussion of this issue \APrefmark{\rWILTEN},
and the main new ingredient in the
light-front renormalization group
is the appearance of arbitrary functions of longitudinal momenta.

The mass terms in $u_2$ provide a simple example.  I refer to any terms
in $u_2$ that do not depend on transverse momenta, regardless of their
dependence on longitudinal momenta, as {\it mass} terms.  For the
purposes of this discussion I simply regard the mass term
that appears with no dependence on longitudinal momentum
in $u_2$ in $H^{\Lambda_0}_{\Lambda_\calN}$ as the {\it physical mass}.  In
reality one must still solve the Schr\"odinger equation with
$H^{\Lambda_0}_{\Lambda_\calN}$ to relate the mass in
$H^{\Lambda_0}_{\Lambda_\calN}$ to an experimentally observed mass, and it
is even possible that there is no experimentally observed mass that
directly corresponds to the mass in $H^{\Lambda_0}_{\Lambda_\calN}$.

While
the mass terms in $H^{\Lambda_0}_{\Lambda_\calN}$
are affected by all of the terms in
$H^{\Lambda_0}_{\Lambda_0}$, if one uses a second-order approximation of
$T$ to construct a Hamiltonian trajectory, any mass term in
$H_{\Lambda_0}^{\Lambda_0}$ appears directly in
$H^{\Lambda_0}_{\Lambda_\calN}$.  Thus, in a second-order analysis, it
is trivial to control the mass term in $H^{\Lambda_0}_{\Lambda_\calN}$.
If we want to produce a Hamiltonian
$H^{\Lambda_0}_{\Lambda_\calN}$ in which the physical mass is zero for
example, we
add a mass `counterterm' to $H^{\Lambda_0}_{\Lambda_0}$ and adjust it
to cancel the entire trajectory of mass shifts that begins with the
shift in Eq. (6.19).  This simplicity is lost in third- and higher-order
analyses, as we will see in Section \rnseven.  On the other hand, we
will find that the coupling constant coherence conditions fix the mass
counterterm to each order in perturbation theory to a value
that yields a massless physical particle in that order of
perturbation theory \APrefmark{\rPERWIL}.

Let me return to the issue of spectator-dependence.
Figure 4 is identical to figure 3a, but I have added a double line to
represent all spectators.  The calculation proceeds almost exactly as
before, except we must include the effects of the spectator momentum.
I use the same variables given in Eq. (6.18), and I assume
that the incoming particle momentum and the spectator momentum are
respectively,

$$(w \calp^+,w \calp^\perp + \tp) \;,\;\;\;\;\; ((1-w) \calp^+, (1-w)
\calp^\perp - \tp) \;. \eqno(6.20)$$

\noindent We also need to know the invariant mass-squared
of the spectator state, which I
assume to be $M^2$.  Even if the
particles are massless, we cannot assume that the invariant mass of the
spectators is zero, of course.
The spectator energy
cancels in all energy denominators when no
interactions occur between spectators.  Such interactions produce
disconnected marginal and irrelevant operators that I do not discuss,
because they are not important in low orders of a perturbative
analysis.
The second-order correction to $u_2$ from the
Hamiltonian diagrams in figure 4 is

$$\eqalign{ \delta v_2 = {g^2 \over 3!} \; \int & \dq \dr \dss
\Bigl[{\tp^2 \over w}-
{\qp^2 \over x} - {\rp^2 \over y} - {\sp^2 \over z}
\Bigr]^{-1} \cr
&(16 \pi^3) \delta(w-x-y-z) \delta^2(\tp-\qp-\rp-\sp) \cr
&\theta\Bigl(\Lambda_0^2- {\tp^2+M^2 \over 1-w} -
{\qp^2 \over x} - {\rp^2 \over y} - {\sp^2
\over z} \Bigr) \cr
&\theta\Bigl({\qp^2 \over x} + {\rp^2 \over y} + {\sp^2 \over z}
+ {\tp^2+M^2 \over 1-w} - \Lambda_1^2\Bigr)
\;.}\eqno(6.21)$$

One can compare this result with Eq. (6.19) to see the effect spectators
have on $\delta v_2$.  In Eq. (6.19) $\delta v_2$ does not depend on the
longitudinal momentum of the incoming boson, or the
transverse momentum of the incoming boson.  In Eq. (6.21) $\delta v_2$
depends on the longitudinal momentum fraction of the incoming boson,
$w$; and it depends
on the relative transverse momentum of this
boson with respect to the rest of the system.

When faced with
any correction to the Hamiltonian, we are supposed to expand the
correction in terms of relevant, marginal and irrelevant variables.  In
this case, this means we are supposed to expand $\delta v_2$ in powers
of the transverse momentum $\tp$.  In addition, we are
supposed to expand in powers of $M^2$.  In the fixed point Hamiltonian
the particles are massless, and
$M^2$ is a function of the transverse momenta of the spectator
particles that goes to zero when these momenta go to zero.  Remember
that masses are treated as perturbations.
The mass terms in $\delta v_2$ are found by setting all
transverse momenta to zero, leading to the relevant part of $\delta
v_2$,

$$\eqalign{ \delta v_{2R} = - {g^2 \over 3!}\; \int & \dq \dr \dss
\Bigl[{\qp^2 \over x} + {\rp^2 \over y} + {\sp^2 \over z}
\Bigr]^{-1} \cr
&(16 \pi^3) \delta(w-x-y-z) \delta^2(\qp+\rp+\sp) \cr
&\theta\Bigl(\Lambda_0^2-
{\qp^2 \over x} - {\rp^2 \over y} - {\sp^2
\over z} \Bigr)
\theta\Bigl({\qp^2 \over x} + {\rp^2 \over y} + {\sp^2 \over z}
- \Lambda_1^2\Bigr)
\;.}\eqno(6.22)$$

\noindent  This expression is identical to the similar limit for Eq.
(6.19), except for one very important difference.  The longitudinal
momentum fractions in the integrand add to $w$ instead of 1.

We need to understand how the mass terms that arise in $u_2$ after
repeated application of $T$ depend on longitudinal momenta, and here
this problem is equivalent to understanding how the mass depends on $w$.
This dependence is easily worked out by changing variables,

$$x'={x \over w} \;,\;y'={y \over w} \;,\;z'={z \over w} \;,\;
\qp'={\qp \over \sqrt{w} \Lambda_0}\;,\;
\rp'={\rp \over \sqrt{w} \Lambda_0}\;,\;\sp'={\sp \over \sqrt{w} \Lambda_0} \;.
\eqno(6.23)$$

\noindent Using these variables, and using the fact that
$\Lambda_1=\Lambda_0/2$, we obtain

$$\eqalign{ \delta v_{2R} = - {g^2 \over 3!}\;w \Lambda_0^2
\; \int & \dqp \drp \dssp
\Bigl[{\qp'^2 \over x'} + {\rp'^2 \over y'} + {\sp'^2 \over z'}
\Bigr]^{-1} \cr
&(16 \pi^3) \delta(1-x'-y'-z') \delta^2(\qp'+\rp'+\sp') \cr
&\theta\Bigl(1-
{\qp'^2 \over x'} - {\rp'^2 \over y'} - {\sp'^2
\over z'} \Bigr)
\theta\Bigl({\qp'^2 \over x'} + {\rp'^2 \over y'} + {\sp'^2 \over z'}
- {1 \over 4} \Bigr)
\;.}\eqno(6.24)$$

\noindent The remaining integral is a finite
number that has no dependence on
any momenta, so we have exactly determined the dependence of the new
mass term on the longitudinal momentum fraction.
The new mass term
is proportional to the longitudinal momentum fraction, unlike the
physical mass which is independent of this fraction!

The next step in the analysis is to complete the calculation of
marginal and irrelevant
operators that arise in Eq. (6.21).
The mass term is the only relevant
operator occurring in Eq. (6.21), and the only marginal operator is the
piece of $\delta v_2$ that is quadratic in the external transverse
momenta, including $M^2$.
Using the new variables in Eq. (6.23), we find

$$\eqalign{ \delta v_2 =  {g^2 \over 3!}\;w \Lambda_0^2
\; \int & \dqp \drp \dssp
\Bigl[{\tp^2 \over w \Lambda_0^2} -
{\qp'^2 \over x'} - {\rp'^2 \over y'} - {\sp'^2 \over z'}
\Bigr]^{-1} \cr
&(16 \pi^3) \delta(1-x'-y'-z') \delta^2({\tp \over \sqrt{w} \Lambda_0} -
\qp'-\rp'-\sp') \cr
&\theta\Bigl(1- {\tp^2+M^2 \over (1-w)\Lambda_0^2}-
{\qp'^2 \over x'} - {\rp'^2 \over y'} - {\sp'^2
\over z'} \Bigr) \cr
&\theta\Bigl({\qp'^2 \over x'} + {\rp'^2 \over y'} + {\sp'^2 \over z'}
+ {\tp^2+M^2 \over (1-w)\Lambda_0^2} - {1 \over 4} \Bigr)
\;.}\eqno(6.25)$$

\noindent
The factor of $\tp^2/(w\Lambda_0^2)$ in the energy denominator
and the factor of $\tp/(\sqrt{w}\Lambda_0)$ in the momentum conserving delta
function each produce a quadratic term in $\delta v_2$ proportional to
$\tp^2$, with no dependence on longitudinal momentum.  This
produces a term in the energy proportional to $\tp^2/(w \calp^+)$.
The terms proportional to $\tp^2+M^2$ in the
cutoffs produce a term in $\delta v_2$ that is proportional to
$w(\tp^2+M^2)/(1-w)$, which then produces a term in the energy
proportional to $(\tp^2+M^2)/[(1-w)\calp^+]$.  The factor $M^2$ is
itself a sum of terms that are each of the form $\qp^2/x$, with $x$
being a longitudinal momentum fraction and $\qp$ being a relative
transverse momentum, if there are no physical masses in the theory.
These corrections are
similar to the fixed point $u_2^*$, but with an important difference; they
do not depend on the total transverse momentum.

Assuming $u_2(q)=\qp^2$ in the Hamiltonian in
Eq. (3.7) and using Jacobi variables $(\qp_i,q^+_i)=(x_i
\calp^+,x_i \calp^\perp+\rp_i)$, we can write the terms involving $u_2$
using projection operators and obtain

$$\eqalign{
\int {d^2\calp^\perp d\calp^+ \over 16\pi^3 \calp^+}\;\Biggl\{
&\; \int {d^2r^\perp_1 dx_1 \over 16 \pi^3 x_1 } \;
(16 \pi^3) \delta^2(\rp_1) \delta(1-x_1)
\;\Biggl[ {\calp^{\perp 2} \over \calp^+}+
{\rp^2_1 \over x_1 \calp^+} \Biggr]\;
|q_1 \rangle \langle q_1 |  \cr
+ &\int {d^2r^\perp_1 dx_1 \over 16 \pi^3 x_1 }
\; \int {d^2r^\perp_2 dx_2 \over 16 \pi^3 x_2 } \;
(16 \pi^3) \delta^2(\rp_1+\rp_2) \delta(1-x_1-x_2) \cr
& \qquad \qquad \qquad
\Biggl[{\calp^{\perp 2} \over \calp^+}+
{\rp^2_1 \over x_1 \calp^+} +{\rp^2_2
\over x_2 \calp^+}\Biggr]
\;|q_1,q_2 \rangle \langle q_1,q_2 | \;\;+\;\;\cdot\cdot\cdot \;\Biggr\}
\;.}\eqno(6.26)
$$

\noindent The corrections to $u_2$ that are of ${\cal O}(\tp^2,M^2)$ from
Eq. (6.25) alter the coefficient of each factor $\rp^2_i/(x_i \calp^+)$
in Eq. (6.26).  These corrections differ from standard wave function
renormalization, because the coefficient of each factor $\calp^{\perp
2}/\calp^+$ in Eq. (6.26) remains $1$, and wave function renormalization
would alter this coefficient also.
These corrections are marginal
operators, and one can include them in a fixed point
Hamiltonian if $u_2$ is allowed to be spectator-dependent, so that it
can depend not only on the momentum of a single particle, but also on
the total momentum of the state.  Eq. (6.24)
proves that
$u_2$ must be spectator-dependent if an invariant-mass transformation is
used.  In the original discussion of Hamiltonians in Section \rnthree, it
was simply assumed that $u_2$ depends only on one momentum, and now we
are finding an example in which the transformation forces us to expand
the original definition of the space of Hamiltonians.  This point is
clarified further below.

In order to control these spectator-dependent corrections to $u_2$ in
the final Hamiltonian, $H^{\Lambda_0}_{\Lambda_\calN}$, we must allow
such terms to appear in
$H^{\Lambda_0}_{\Lambda_0}$.  Since these counterterms
are part of $\delta H_l$, they do not modify the second-order behavior
of the transformation, but they do enter at third order.
If the scalar particles
appear as asymptotic particles (\ie, are not confined), we need to
precisely cancel these corrections to $u_2$ to obtain the
appropriate dispersion relation for the physical scalar particles,
as we are forced to introduce mass counterterms to cancel the
corrections found in Eq. (6.24).  Just as it is trivial to cancel any
mass that arises in a second-order analysis, it is trivial to cancel
these marginal corrections to $u_2$.  Again, these corrections may have
nontrivial effects in a third-order analysis.

The analysis of $u_4$ has not been completed,
and we must evaluate the remaining
Hamiltonian diagrams in figure 3b.  It is
convenient to use the variables

$$\eqalign{
&(p_i^+,\pp_i) = (x_i \calp^+, x_i \calp^\perp+\rp_i)
\;,\;\;\; (k_i^+,\kp_i) = (y_i \calp^+, y_i \calp^\perp+\sp_i)\;.}
\eqno(6.27)$$

\noindent
I am only
interested in the correction to the marginal part of $u_4$, so these
diagrams are evaluated with the external transverse momenta and the
total transverse momentum set to
zero.  It is only necessary to explicitly
evaluate the second and sixth Hamiltonian diagrams in figure 3b, as
others are simply related to these two.

The second
Hamiltonian diagram in figure 3b,
combining both second-order terms in Eq. (4.14) involving incoming and
outgoing energy, leads to a correction of the marginal part of $\delta
v_4$,

$$\eqalign{ \delta v_{4M} = - {g^2 \over 2}\; \theta(x_2-x_4)
\; \int
& {d^2s^\perp_1 dy_1 \over 16 \pi^3 y_1}\; \int {d^2s^\perp_2 dy_2 \over
16 \pi^3 y_2}\; (16\pi^3)\delta(x_2-x_4-y_1-y_2)\delta^2(\sp_1+\sp_2)
\cr
&
\theta\bigl(\Lambda_0^2-{\sp^2_1 \over y_1}-{\sp^2_2 \over y_2}\bigr)
\theta\bigl({\sp^2_1 \over y_1}+{\sp^2_2 \over y_2}-
\Lambda_1^2\bigr)
\Bigl[{\sp^2_1 \over y_1}
+ {\sp^2_2 \over y_2} \Bigr]^{-1}
\;.}\eqno(6.28)$$

\noindent
To simplify this expression and remove dependence on external
momenta from the integrand, we can change variables,

$$y_i=(x_2-x_4)z_i \;\;,\;\;\sp_i=\sqrt{x_2-x_4}\Lambda_0\qp_i
\;.\eqno(6.29)$$

\noindent This leads to

$$\eqalign{ \delta v_{4M} = &- {g^2 \over 2}\; \theta(x_2-x_4)
\; \int
{d^2q^\perp_1 dz_1 \over 16 \pi^3 z_1}\; \int {d^2q^\perp_2 dz_2 \over
16 \pi^3 z_2}\; (16\pi^3)\delta(1-z_1-z_2)\delta^2(\qp_1+\qp_2)
\cr
& \qquad \qquad \qquad
\theta\bigl(1-{\qp^2_1 \over z_1}-{\qp^2_2 \over z_2}\bigr)
\theta\bigl({\qp^2_1 \over z_1}+{\qp^2_2 \over z_2}-
{1 \over 4}\bigr)
\Bigl[{\qp^2_1 \over z_1}
+ {\qp^2_2 \over z_2} \Bigr]^{-1} \cr
= &-g^2\;{ln(2) \over 16 \pi^2}\;\theta(x_2-x_4)
\;.}\eqno(6.30)$$

This correction is of exactly the same form as the correction to the
marginal part of $u_4$ from the first
Hamiltonian diagram in figure 3b. The
correction from the third diagram in figure 3b is identical to the
correction in Eq. (6.30), the only difference being that the third
diagram survives when $x_4>x_2$.  The fourth and fifth diagrams contribute
the same amount to the marginal part of $\delta v_4$ as the second and
third diagrams.
The total
correction to the 2-particle $\rightarrow$ 2-particle part of $u_4$ from
the first five diagrams in figure 3b
leads to the second-order transformation,

$$g \rightarrow g - {3 \;ln(2) \over 16 \pi^2} \;g^2
\;.\eqno(6.31)$$

It is a straightforward exercise to compute changes in the irrelevant
parts of $u_4$ by allowing the external transverse momenta to be nonzero
and expanding in powers of these momenta.  It is also straightforward to
determine the effects of spectators on the change of $u_4$.  I return
to the discussion of irrelevant parts of $u_4$ below. Note
that spectators apparently have no effect on the marginal part of $u_4$
for the simple vertex in Eq. (6.3).
The transverse momenta of the spectators are set to zero when computing the
marginal part of $u_4$, so the cutoffs in Eq. (6.30) are not affected.
The only effect is in the momentum conserving delta functions, and using
variables similar to those in Eq. (6.23) one explicitly recovers Eq.
(6.30) unchanged.

We will see later that the marginal part of $u_4$ is inevitably
spectator-dependent.  Higher-order corrections produce
spectator-dependence, and Lorentz covariance and cluster decomposition
require spectator-dependence.  We must adjust the Hamiltonian so that
all observers in frames related to one another by rotations obtain
covariant results, and so that systems of particles that are
not causally connected do not affect one another.

The last six
Hamiltonian diagrams in figure 3b lead to a correction of the marginal
1-particle $\rightarrow$ 3-particle and 3-particle $\rightarrow$
1-particle parts of $u_4$ identical to Eq. (6.31).

The first diagram in figure 3c does not occur, because the intermediate
state always has less energy than the external states, which is not
allowed in a second-order correction.  The second and
third Hamiltonian
diagrams are allowed and lead to the irrelevant operator $u_6$.
There are additional contributions to $u_6$ that I do not show.
The cutoffs require that the external states have an energy below the
cutoff while the intermediate state has an energy above the cutoff.
Moreover, momentum conservation completely determines the momentum of the
single internal particle line.
Using the coordinates
$(p_i^+,\pp_i)=(x_i \calp^+,x_i \calp^\perp+\rp_i)$ and ignoring
spectator effects, one part of the contribution to
$\delta u_6$ from the second diagram in figure 3c is

$$\eqalign{
\delta v_6 = {g^2 \over x_2^+ + x_3^+ - x_6^+}
\;&\Bigl[{\rp^2_2 \over x_2^+}+{\rp^2_3 \over x_3^+}-
{\rp_6^2 \over x_6^+} -
{(\rp_2+\rp_3-\rp_6)^2 \over x_2^+ + x_3^+ - x_6^+} \Bigr]^{-1} \cr
&\theta\bigl(\Lambda_0^2-{\rp^2_1 \over x_1^+}-{\rp_6^2 \over x_6^+}-
{(\rp_2+\rp_3-\rp_6)^2 \over x_2^+ + x_3^+ - x_6^+}\bigr) \cr
&\theta\bigl({\rp^2_1 \over x_1^+}+{\rp_6^2 \over x_6^+}+
{(\rp_2+\rp_3-\rp_6)^2 \over x_2^+ + x_3^+ - x_6^+}- \Lambda_1^2\bigr) \cr
&\theta\bigl(\Lambda_1^2-{\rp^2_1 \over x_1^+}-{\rp_2^2 \over x_2^+}-
{\rp_3^2 \over x_3^+}\bigr)
\theta\bigl(\Lambda_1^2-{\rp^2_4 \over x_4^+}-{\rp_5^2 \over x_5^+}-
{\rp_6^2 \over x_6^+}\bigr)
\;.} \eqno(6.32)$$

\noindent Here I have included the cutoffs associated with the external
lines, because they are important for further analysis of this term.

At this point we should try to proceed by expanding this operator in terms
of increasingly irrelevant operators.  All terms in $u_6$ are
irrelevant if inverse powers of transverse momenta do not arise, but
this correction seems problematic.
We should obtain the leading correction by letting the
external transverse momenta go to zero.  However in this limit the
energy denominator vanishes and the cutoffs go to
zero, forcing us to analyze $\infty \times 0$.
I assume that all external longitudinal momenta remain finite as the
external transverse momenta approach zero.
The leading correction to $u_6$ in this limit becomes

$$\delta v_6 \rightarrow {g^2 \over \kp^2}
\theta\bigl(\Lambda_0^2-{\kp^2 \over y}\bigr)
\theta\bigl({\kp^2 \over
y}- \Lambda_1^2\bigr) \;, \eqno(6.33)$$

\noindent where $\kp \rightarrow 0$, and $y=k^+/\calp^+$.  Here $\kp$ is
the transverse momentum carried by the internal boson line, and $y$ is
its longitudinal momentum fraction.

To analyze this distribution, let us integrate it with a smooth function
of $y$, $f(y)$.  This leads to the integral

$$\eqalign{ &{g^2 \over \kp^2}
\int_0^1 dy \;f(y) \;\theta\bigl(\Lambda_0^2-{\kp^2
\over y}\bigr)\;\theta\bigl({\kp^2 \over y}- \Lambda_1^2\bigr) \cr
= &{g^2 \over \Lambda_0^2}\;\int_0^{\Lambda_0^2/\kp^2} dz\; f\Bigl(
{\kp^2 \over \Lambda_0^2}z\Bigr)\;\theta(z-1)\;\theta(4-z) \cr
\rightarrow & \qquad\qquad {3 g^2 \over \Lambda_0^2}\; f(0)
\;.}\eqno(6.34)$$

\noindent Of course I have used $\Lambda_1=\Lambda_0/2$ again, and in
the last line I have finally completed the limit in which all external
transverse
momenta are taken to zero.  Thus we see that the distribution that
corresponds to the leading correction to $u_6$ is a delta function in
the longitudinal momentum transfer, and as expected the coefficient of
the leading correction to $u_6$ is inversely proportional to
$\Lambda_0^2$, as an irrelevant operator should be.  No long range
transverse interactions are produced by the elimination of high energy
states, and inverse powers of transverse momenta do not appear when one
expands the second-order transformation
in terms of relevant, marginal and irrelevant
operators.  The appearance of delta functions of longitudinal momentum
transfer may have interesting consequences, but they are not important
in low orders of a perturbative analysis.

In figure 3d two of the disconnected
Hamiltonian diagrams that affect $u_8$ are
displayed.  While such disconnected diagrams cancel when there are no
cutoffs on the intermediate energies, this cancellation does not occur
when the cutoffs are in place.  It is possible for the incoming and
outgoing states to have energies just below the cutoff, while the
intermediate states have energies just above the cutoff.  I do not go
through a detailed analysis, because the correction is again irrelevant
and local in the transverse direction,
in this case being proportional to $1/\Lambda_0^4$.

This completes the initial analysis of the second-order transformation for a
massless Hamiltonian with the simple interaction given in Eq. (6.3).
There are two remaining topics in this Section.
First, I want to study repeated second-order transformations.
Second, I want to discuss Lorentz covariance and cluster
decomposition in second-order perturbation theory, and introduce
coupling coherence \APrefmark{\rPERWIL,\rOEHONE-\rKRAUS}.

Let us assume that the second-order behavior of the invariant-mass
transformation is a reasonable approximation of the complete
transformation.  To compute an example trajectory of Hamiltonians,
$H^{\Lambda_0}_{\Lambda_n}$, we can choose $H^{\Lambda_0}_{\Lambda_0}$
to be,

$$\eqalign{ H^{\Lambda_0}_{\Lambda_0} =&
\int {d^2\calp^\perp d\calp^+ \over 16\pi^3 \calp^+}\;\Biggl\{ \cr
&\qquad \; \int {d^2r^\perp_1 dx_1 \over 16 \pi^3 x_1 } \;
(16 \pi^3) \delta^2(\rp_1) \delta(1-x_1)
\;\Biggl[ {\calp^{\perp 2} \over \calp^+}+
(1+\xi_0) {\rp^2_1 \over x_1 \calp^+}+\mu_0^2 \Biggr]\;
|q_1 \rangle \langle q_1 |  \cr
&\qquad + \;\int {d^2r^\perp_1 dx_1 \over 16 \pi^3 x_1 }
\; \int {d^2r^\perp_2 dx_2 \over 16 \pi^3 x_2 } \;
(16 \pi^3) \delta^2(\rp_1+\rp_2) \delta(1-x_1-x_2) \cr
& \qquad \qquad \Biggl[{\calp^{\perp 2} \over \calp^+}+
(1+\xi_0) {\rp^2_1 \over x_1 \calp^+} + (1+\xi_0) {\rp^2_2
\over x_2 \calp^+}+2\mu_0^2\Biggr]
\;|q_1,q_2 \rangle \langle q_1,q_2 | \;\;+\;\;\cdot\cdot\cdot \;\Biggr\}
\cr
+&{g_0 \over 6} \int \dqt_1\; \dqt_2\; \dqt_3\; \dqt_4 \; (16 \pi^3)
\delta^3(q_1+q_2+q_3-q_4)
\; a^\dagger(q_1) a^\dagger(q_2)
a^\dagger(q_3)
a(q_4)  \cr
+&{g_0 \over 4} \int \dqt_1\; \dqt_2\; \dqt_3\; \dqt_4 \; (16 \pi^3)
\delta^3(q_1+q_2-q_3-q_4)
\; a^\dagger(q_1)
a^\dagger(q_2) a(q_3) a(q_4)  \cr
+&{g_0 \over 6} \int \dqt_1\; \dqt_2\; \dqt_3\; \dqt_4 \; (16 \pi^3)
\delta^3(q_1-q_2-q_3-q_4)
\; a^\dagger(q_1)
a(q_2) a(q_3) a(q_4)
\;.}\eqno(6.35)
$$

\noindent  Note that in the critical Gaussian fixed point Hamiltonian,
$\xi$, $\mu$, and $g$ are all zero.
This Hamiltonian contains a complete set of relevant and
marginal operators required for an approximate
second-order analysis.  In this example there
is no sector-dependence in the marginal part of $u_4$, so we are able to
write the marginal interaction without using the awkward projection
operators that are required to display $u_2$.
When the second-order
invariant-mass transformation is applied to this Hamiltonian, and
irrelevant operators are dropped at every stage, the
resultant Hamiltonian contains no new relevant or marginal operators.
In this case
the only effect on the relevant and marginal operators is to change
the constants $\xi_0$, $\mu_0$, and $g_0$.  A complete second-order
analysis would require us to consider more general interactions than
shown in Eq. (6.3), and to retain irrelevant operators.
Irrelevant operators typically produce new relevant and
marginal operators.  Dropping irrelevant operators,
we can write the relevant
and marginal operators in $H^{\Lambda_0}_{\Lambda_n}$
as in Eq. (6.35), using the constants $\xi_n$, $\mu_n$, and
$g_n$.

The approximate second-order transformation can be
summarized by the equations,

$$g_{n+1} = g_n - c_g\;g_n^2
\;, \eqno(6.36)$$

$$\xi_{n+1} = \xi_n + c_\xi \;g_n^2 \;, \eqno(6.37)$$

$$\mu_{n+1}^2 = 4 \mu_n^2 - c_\mu
\;g_n^2\;\Lambda_0^2 \;.
\eqno(6.38)$$

\noindent
Note that if $g_n$ is small, the initial
assumption that the second-order corrections are small in comparison to
the linear corrections is consistent.
I have shown that the constants $c_g$ and $c_\mu$ are
positive, but have not analyzed $c_\xi$.  It is fairly easy to see from
Eq. (6.25) that $c_\xi$ is also positive.
If $g_n$ is
sufficiently small, these equations should be a reasonable
crude approximation
to the entire transformation.

Wilson has discussed how one must solve
such equations so that errors are controlled \APrefmark{\rWILTEN},
and I have already
repeated some of his discussion in Section \rntwo.  Eqs.
(6.36)-(6.38) are much simpler than the general case discussed by
Wilson, and they are readily solved for large $\calN$.
For a massless
scalar theory, $\mu_0^2$ and $\xi_0$ can be adjusted so that
the physical particle has the dispersion relation of a massless
particle.  If one
insists on actually computing the value of $\mu_0^2$ required to obtain
a specific value of $\mu_\calN^2$,  one finds that $\mu_0^2$ must be
controlled to an accuracy of about 1 part in $4^{-\calN}$.  In practice, one
is never interested in $\mu_0^2$ directly, and the rest of the
calculation should be adjusted so that no equations depend on the precise
value of $\mu_0^2$.  It is easiest to fix $\mu_\calN^2$ and solve Eq.
(6.38) towards decreasing $n$.
$\xi_\calN$ should be close to $0$, and $\xi_0$ is adjusted to achieve
this result.  Through second-order in $\delta H^2$, the equation
for $g_n$ decouples from the equations for $\xi_n$ and $\mu_n$; and the
first step in solving the complete set is to solve Eq. (6.36).
As successive
transformations are applied, Eq. (6.36) shows that $g$ decreases.  We
must start with a sufficiently small value of $g$ for perturbation
theory to be reasonable.  An accurate solution of Eq. (6.36) can easily
be constructed by iteration.  A reasonable approximation over any finite
segment of the trajectory is

$$g_n={g_m \over 1+c_g (n-m) g_m }={g_m \over
1+{c_g \over ln(2)} ln\Bigl({\Lambda_m \over \Lambda_n}\Bigr)
g_m }={g_m \over 1+{3 \over 16\pi^2} ln\Bigl({\Lambda_m \over
\Lambda_n}\Bigr) g_m } \;, \eqno(6.39)$$

\noindent where in the last step I have used the result for $c_g$ in Eq.
(6.31).
The error in this approximation grows as $|n-m|$ becomes large.
It is interesting to note that the factor of $ln(2)$ in $c_g$,
which
came from the choice $\Lambda_1=\Lambda_0/2$, drops out of this final
result.
The result is well-known. In the renormalized
Hamiltonian, which as discussed above is obtained by allowing $\calN
\rightarrow \infty$, $g=0$.  This analysis is not complete, of course,
because I have only considered the case where $g_0$ is small;
however, I am only interested in showing that a perturbative analysis
may be possible.

This brings me to the final subject for this Section.  Up to this point,
there has been no discussion of how Lorentz covariance and cluster
decomposition, both of which are violated by an invariant-mass cutoff,
are restored in physical predictions.
Perhaps the most
important observation is that if one succeeds in restoring these
properties in predictions with one cutoff, they are restored for
all cutoffs, because the renormalization group is designed to
preserve the matrix
elements of observables as cutoffs are changed.

I give three simple examples in second-order perturbation theory.
The first example is
the dispersion relation for a single boson in the presence of
spectators.  I compute the second-order
contribution to the Green's function
with the external propagators removed; which is
also the second-order shift in the invariant-mass-squared of the state
when $\epsilon$ (see
Appendix A) is chosen to be the state's on-shell free energy.
If this invariant-mass-squared
shift does not
transform as a scalar under Lorentz transformations, and/or it depends on
the spectators, we must add
counterterms to restore Lorentz covariance and cluster decomposition.
The second example is the boson-boson scattering amplitude; for which
the second-order correction comes from diagrams identical to the first
five in
figure 3b, with spectators added.
This amplitude should be manifestly covariant,
depending only on the invariant-mass of the two bosons that scatter from
one another, with no dependence on spectators.  If these conditions are
not satisfied, we must again add counterterms.  Finally, I list the
correction to the one-boson to three-boson Green's function, given by
the fifth, sixth, and seventh diagrams in figure 3b.

The main differences between second-order perturbation theory diagrams
and second-order Hamiltonian diagrams are the ranges of integration
and the energy denominators.  In
Hamiltonian diagrams the range of integration
is bounded by an upper and lower cutoff, and the energy denominator
always involves the on-mass-shell energies of incoming, outgoing, or
intermediate states.
In time-ordered
perturbation theory only the upper cutoff appears, and the energy in the
denominator is arbitrary.  We have already
seen that a `mass' term appears in $u_2$ with the wrong dispersion
relation when there are spectators, Eq. (6.24);
and we will find a mass with the
wrong dispersion relation when we study the invariant-mass of a boson in
time-ordered perturbation theory.  From the point of view of perturbation
theory, we need to add a counterterm that exactly cancels this mass term
or we do not obtain an invariant-mass shift that transforms
like a scalar.  The counterterm must completely cancel
the mass shift,
and I show that one obtains the same counterterm from the
renormalization group Eqs. (6.36) and (6.38) using the condition that
$\mu_n$ is a function of $g_n$ with no further dependence on $n$.  This
is a coupling constant coherence condition.

The second-order contribution to the boson invariant-mass
corresponding to figure 4 is readily determined using the diagrammatic
rules in Appendix A.
I assume that
the total longitudinal momentum is $\calp^+$ and that the total
transverse momentum is $\calp^\perp$, with
$\epsilon=\calp^-={\calp^{\perp 2} \over \calp^+}$
for massless bosons.
I assume that there are spectators, with all relevant
momenta given in Eq. (6.20).  The invariant-mass of a state with a
boson in the
presence of a massless spectator consists of a kinetic energy term and a
mass term for the boson itself.
At this point I am only interested in the
violation of Lorentz covariance coming from the mass term that
appears in the shift, so I set the relative transverse
momentum $\tp=0$ and assume that
the invariant-mass of the spectators is zero.  In other words, I set all
relative
transverse momenta equal to zero.  In this case, using the Jacobi
variables defined in Eq. (6.18), the boson mass shift is

$$\eqalign{- {g^2 \over 3!} \; \int & \dq \dr \dss
\Bigl[
{\qp^2 \over x} + {\rp^2 \over y} + {\sp^2 \over z}
\Bigr]^{-1} \cr
&(16 \pi^3) \delta(w-x-y-z) \delta^2(\qp+\rp+\sp)
\; \theta\Bigl(\Lambda_0^2-
{\qp^2 \over x} - {\rp^2 \over y} - {\sp^2
\over z} \Bigr)
\;.}\eqno(6.40)$$

\noindent  Lorentz covariance and cluster decomposition require that
this shift be a constant, independent of all longitudinal momenta.

Following the
same steps leading to Eq. (6.24), we find that this mass shift can be
written as

$$\eqalign{ - {g^2 \over 3!}\;w \Lambda_0^2
\; \int & \dqp \drp \dssp
\Bigl[{\qp'^2 \over x'} + {\rp'^2 \over y'} + {\sp'^2 \over z'}
\Bigr]^{-1} \cr
&(16 \pi^3) \delta(1-x'-y'-z') \delta^2(\qp'+\rp'+\sp')
\; \theta\Bigl(1-
{\qp'^2 \over x'} - {\rp'^2 \over y'} - {\sp'^2
\over z'} \Bigr)
\;.}\eqno(6.41)$$

\noindent  This can be related to the constant $c_\mu$ that appears in
the renormalization group Eq. (6.38), and one finds that the mass shift
is

$$-{c_\mu g^2 \over 3}\;w \Lambda_0^2\;,\eqno(6.42)$$

\noindent where we can use Eq. (6.24), to show that

$$\eqalign{ c_\mu= {4 \over 3!}\;
\; \int & \dqp \drp \dssp
\Bigl[{\qp'^2 \over x'} + {\rp'^2 \over y'} + {\sp'^2 \over z'}
\Bigr]^{-1} \cr
&(16 \pi^3) \delta(1-x'-y'-z') \delta^2(\qp'+\rp'+\sp') \cr
&\theta\Bigl(1-
{\qp'^2 \over x'} - {\rp'^2 \over y'} - {\sp'^2
\over z'} \Bigr)
\theta\Bigl({\qp'^2 \over x'} + {\rp'^2 \over y'} + {\sp'^2 \over z'}
- {1 \over 4} \Bigr)
\;.}\eqno(6.43)$$

\noindent To obtain this result for $c_\mu$ one must remember that
$\delta u_4$ contains an extra factor of four not found in Eq. (6.24)
that results from the rescaling part of the transformation.  To go from
Eq. (6.41) to Eq. (6.42), note that the integral in Eq. (6.41) can be
written as an infinite sum of integrals with upper and lower cutoffs,
with each successive cutoff being 1/4 the previous cutoff.  In each of these
integrals one can rescale momenta in exactly the manner used to obtain
Eq. (6.24), leading to a sum $1+1/4+1/16+\cdot\cdot\cdot=4/3$.

The factor of $w$ that appears in Eq. (6.42) shows that the mass shift
is neither covariant nor spectator-independent, and a mass counterterm
must be added to exactly cancel this shift.

Return to Eqs. (6.36) and
(6.38) and ask whether it is possible for $\mu_n^2$ to be a
perturbative function of
$g_n$ with no further dependence on $n$.  In general we can choose any
initial value for $\mu_0$ and solve Eq. (6.38) to find $\mu_n$; but if
we assume that $\mu_n^2=(\alpha g_n+\beta
g_n^2+\cdot\cdot\cdot)\Lambda_0^2$, and substitute this into Eq. (6.38),
using Eq. (6.36), we find $\alpha=0$, $\beta=c_\mu/3$.  To this order we
have a unique result, $\mu_n^2=(c_\mu/3)\; g_n^2\;\Lambda_0^2$.
Higher order
terms are not determined, because Eqs. (6.36) and (6.38) are altered
at $\order(g^3)$.  For this choice of $\mu_n^2$,
$\mu_0^2$ is uniquely determined,

$$\mu_0^2={c_\mu g_0^2 \over 3}\;\Lambda_0^2+\order(g_0^3)\;.\eqno(6.44)$$

\noindent The factor of $w$ found in Eq. (6.42) is included in the
definition of the term in which $\mu_n$ appears, as seen in Eq. (6.35);
therefore, this value of $\mu_0^2$ precisely cancels the entire
mass shift found in second-order perturbation theory and acts to restore
covariance and cluster decomposition.  In other words, with no direct
reference to these properties, one can use the renormalization group
and the coupling constant coherence conditions to
remove the violations caused by the invariant-mass cutoff.

Let me next consider the boson-boson scattering amplitude corresponding
to the first diagram in figure 3b, with spectators added.  I assume that
the total longitudinal and transverse momenta of the two bosons that scatter
are given by the first momentum in Eq. (6.20), with the spectator
momentum being the second momentum.  For simplicity I assume that
the invariant-mass of the spectators is zero. For the
bosons that scatter I use the
Jacobi variables,

$$\eqalign{
&(k_1^+,\kp_1) = (x w\calp^+, x (w\calp^\perp+\tp)+\sp), \cr
&(k_2^+,\kp_2) = ((1-x)w \calp^+, (1-x)(w \calp^\perp+\tp)-\sp)\;.}
\eqno(6.45)$$

\noindent  Letting $\epsilon=(\calp^{\perp 2}+M^2)/\calp^+$, the
scattering amplitude is

$$\eqalign{ & {g^2 \over 2w}\;  \int
 \dsx  \Bigl[M^2-{\tp^2 \over w(1-w)}
- {\sp^2 \over wx(1-x)} +i0_+\Bigr]^{-1} \cr
&\qquad\qquad\qquad\qquad
\theta\Bigl(\Lambda_0^2-{\tp^2 \over w(1-w)}-{\sp^2 \over wx(1-x)}\Bigr)
\;.}\eqno(6.46)$$

\noindent The real part of this amplitude is

$${g^2 \over 32 \pi^2} ln\Biggl({|M^2-{\tp^2 \over w(1-w)}|
\over \Lambda_0^2 -M^2}\Biggr) \;.
\eqno(6.47)$$

Remember that $M^2$
is the invariant-mass of the
entire state, including the spectators.  The invariant-mass of the
two-boson subsystem is

$$(p_1+p_2)^\mu (p_1+p_2)_\mu=w \Bigl(M^2-{\tp^2 \over w(1-w)}\Bigr)
\;.\eqno(6.48)$$

\noindent The
boson-boson scattering amplitude is
neither covariant nor spectator-independent.  For massless bosons with

$$\eqalign{&(p_1^+,\pp_1) = (y w\calp^+, y(w\calp^\perp+\tp)+\rp), \cr
&(p_2^+,\pp_2) = ((1-y)w \calp^+, (1-y)(w \calp^\perp+\tp)-\rp)\;,}
\eqno(6.49)$$

\noindent it is also easily seen that

$$(p_1+p_2)^\mu (p_1+p_2)_\mu={\rp^2 \over y(1-y)} \;.\eqno(6.50)$$

\noindent The on-shell
scattering amplitude blows up in perturbation theory as
the relative transverse momentum between the scattering bosons
goes to zero, but the important
observation is that counterterms must be added to restore covariance and
cluster decomposition.  To discuss the counterterms, I set $\tp=0$
for simplicity.
To determine what counterterms are required,
expand the amplitude in Eq. (6.47), obtaining

$${g^2 \over 32 \pi^2}\Biggl[ln\Biggl( {\rp^2 \over y(1-y)\Lambda_0^2}
\Biggr) \;-\;ln(w) \;+\; {\rp^2 \over wy(1-y)\Lambda_0^2} \;+\;
\cdot\cdot\cdot \Biggr]  \;.\eqno(6.51)$$

The first term is covariant and spectator-independent, and after the
subtraction of a constant associated with coupling renormalization it
yields the correct result.  All remaining terms must be canceled by
counterterms.  The counterterms are part of $u_4$, and it is only the
second term in the series that affects the marginal part of $u_4$.  In
this case we find that covariance and cluster decomposition requires us
to depart from the simple marginal interaction in Eq. (6.3), and use

$$\tg(p_i^+)=g+{g^2 \over 32 \pi^2}\;ln\Biggl(
{p_1^++p_2^+\over \calp^+}\Biggr)+\cdot\cdot\cdot \;.
\eqno(6.52)$$

\noindent Not only are functions of longitudinal momentum allowed in
the marginal part of $u_4$, they are required by covariance and cluster
decomposition when we use an invariant-mass cutoff.  Calculations of the
remaining diagrams in figure 3b reveal additional logarithmic
corrections to the vertex, but nothing qualitatively new is revealed.
Letting $p_3^+=z w \calp^+$ and $p_4^+=(1-z) w \calp^+$,
the real part of the complete scattering amplitude can be written

$${g^2 \over 32 \pi^2}\Biggl[ ln\Biggl({(p_1+p_2)^2 \over
w(\Lambda_0^2-M^2)}\Biggr)+ln\Biggl({(p_1-p_3)^2 \over
w |y-z| (\Lambda_0^2-M^2)}\Biggr)+ln\Biggl({(p_2-p_3)^2 \over
w |1-y-z| (\Lambda_0^2-M^2)}\Biggr) \Biggr] \;. \eqno(6.53)$$

\noindent The additional counterterms required to restore covariance and
cluster decomposition to the entire amplitude are easily determined.

We can think of the corrections to $u_4$
as new marginal and irrelevant variables, and as such we expect to find
renormalization group equations that show how their strengths change
with the cutoff.  On the other hand, from the point of view of
perturbation theory the strengths of these counterterms are determined by
$g$; and they change with the cutoff only because $g$ changes with the
cutoff.

It is straightforward to compute the one-boson to three-boson Green's
functions corresponding to the diagrams in figure 3b, and the real part
is,

$${g^2 \over 32 \pi^2}\Biggl[ ln\Biggl({(p_2+p_3)^2 \over
w(x+y)(\Lambda_0^2-M^2)}\Biggr)+ln\Biggl({(p_2+p_4)^2 \over
w(x+z)(\Lambda_0^2-M^2)}\Biggr)+ln\Biggl({(p_3+p_4)^2 \over
w(y+z)(\Lambda_0^2-M^2)}\Biggr) \Biggr] \;. \eqno(6.54)$$

\noindent Here I have used $p_1^+=w \calp^+$, $p_2^+=w x \calp^+$,
$p_3^+=w y \calp^+$, and $p_4^+=w z \calp^+$.
Again, we find that covariance and cluster decomposition are
violated, and that counterterms must be added to $u_4$ to subtract these
violations.

Next I want to use the coupling constant coherence conditions to compute the
complete set of $\order(g^2)$ counterterms.
Let us first consider the generic problem of determining the strength
of an irrelevant variable from its renormalization group equation, using the
condition that it can depend on the cutoff only through its
perturbative dependence
on $g_n$.  The generic equation for an irrelevant variable can be
written,

$$w_{n+1}=\Bigl({1 \over 4}\Bigr)^{n_w} w_n-
c_w g_n^2+\order(g_n^3) \;,
\eqno(6.55)$$

\noindent where $n_w$ is an integer determined by the transverse
dimension of the operator.  Note that the $\order(g_n^3)$ corrections
include corrections to the Hamiltonian coming from diagrams that are
third-order in the original interaction in Eq. (6.3),
as well as corrections coming from diagrams that include one original
vertex and one counterterm vertex.  The fact that tadpoles are
eliminated when zero-modes are dropped is used here, because an
$\order(g^2)$ counterterm produces $\order(g^3)$ corrections when
zero-modes are removed.  The assumption that all counterterms are at
least $\order(g^2)$ must be justified \apost.
It is easy to see that Eq (6.55) implies that an expansion of $w_n$ in
powers of $g_n$ should start at second order.  Assuming that $w_n=\omega
g_n^2+\order(g_n^3)$, and using Eq. (6.36), we find,

$$\omega={c_w \over 1 - \bigl({1 \over 4}\bigr)^{n_w}} \;.\eqno(6.56)$$

Thus, we find that the second-order transformation fixes the
strength of all irrelevant operators when we insist that they run only
with the coupling.  To illustrate the consequences of Eq. (6.56),
consider $\delta u_4$ given by Eq. (6.15).  Eq. (6.15) yields the values of
$c_w$ for an infinite number of irrelevant operators.  Using Eq. (6.56)
we find that the Hamiltonian must contain the set of irrelevant operators,

$$\eqalign{
w_4&={g^2 \over 64 \pi^2}\;ln\Bigl(1+{\rp^2 \over y(1-y)\Lambda_0^2}
\Bigr) \cr
&={g^2 \over 4} \; \int \dsx \Biggl\{
\Bigl[{\rp^2 \over y(1-y)} - {\sp^2 \over x(1-x)}\Bigr]^{-1} - \cr
&\qquad\qquad\qquad \qquad\qquad\qquad \qquad
\Bigl[{\sp^2 \over x(1-x)}\Bigr]^{-1} \Biggr\}\;
\theta\Bigl({\sp^2 \over x(1-x)}-\Lambda_0^2\Bigr) \;.}\eqno(6.57)$$

\noindent All momenta are defined in Eq. (6.4). There is another
counterterm coming from the first Hamiltonian diagram in figure 3b, with
the momenta in Eq. (6.57) being replaced by the momenta of the outgoing
particles; as well as additional counterterms from the remaining
Hamiltonian diagrams in figure 3b.  However, one can easily determine
the integral form of the counterterms in each case.  Notice that the
integrand in Eq. (6.57) is similar to the integrand in Eq. (6.11).
There is
a subtraction of the latter integrand with the external transverse momenta
set to zero to remove the marginal piece, and the step functions are
altered so that it is intermediate energies {\it above} the upper cutoff
that are included, rather than energies between the upper and lower
cutoffs.  In addition to having a simple universal form, the
integral representation of the irrelevant `counterterms' will prove convenient
in Section \rnseven, when I group a second-order contribution that
includes this counterterm and the original interaction in Eq. (6.3) with
a third-order contribution from the original interaction alone.  In
diagrammatic terms, I group the one-loop correction that contains a vertex
counterterm with a two-loop contribution to the running Hamiltonian; and
the integral representation of the counterterm allows this regrouping to
be performed directly in the two-loop integrand.

The coupling constant coherence
conditions placed on the irrelevant operators require that to
$\order(g^2)$ they must be invariant under the action of the full
transformation.  In the first part of a transformation one lowers the
cutoff, and generates new $\order(g^2)$ irrelevant operators.  When these new
irrelevant operators are added to the old irrelevant operators, the
resultant operators must be identical in form to the old irrelevant
operators, with the only change being the replacement of $\Lambda_0$
with $\Lambda_1$.  After the scaling part of the transformation is
completed, the complete set of irrelevant operators returns to exactly
its original form; in this case, the logarithm in Eq. (6.57) is exactly
reproduced.  Higher order corrections to the irrelevant operators insure
that the $g_n^2$ coefficient runs correctly, so that it is indeed the
correct running coupling that appears, and they generate
new irrelevant operators of $\order(g_n^3)$ and higher.

As a final example of an irrelevant counterterm, consider the correction
to $u_6$ resulting from Eq. (6.32).  All operators in $u_6$ are
irrelevant, and if $u_6$ runs only because the coupling in Eq. (6.3)
runs, the same reasoning used above implies that $u_6$ must contain,

$$\eqalign{
{g^2 \over x_2^+ + x_3^+ - x_6^+}
\;&\Bigl[{\rp^2_2 \over x_2^+}+{\rp^2_3 \over x_3^+}-
{\rp_6^2 \over x_6^+} -
{(\rp_2+\rp_3-\rp_6)^2 \over x_2^+ + x_3^+ - x_6^+} \Bigr]^{-1} \cr
&\theta\bigl({\rp^2_1 \over x_1^+}+{\rp_6^2 \over x_6^+}+
{(\rp_2+\rp_3-\rp_6)^2 \over x_2^+ + x_3^+ - x_6^+}- \Lambda_0^2\bigr) \cr
&\theta\bigl(\Lambda_0^2-{\rp^2_1 \over x_1^+}-{\rp_2^2 \over x_2^+}-
{\rp_3^2 \over x_3^+}\bigr)
\theta\bigl(\Lambda_0^2-{\rp^2_4 \over x_4^+}-{\rp_5^2 \over x_5^+}-
{\rp_6^2 \over x_6^+}\bigr)
\;.} \eqno(6.58)$$

\noindent Comparison of this term with the correction to $u_6$ resulting
from the second Hamiltonian diagram in figure 3c, Eq. (6.32), shows that
the second-order tree level counterterms are easily determined.

Return to the calculation of the boson-boson scattering amplitude, and
add the above irrelevant counterterms to $u_4$. The real part of the
scattering amplitude becomes,

$${g^2 \over 32 \pi^2}\Biggl[ ln\Biggl({(p_1+p_2)^2 \over
w\Lambda_0^2}\Biggr)+ln\Biggl({(p_1-p_3)^2 \over
w |y-z| \Lambda_0^2}\Biggr)+ln\Biggl({(p_2-p_3)^2 \over
w |1-y-z| \Lambda_0^2}\Biggr) \Biggr] \;. \eqno(6.59)$$

\noindent It should be obvious from this expression that covariance and
cluster decomposition are still violated, because of the longitudinal
momentum fractions appearing in the logarithms; however, these properties can
now be restored by marginal operators alone.  In other words, coupling
coherence leads to irrelevant operators that
restore covariance and cluster decomposition to the irrelevant part of
the scattering amplitude.  Adding the appropriate irrelevant operator
contributions to the one-boson to three-boson Green's function yields,

$${g^2 \over 32 \pi^2}\Biggl[ ln\Biggl({(p_2+p_3)^2 \over
w(x+y)\Lambda_0^2}\Biggr)+ln\Biggl({(p_2+p_4)^2 \over
w(x+z)\Lambda_0^2}\Biggr)+ln\Biggl({(p_3+p_4)^2 \over
w(y+z)\Lambda_0^2}\Biggr) \Biggr] \;. \eqno(6.60)$$

\noindent Violations of Lorentz covariance and cluster decomposition in
the irrelevant part of this Green's function are also removed.

Let us now consider the generic problem of determining the strength
of a marginal variable from its renormalization group equation, using the
condition that it can depend on the cutoff only through its dependence
on $g_n$.  To simplify the presentation let me simply state that we must
simultaneously consider
dependent marginal operators of $\order(g_n^2)$, to
which I collectively
refer as $h_n$, and dependent marginal operators
of $\order(g_n^3)$, to which I collectively refer as $j_n$.  The
reason that both are required will become apparent below.  We have
already seen that no new marginal operators are produced in the
$\order(g_n^2)$ Hamiltonian diagrams,
so the generic equations for these marginal variables can be
written,

$$h_{n+1}=h_n-c_h g_n h_n-d_h g_n^3+\order(g_n^4)\;,
\eqno(6.61)$$

$$j_{n+1}=j_n-c_j g_n h_n-d_j g_n^3+\order(g_n^4)\;.
\eqno(6.62)$$

\noindent In principle,
there should be terms of $\order(g_n w_n)$ in each of
these equations;  which result from second-order (\ie, one-loop)
corrections that contain an irrelevant vertex in addition to the
original vertex in Eq. (6.3).  Since the irrelevant counterterms have
already been uniquely determined to $\order(g_n^2)$, I group these
one-loop corrections
with corrections that are third-order
(\ie, two-loop) in the original vertex for simplicity.  Thus the final
terms in these equations,
$d_h g_n^3$ and $d_j g_n^3$, result from a sum of two-loop
diagrams and irrelevant counterterm insertions in one-loop diagrams.  It
is straightforward, but tedious, to separately display these terms in
Eqs. (6.61) and (6.62).  It is far more tedious to separately compute
these one-loop and two-loop diagrams in closed form.
Fortunately, there is no need to
do so.

By assumption,

$$h_n=\alpha g_n^2+\beta g_n^3\;,\eqno(6.63)$$

$$j_n=\gamma g_n^3\;,\eqno(6.64)$$

\noindent with higher order terms being suppressed.
It is straightforward to confirm that there are two types of solution to
Eq. (6.61).  If $d_h=0$ and $\alpha \ne 0$, we must have

$$c_h=2 c_g\;;\eqno(6.65)$$

\noindent where $c_g$ is shown in Eq. (6.36). If $d_h \ne 0$, then

$$\alpha={d_h \over 2 c_g-c_h} \;.\eqno(6.66)$$

\noindent $\beta$ is not determined by Eq. (6.61), but it is clear that
one must perform a third-order renormalization group calculation and
determine $d_h$ to fix $\alpha$, even though this is the coefficient of
the $\order(g_n^2)$ piece of the marginal operator.

$\gamma$ is not fixed by Eq. (6.62), but if a marginal operator
arises in the third-order behavior of the transformation, Eq. (6.62)
shows that the strength of this operator is $\order(g_n^3)$, as opposed
to $\order(g_n^2)$, only if

$$\alpha c_j+d_j=0 \;.\eqno(6.67)$$

\noindent  This condition must be met if an operator appears in the
third-order analysis, but is not $\order(g_n^2)$.  However, since $\gamma$
is not fixed, there is no guarantee that $\gamma \ne 0$; so the
appearance of the operator in the third-order behavior of the
transformation does not insure its appearance in the Hamiltonian.  The
important point to observe here is that Eq. (6.67) requires $\alpha \ne
0$ if $d_j \ne 0$.  This means that if we find any new marginal operators
in the third-order analysis, at least one new marginal operator is
$\order(g_n^2)$.

This generic analysis shows that there are two types of marginal
operator that can arise with strength of
$\order(g_n^2)$.  One type is explicitly
produced by the third-order behavior of the transformation, after the
irrelevant counterterms have been properly included to this order,
and this type
cannot be missed.  The second type of marginal operator does not appear
in the subtracted
third-order analysis; but it must exactly reproduce itself in a
second-order analysis when combined with the interaction in Eq. (6.3),
with the strength fixed by the condition $c_h=2 c_g$.  I proceed no
further with this analysis in this Section.   In the next
Section I show that the logarithmic functions of longitudinal momentum
fractions required to restore Lorentz covariance and cluster
decomposition are indeed solutions to Eqs. (6.61) and (6.62).


\bigskip
\noindent {\bf \rnseven. Third- and Higher-Order Behavior Near Critical
Gaussian Fixed Points}
\medskip

Third-order behavior of a transformation near the Gaussian fixed point
can be computed using the
final terms in Eq. (4.14), and the additional terms required to study
higher-order behavior are readily computed, with rapidly increasing
algebraic complexity.

There are two issues I want to address in this Section.  First, I want to
complete the calculation of corrections to the marginal part of $u_4$
initiated in the last Section; and show that the coupling constant
coherence
conditions lead to the marginal counterterms required to restore Lorentz
covariance and cluster decomposition to the four-point functions in
second-order perturbation theory.  Second, I want to discuss the errors
one encounters when various approximate analyses are performed.
The latter
discussion is not intended to be rigorous or complete.
The renormalization group offers the possibility of overcoming some
serious logical flaws in `old-fashioned perturbation theory'
\APrefmark{\rSCH,\rDIRTHR,\rBROWN}, a fact that
does not yet receive sufficient attention in text books.
The perturbative renormalization
group may allow one to effect renormalization so that the
perturbative expansions encountered at every stage of a calculation
involve only small coupling constants.
The caveat is that the
running coupling(s) must remain small over the entire range of scales
directly encountered in the perturbative portion of the calculation.
In `old-fashioned' perturbation theory
the bare parameters always diverge, and one must simply follow
renormalization recipes \APrefmark{\rSCH}
without being distracted by intermediate
expansions in powers of divergent constants.
We will find that the
canonical running coupling constant in the scalar theory remains small if
it is small in $H_{\Lambda_0}^{\Lambda_0}$;
however, there are
logarithms of longitudinal momentum fractions that appear in the
Hamiltonian, leading to effective couplings that diverge at small
longitudinal momentum transfer.

There are severe limitations to how much can be learned from a
perturbative renormalization group study of scalar field theory,
because it is not asymptotically free and there is no
possibility of the perturbative analysis being complete
\APrefmark{\rWILNINE}.  It
is not possible to construct a trajectory of renormalized Hamiltonians
using a perturbative renormalization group if the theory is not
asymptotically free.
For the examples in this Section, I simply
assume that the coupling constant in $H^{\Lambda_0}_{\Lambda_0}$, $g_0$,
is small;
in which case Eq. (6.36) reveals that it
should remain small in every Hamiltonian on a renormalization group
trajectory.  This analysis is naive; however, my
interest is the development of the perturbative formalism.

Let me turn now to the third order corrections to the part of $u_4$ that
governs one-boson to three-boson transitions.  This part of $u_4$ is
chosen rather than the two-boson to two-boson transition because fewer
Hamiltonian diagrams must be computed.
The requisite Hamiltonian diagrams are shown in figure
5, where I have suppressed the arrows that indicate the energies
appearing in the energy denominators.  Each diagram in figure 5 actually
corresponds to four Hamiltonian diagrams, which differ in their energy
denominators and in the distribution of cutoffs, as seen in the four
third-order terms in Eq. (4.14).
As discussed
in the last Section, I want to group all $\order(\delta H^2)$ terms
that are $\order(g^3)$ because they contain one
$\order(g)$ marginal vertex and
one $\order(g^2)$ irrelevant
vertex, with the $\order(\delta H^3)$ terms that are $\order(g^3)$
because they contain three $\order(g)$ marginal vertices.
This comment is most easily
understood after studying the results below.  The bracketed expressions
in figure 5 indicate diagrams in which the momenta of the outgoing
particles are permuted, with the permutation being listed in the
brackets.

I only want to compute the third-order corrections to the marginal
part of $u_4$, so I set all incoming and outgoing transverse momenta to
zero.  Since I am studying the critical theory, this means that the
initial and final energies are zero, which considerably simplifies
most of the energy denominators.
I do not display spectators because they have no direct effect on
$u_4$.  One can always choose variables in which $\delta v_4$ is
independent of the longitudinal momenta of the spectators, even
though cluster decomposition is
restored by marginal counterterms that explicitly depend on these
momenta.

Let me begin with the Hamiltonian diagrams in figure 5a.  There are four
diagrams, and each makes an identical contribution to the marginal part
of $\delta v_4$.  Using Eq. (4.14) and the interaction in Eqs.
(6.2)-(6.3), one finds,

$$\eqalign{\delta v_{4M} =
&-{1 \over 2}\; {g^3 \over 3!\; p^+} \int \dkt_1 \dkt_2 \dkt_3 \;(16 \pi^3)
\delta^3(p-k_1-k_2-k_3)\cr
&\qquad\qquad\qquad\qquad (k_1^-+k_2^-+k_3^-)^{-2}\;
\Theta_{high}(k_1^-+k_2^-+k_3^-)\;; } \eqno(7.1)$$

\noindent where $\dkt$ is defined in Eq. (3.4), $k^-=\kp^2/k^+$, and
$p^+$ is the longitudinal momentum entering the loops. This last factor
results from the presence of an internal line that is not part of a
loop.  Only one of the
four terms in Eq. (4.14) survives here, because only the intermediate
state that includes the boson loops can be a high energy state, since the
initial and final states must both be low energy states.
The first factor of $-1/2$ is seen in Eq. (4.14); while $3!$ is a
symmetry factor.  I have introduced,

$$\Theta_{high}(K^-)=\theta\Bigl({\Lambda_0^2 \over \calp^+}
-K^-\Bigr)\; \theta\Bigl(K^--{\Lambda_1^2 \over \calp^+}
\Bigr) \;,\eqno(7.2)$$

\noindent which projects onto states of high energy;
and later I also need,

$$\Theta_{low}(K^-)=\theta\Bigl({\Lambda_1^2 \over \calp^+}-K^-\Bigr) \;,
\eqno(7.3)$$

\noindent which projects onto states of low energy.  $\calp^+$ is the
total longitudinal momentum.
I have not displayed the low energy cutoffs associated with external lines.

The integrals in Eq. (7.1) are most easily evaluated by introducing
Jacobi coordinates; for example, $p^+=w\calp^+$,
$k_1^+=xyp^+$, $\kp_1=\sqrt{w}\Lambda_0(x\bs+\br)$, $k_2^+=(1-x)yp^+$,
$\kp_2=\sqrt{w}\Lambda_0((1-x)\bs-\br)$, $k_3^+=(1-y)p^+$,
$\kp_3=-\sqrt{w}\Lambda_0 \bs $.  In these coordinates,

$$\eqalign{\delta v_{4M} =
& -{g^3 \over 12} \int {d^2r \over 16 \pi^3}
{d^2s \over 16 \pi^3} \int_0^1 dx \int_0^1 dy \; {1 \over x(1-x)y(1-y)}
\cr
&\qquad\qquad\qquad\qquad \Bigl({\br^2 \over yx(1-x)}+{\bs^2 \over
y(1-y)}\Bigr)^{-2}\;\Theta_{high}\Bigl({\br^2 \over yx(1-x)}+{\bs^2
\over y(1-y)}\Bigr) \;,}\eqno(7.4)$$

\noindent where now,

$$\Theta_{high}\Bigl({\bt^2 \over z}\Bigr)=\theta\Bigl(1-{\bt^2 \over
z}\Bigr)\;\theta\Bigl({\bt^2 \over z}-\eta\Bigr) \;,\eqno(7.5)$$

\noindent and,

$$\Theta_{low}\Bigl({\bt^2 \over z}\Bigr)=\theta\Bigl(\eta-{\bt^2
\over z}\Bigr) \;.\eqno(7.6)$$

\noindent I have introduced $\eta=\Lambda_1^2/\Lambda_0^2$ because
this ratio appears repeatedly.
The appropriate definition of $\Theta_{high}$ and
$\Theta_{low}$ is always apparent from context. All
integrals are evaluated using Jacobi coordinates, so the definitions in
Eqs. (7.5) and (7.6) are always used to explicitly evaluate a
correction.

The
remaining integrals are readily completed, and one finds that each
Hamiltonian diagram in figure 5a contributes,

$$\delta v_{4M} = \;-{g^3 \over 24\; (16 \pi^2)^2} \;ln\Bigl( {1
\over \eta}\Bigr) \;. \eqno(7.7)$$

\noindent Of course, the entire correction $\delta v_4$ is extremely
complicated, but the marginal part of $\delta v_4$ is simple.  The
rescaling does not change this marginal term, so Eq. (7.7) immediately
yields the correction to the marginal part of $u_4$ from each diagram in
figure 5a.

Next consider the diagrams in figure 5b.  The first diagram in figure 5b
leads to,

$$\eqalign{\delta v_{4M} =
&{g^3 \over 2} \int \dkt_1 \dkt_2 \dkt_3 \dkt_4 \;(16
\pi^3)\delta^3(p_1-p_4-k_3-k_4)\;(16 \pi^3)\delta^3(p_2-k_1-k_2-k_3) \cr
& \Biggl\{
{}~(k_3^-+k_4^-)^{-1} \;(k_1^-+k_2^-+k_3^-)^{-1}\;
\Theta_{high}(k_3^-+k_4^-) \;\Theta_{high}(k_1^-+k_2^-+k_3^-) \cr
& ~-{1 \over 2} (k_3^-+k_4^-)^{-1} \;
(k_4^--k_1^--k_2^-)^{-1}\; \Theta_{high}(k_3^-+k_4^-)
\;\Theta_{low}(k_1^-+k_2^-+k_3^-) \cr
& ~-{1 \over 2} (k_1^-+k_2^-+k_3^-)^{-1}\;
(k_1^-+k_2^--k_4^-)^{-1}\;\Theta_{low}(k_3^-+k_4^-)
\;\Theta_{high}(k_1^-+k_2^-+k_3^-)\Biggr\} \;.} \eqno(7.8)$$

\noindent The first two of the four third-order terms in Eq. (4.14)
combine to give the first term in the integrand.  The third term in the
integrand is zero because it is not possible for the energy of the first
intermediate state in figure 5b to have higher energy than the second
intermediate state.  I do not display terms that vanish in this manner
below.

While it is possible to evaluate this integral, it is convenient to
group it with the one-loop diagrams that contain all irrelevant
counterterms that result from sub-diagrams (\eg, nested loops) in figure
5b.  In figure 6a I show the first diagram in figure 5b added to a new
one-loop diagram in which there is a new four-boson vertex.
The new vertex is a sum of irrelevant operators
that must be added to the Hamiltonian to satisfy the coupling constant
coherence
conditions, and figure 6b shows the original diagram that led to their
addition.  The derivation of irrelevant operators is discussed in the
last Section.  Using an integral representation for the appropriate
irrelevant counterterms, the one-loop correction to the Hamiltonian shown
in figure 6a is,

$$\eqalign{\delta v_{4M} =
&{g^3 \over 2} \int \dkt_1 \dkt_2 \dkt_3 \dkt_4 \;(16\pi^3)
\delta^3(p_1-p_4-k_3-k_4) \; (k_3^-+k_4^-)^{-1}\;
\Theta_{high}(k_3^-+k_4^-) \cr
& \Biggl\{{1 \over 2} \; (16 \pi^3)\delta^3(p_2-k_1-k_2-k_3) \cr
&\qquad\qquad
\Bigl[(k_1^-+k_2^-+k_3^-)^{-1}+(k_1^-+k_2^--k_4^-)^{-1}
\Bigr] \;
\Theta_{super}(k_1^-+k_2^-+k_3^-) \cr
&-(16 \pi^3)\delta^2(\kp_2+\kp_3)\delta(p_2^+-k_1^+-k_2^+-k_3^+)\;
(k_1^-+k_2^-)^{-1}
\; \Theta_{super}(k_1^-+k_2^-) \Biggr\} \;.}\eqno(7.9)$$

\noindent The first two terms in the integrand have identical energy
denominators to the second-order Hamiltonian diagram in figure 6b that
leads to these counterterms, while the third term cancels the marginal
part of the counterterm and insures that it is composed of
irrelevant operators only.  I have introduced a new cutoff function,

$$\Theta_{super}(K^-)=\theta\Bigl(K^--{\Lambda_0^2 \over p^+}\Bigr)
\;.\eqno(7.10)$$

\noindent This cutoff is not really associated with an intermediate
state, because it is part of the integral representation of an
irrelevant operator that is added to the Hamiltonian, not part of an
operator that is induced by a transformation.  However, one can think of
an intermediate state in which the energy lies above all cutoffs; and
say that the counterterm directly implements a complete set of irrelevant
interactions that would have been provided by the exchange of particles
whose energy is above the cutoffs placed on the Hamiltonian.

I want to emphasize that the four-boson vertex nested in the one-loop
diagram in figure 6a, while complicated, is uniquely determined by
coupling coherence.  If this vertex is not in the Hamiltonian,
there are variables that explicitly
run with the cutoff other than the coupling
in Eq. (6.3).

To proceed we need to define Jacobi coordinates that satisfy
the delta function constraints, and then combine the
integrals remaining in Eq. (7.8) with those in Eq. (7.9).  The
calculation from this point is complicated only because of tedious
calculus, which I do not detail.  Perhaps the most difficult part of
the calculation is keeping track of the various step function cutoffs,
and finding how to regroup terms at appropriate stages of the calculation
so that nested integrals lead to simple analytical results.
These problems make the calculation more difficult than a
simple two-loop Feynman
diagram calculation, but this should be no surprise.

If we let
$p_2^+=w(p_1^+-p_4^+)$, so that $p_3^+=(1-w)(p_1^+-p_4^+)$, the complete
result from the diagrams in figure 6a is,

$$\delta v_{4M} = {g^3 \over 2\;(16 \pi^2)^2}\; \Biggl\{ ln\Bigl(
{1 \over \eta}\Bigr)\;(1-w)\;ln(1-w)+{1 \over 2}\Biggl(ln\Bigl( {1
\over \eta}\Bigr)\Biggr)^2\;w \Biggr\} \;.\eqno(7.11)$$

\noindent This result needs to be added to those from the remaining
diagrams in figure 5b, with the one-loop corrections that are analogous
to Eq. (7.9) being added.  The second diagram in figure 5b leads to a
correction that is identical in form to Eq. (7.11), but with $w$ and
$1-w$ interchanged.  Letting $p_2^+=x p_1^+$, $p_3^+=y p_1^+$, and
$p_3^+=z p_1^+$; the complete sum of terms in figure 5b, with the
appropriate one-loop counterterm diagrams added, is,

$$\eqalign{ \delta v_{4M} = &{g^3 \over 2\;(16 \pi^2)^2}\;
\Biggl\{ {3 \over 2} \Biggl(ln\Bigl( {1 \over \eta}\Bigr)\Biggr)^2 +
\cr & \qquad ln\Bigl({1 \over \eta}\Bigr)\;\Biggl[ {x \over 1-y}\;
ln\Bigl({x \over 1-y}\Bigr)+{x \over 1-z}\;ln\Bigl({x \over
1-z}\Bigr)+{y \over 1-z}\;ln\Bigl({y \over 1-z}\Bigr) \cr
&\qquad\qquad\qquad + {y \over 1-x}\;
ln\Bigl({y \over 1-x}\Bigr)+{z \over 1-x}\;ln\Bigl({z \over
1-x}\Bigr)+{z \over 1-y}\;ln\Bigl({z \over 1-y}\Bigr) \Biggr]
\Biggr\} \;. } \eqno(7.12)$$

The first diagram in figure 5c is redisplayed in figure 7a, with the
one-loop correction that contains the appropriate irrelevant counterterm.
Figure 7b displays the sub-diagram from which the irrelevant counterterm
results, and one sees that in this case the counterterm is part of
$u_6$.  The sum of both diagrams is,

$$\eqalign{\delta v_{4M} &=
{g^3 \over 2} \int \dkt_1 \dkt_2 \dkt_3 \dkt_4 \;(16
\pi^3)\delta^3(p_1-k_1-k_2-k_3)\;(16 \pi^3)\delta^3(p_2-k_1-k_2-k_4) \cr
& \Biggl\{
{}~(k_1^-+k_2^-+k_3^-)^{-1}\;(k_1^-+k_2^-+k_4^-)^{-1} \;
\Theta_{high}(k_1^-+k_2^-+k_3^-) \;\Theta_{high}(k_1^-+k_2^-+k_4^-) \cr
& ~-{1 \over 2} (k_1^-+k_2^-+k_4^-)^{-1} \;
(k_4^--k_3^-)^{-1}\; \Theta_{low}(k_1^-+k_2^-+k_3^-)
\;\Theta_{high}(k_1^-+k_2^-+k_4^-) \cr
& ~+{1 \over 2} (k_1^-+k_2^-+k_3^-)^{-1}\;\Bigl[
(k_1^-+k_2^-+k_4^-)^{-1}+(k_4^--k_3^-)^{-1}\Bigr] \cr
&\qquad\qquad\qquad\qquad\qquad \Theta_{high}(k_1^-+k_2^-+k_3^-)
\;\Theta_{super}(k_1^-+k_2^-+k_4^-)\Biggr\} \;.} \eqno(7.13)$$

\noindent The first two terms in the integrand result from the
third-order corrections to the Hamiltonian in Eq. (4.14), while the last
two terms result from the second-order corrections in Eq. (4.14) with
one of the interactions being an irrelevant operator.  The steps required
to evaluate this integral are identical to those required above, and the
result is,

$$\delta v_{4M} = - {g^3 \over 2\;(16 \pi^2)^2}\;ln\Bigl({1 \over
\eta}\Bigr)\;ln(1-x) \;,\eqno(7.14)$$

\noindent where $p_2^+=x p_1^+$.

The complete set of diagrams in figure 5c, with the accompanying
one-loop corrections, yield,

$$\delta v_{4M} = - {g^3 \over 2\;(16 \pi^2)^2}\;ln\Bigl({1 \over
\eta}\Bigr)\Bigl[ln(1-x)+ln(1-y)+ln(1-z)\Bigr] \;, \eqno(7.15)$$

\noindent where I have again chosen $p_2^+=x p_1^+$, $p_3^+=y p_1^+$,
and $p_4^+=z p_1^+$.

The first diagram in figure 5d is redisplayed in figure 8a, along with
two one-loop corrections with which it must be grouped.  In this case
there are two sub-diagrams that lead to irrelevant counterterms, as shown
in figures 8b and 8c.  After using coupling coherence to
uniquely determine the irrelevant vertices in the one-loop diagrams, one
finds that the diagrams in figure 8a yield,

$$\delta v_{4M} =  {g^3 \over 2\;(16 \pi^2)^2}\;\Biggl[ {1 \over
2}\;\Biggl(ln\Bigl({1 \over \eta}\Bigr)\Biggr)^2 - {x \over 1-x}\;
ln(x) \Biggr] \;.\eqno(7.16)$$

The complete set of diagrams in figure 5d, with the accompanying
one-loop corrections, yield

$$\delta v_{4M} = - {g^3 \over 2\;(16 \pi^2)^2}\;\Bigl[ {3 \over
2}\; \Biggl(ln\Bigl({1 \over \eta}\Bigr)\Biggr)^2 + {x \over 1-x}\;
ln(x) + {y \over 1-y}\; ln(y) + {z \over 1-z}\; ln(z) \Bigr] \;.
\eqno(7.17)$$

The final set of two-loop contributions, along with the accompanying
$\order(g^3)$ one-loop contributions, to the marginal part of $\delta
v_4$ are shown in figure 9a.  These diagrams yield

$$\delta v_{4M} = {3 g^2 \over 4\;(16 \pi^2)} \; \Biggl(ln\Bigl(
{1 \over \eta}\Bigr)\Biggr)^2 \;. \eqno(7.18)$$

At this point we have all $\order(g^3)$ contributions to the
renormalization of the marginal one-boson to three-boson part of $u_4$,
except for one-loop contributions that involve one of the
$\order(g^2)$ marginal operators that we must find using coupling
coherence.  For the benefit of later calculations it is convenient to
list the complete result using coordinates in which $p_i^+=x_i\calp^+$.
The complete set of two-loop contributions to the marginal one-boson to
three-boson part of
$\delta u_4$, along with all $\order(g^3)$ one-loop contributions that
result from the $\order(g^2)$ irrelevant operators determined by
coupling coherence, yields:

$$\eqalign{ \delta u_{4M} = &{g^3 \over 2\;(16 \pi^2)^2} \Biggl\{
-{1 \over 3}\;ln\Bigl({1 \over \eta}\Bigr) + {9 \over 2} \;
\Biggl(ln\Bigl({1 \over \eta}\Bigr)\Biggr)^2 \cr
&\qquad-2 ln\Bigl({1 \over \eta}\Bigr) \Bigl[
ln\Bigl(x_1-x_2\Bigr)+\;ln\Bigl(x_1-x_3\Bigr)+\;ln\Bigl(x_1-x_4\Bigr)
-3\;ln\Bigl(x_1\Bigr)\Bigr] \cr
&\qquad-ln\Bigl({1 \over \eta}\Bigr)\;\;
\Bigl[ \Bigl({x_2 \over x_1-x_2}- {x_2 \over
x_1-x_3}-{x_2 \over x_1-x_4}\Bigr)\;ln\Bigl({x_2 \over x_1}\Bigr) \cr
&\qquad\qquad\qquad\qquad\qquad
+ \Bigl({x_3 \over x_1-x_3}- {x_3 \over x_1-x_2}-{x_3
\over x_1-x_4}\Bigr)\;ln\Bigl({x_3 \over x_1}\Bigr)\cr
&\qquad\qquad\qquad\qquad\qquad\qquad
+\Bigl({x_4 \over x_1-x_4}- {x_4 \over x_1-x_2}-{x_4 \over
x_1-x_3}\Bigr)\;ln\Bigl({x_4 \over x_1}\Bigr)
\Bigr]\Biggr\} \;.} \eqno(7.19)$$

This result includes four separate functions of longitudinal momenta
that must be considered in the renormalization group analysis, each
symmetric under the interchange of outgoing momenta.  Any one or all of
these may
appear in $u_4$ at $\order(g_n^2)$.  At least one of them must appear at
this order if the Hamiltonian satisfies the coupling constant coherence
conditions, as was demonstrated at the end of Section \rnsix, so there
are several possibilities that should be studied.

A similar calculation that is slightly more tedious leads to the two-loop
contributions to the
two-boson to two-boson marginal part
of $\delta u_4$.  The complete set of two-loop contributions to this
part of $\delta u_4$, along with all $\order(g^3)$ one-loop
contributions that result from the $\order(g^2)$ irrelevant operators
determined by coupling coherence, yields:

$$\eqalign{ \delta u_{4M} = &{g^3 \over 2\;(16 \pi^2)^2} \Biggl\{
-{1 \over 3}\;ln\Bigl({1 \over \eta}\Bigr) + {9 \over 2} \;
\Biggl(ln\Bigl({1 \over \eta}\Bigr)\Biggr)^2 \cr
&\qquad
-2\; ln\Bigl({1 \over \eta}\Bigr) \Bigl[ln(|x_1-x_3|)+ln(|x_1-x_4|)
-2\; ln(x_1+x_2)
\Bigr] \cr
&\qquad+ ln\Bigl({1 \over \eta}\Bigr) \Bigl[ \Bigl(
{x_1 \over x_1+x_2}+{x_1 \over x_1-x_3}+
{x_1 \over x_1-x_4}\Bigr)ln\Bigl({x_1 \over x_1+x_2}\Bigr) \cr
&\qquad\qquad\qquad\qquad+\Bigl(
{x_2 \over x_1+x_2}+{x_2 \over x_2-x_3}+{x_2 \over x_2-x_4}\Bigr)
ln\Bigl({x_2 \over x_1+x_2}\Bigr) \cr
&\qquad\qquad\qquad\qquad\qquad
+ \Bigl({x_3 \over x_3+x_4}+{x_3 \over x_3-x_1}+
{x_3 \over x_3-x_2}\Bigr)ln\Bigl({x_3 \over x_3+x_4}\Bigr)
\cr &\qquad\qquad\qquad\qquad\qquad\qquad
+\Bigl({x_4 \over x_3+x_4}+{x_4 \over x_4-x_1}+
{x_4 \over x_4-x_2}\Bigr)ln\Bigl({x_4 \over x_3+x_4}\Bigr) \Bigr]
\Biggr\} \;.} \eqno(7.20)$$

\noindent Here I use the momenta in figure 3b, again choosing
$p_i^+=x_i \calp^+$.  There are four functions of longitudinal
momenta that appear in this correction to the marginal operator, all of
which must be included in the renormalization group analysis.

A complete renormalization group analysis of the marginal operator
requires us to at least
introduce all of the functions of longitudinal momentum
fractions appearing above in the marginal part of $\delta
u_4$, which I
indicate as $\tg(x_1,x_2,x_3,x_4)$.  We should distinguish between the
two-boson to two-boson and the one-boson to three-boson parts of this
operator.  The three-boson to one-boson interaction is not independent
because the Hamiltonian is Hermitian.  Thus, we are led to consider,

$$\eqalign{ \tg^{(2 \rightarrow 2)}&(x_i) =
g+h^{(1)} \Bigl[ ln(|x_1-x_3|)
+ln(|x_1-x_4|)\Bigr] +h^{(2)}\; ln(x_1+x_2) \cr
&+ j^{(1)} \Bigl[ {x_1 \over x_1+x_2}\;ln\Bigl({x_1 \over x_1+x_2}\Bigr)
+{x_2 \over x_1+x_2}\;ln\Bigl({x_2 \over x_1+x_2}\Bigr) \cr
&\qquad\qquad +
{x_3 \over x_3+x_4}\;ln\Bigl({x_3 \over x_3+x_4}\Bigr) +
{x_4 \over x_3+x_4}\;ln\Bigl({x_4 \over x_3+x_4}\Bigr) \Bigr] \cr
&+ j^{(2)} \Bigl[\Bigl( {x_1 \over x_1-x_3}+{x_1 \over
x_1-x_4}\Bigr)ln\Bigl({x_1 \over x_1+x_2}\Bigr) + \Bigl( {x_2 \over
x_2-x_3}+{x_2 \over x_2-x_4}\Bigr) ln\Bigl({x_2 \over x_1+x_2}\Bigr) \cr
&+\qquad
\Bigl({x_3 \over x_3-x_1}+{x_3 \over x_3-x_2}\Bigr)ln\Bigl({x_3 \over
x_3+x_4}\Bigr) + \Bigl( {x_4 \over x_4-x_1}+ {x_4 \over
x_4-x_2}\Bigr)ln\Bigl({x_4 \over x_3+x_4}\Bigr) \Bigr]
\;,} \eqno(7.21)$$

$$\eqalign{ \tg^{(1 \rightarrow 3)}&(x_i) =
g+h^{(3)} \Bigl[ ln(x_1-x_2) +
ln(x_1-x_3)+ln(x_1-x_4)\Bigr] + j^{(3)}\;ln(x_1) \cr
&+j^{(4)} \Bigl[{x_2 \over x_1-x_2}\;ln\Bigl({x_2 \over x_1}\Bigr) +
{x_3 \over x_1-x_3}\;ln\Bigl({x_3 \over x_1}\Bigr) +
{x_4 \over x_1-x_4}\;ln\Bigl({x_4 \over x_1}\Bigr)\Bigr] \cr
&-j^{(5)} \Bigl[ \Bigl({x_2 \over x_1-x_3}+{x_2 \over
x_1-x_4}\Bigr)\;ln\Bigl({x_2 \over x_1}\Bigr) +
\Bigl( {x_3 \over x_1-x_2} + {x_3 \over x_1-x_4}\Bigr)\;ln\Bigl({x_3
\over x_1}\Bigr) \cr
&\qquad\qquad\qquad\qquad+
\Bigl( {x_4 \over x_1-x_2}+ {x_4 \over x_1-x_3}\Bigr)\;ln\Bigl({x_4
\over x_1}\Bigr) \Bigr]
\;,} \eqno(7.22)$$

\noindent I have simply assumed that the constant, $g$, appearing in
each of these terms is the same.
This issue does not need
to be resolved until one computes the $\order(g^3)$ corrections to the
renormalization group equation for the running coupling, so I do not
pursue it further.  I have also simply assumed that some of the
operators appearing in $\tg^{(2 \rightarrow 2)}$ and $\tg^{(1
\rightarrow 3)}$ are $\order(g^2)$, with their couplings being
$h^{(i)}$; while other operators are assumed to be at least
$\order(g^3)$, with their couplings being $j^{(i)}$.  These assumptions
are justified \apost, and I do not try to prove that this solution
is unique.  I believe that the solution is unique, but a proof exceeds
my patience.

To proceed further we must generalize Eqs. (6.61) and (6.62) to allow
for additional operators.  The complete equations are,

$$h^{(k)}_{n+1}=h^{(k)}_n-\sum_{l=1}^3\; c_h^{(k,l)}\;g_n\;h_n^{(l)}
-d_h^{(k)}\;g_n^3+\order(g_n^4) \;,\eqno(7.23)$$

$$j^{(k)}_{n+1}=j^{(k)}_n-\sum_{l=1}^3\; c_j^{(k,l)}\;g_n\;h_n^{(l)}
-d_j^{(k)}\;g_n^3+\order(g_n^4) \;.\eqno(7.24)$$

\noindent Eqs. (7.19) and (7.20) indicate,

$$d_h^{(1)}=-{1 \over 2}\;d_h^{(2)}=d_h^{(3)}={ln(1/\eta) \over (16
\pi^2)^2} \;,\eqno(7.25)$$

$$d_j^{(1)}=d_j^{(2)}={1 \over 6}\;d_j^{(3)}=-d_j^{(4)}=-d_j^{(5)}=
-{ln(1/\eta) \over 2\;(16 \pi^2)^2} \;.\eqno(7.26)$$

\noindent  The calculation of the various coefficients $c_h^{(k,l)}$ and
$c_j^{(k,l)}$ requires us to evaluate a set of second-order Hamiltonian
diagrams in which one vertex comes from the constant $g$, and the second
vertex comes from the appropriate marginal operator.  Thus, to compute
$c_j^{(1,2)}$, for example, we need to include the vertex corresponding
to the operator whose strength is $h^{(2)}$ and determine the part of
the resultant one-loop integral that leads to a function of longitudinal
momenta that exactly matches the function in the operator with strength
$j^{(1)}$.  These calculations are straightforward, and the results are,

$$c_h^{(1,3)}={1 \over 2}\;c_h^{(2,1)}=c_h^{(2,2)}=
{1 \over 4}\;c_h^{(2,3)}=2\;c_h^{(3,2)}=2\;c_h^{(3,3)}={ln(1/\eta)
\over 16\pi^2} \;,\eqno(7.27)$$

$$c_j^{(1,1)}=c_j^{(2,3)}={1 \over 3}\;c_j^{(3,1)}={1 \over 3}\;
c_j^{(3,3)}=-c_j^{(4,3)}=-c_j^{(5,1)}= {ln(1/\eta) \over 16 \pi^2}
\;.\eqno(7.28)$$

\noindent  All others are zero.

When one solves Eqs. (7.23) and (7.24) using these coefficients, it is
found that each coupling $j^{(k)}$ is $\order(g_n^3)$ and that,

$$h^{(1)}_n=h^{(2)}_n=h^{(3)}_n={1 \over 2\;(16\pi^2)}\;g_n^2
+\order(g_n^3)\;.\eqno(7.29)$$

\noindent These results correspond exactly with the counterterms
required to restore Lorentz covariance and cluster decomposition to the
boson-boson scattering amplitude and to the one-boson to three-boson
Green's function.  Thus, the coupling constant coherence conditions lead
to a solution of the complete renormalization group equations that
yields covariant results with cluster decomposition despite the fact
that the cutoff violates these conditions.

An exact renormalization group analysis apparently leads to
cutoff-independent results with a cutoff Hamiltonian, and coupling
constant coherence apparently allows one to find Hamiltonians that
produce Lorentz covariant results.  This is useless, however, unless one
can make approximations with bounded errors.   Moreover, if one simply
wants to
use renormalization group improved perturbation theory in powers of the
canonical coupling, it should be clear from the above calculations that
the light-front renormalization group is not the simplest tool at one's
disposal.

It is not obvious how one should estimate `errors'.  One might consider
fixing the initial Hamiltonian, and measuring the errors by computing
differences in the results produced by the final Hamiltonian on a
trajectory in comparison to those from the initial Hamiltonian.  This is
not a good measure, because in general we do not know the
appropriate initial Hamiltonian nor can we compute with it,
and one of the goals of the
renormalization group is to formulate physical problems so that it is
never necessary to explicitly deal with the initial Hamiltonian.
In order to produce a
meaningful discussion of errors, we should consider how a `typical'
renormalization group calculation proceeds, and study how observables
change as the approximation is systematically improved.  This discussion
was initiated at the end of Section \rntwo, and some basic issues were
discussed in Section \rnsix; however, the results computed in this
Section have a dramatic effect on the analysis of errors in the
light-front renormalization group.

In Wilson's perturbative Euclidean renormalization group
calculations,\APrefmark{\rWILNINE} it is necessary to fix all relevant
and marginal
couplings at the lowest cutoff,  and any irrelevant couplings at an
upper cutoff.  The upper cutoff is not chosen to be infinite, but should
be sufficiently large that all final results are insensitive to
the boundary values chosen for the irrelevant variables.  It should not
be so large that intolerable errors accumulate over the trajectory.  One
must approximate the marginal couplings over
the entire trajectory, and use an iterative algorithm to compute the
trajectory.  The output is the irrelevant couplings at
the lower cutoff, because all relevant and marginal couplings
at this cutoff must be input.  This is an extremely powerful procedure
when there are a finite number of relevant and marginal couplings,
because it allows one to obtain an accurate approximation to the
endpoint of a complete renormalized trajectory by inputting a finite
number of boundary values \APrefmark{\rWILTEN}.
However, we have seen that the light-front
renormalization group requires an infinite number of relevant and
marginal operators.  This means that a light-front renormalization group
calculation requires a new type of approximation not considered in the
Euclidean renormalization group. One must approximate the
boundary conditions placed on the relevant and marginal variables at the
lowest cutoff.

There are three obvious approximations one should consider.  First, one can
approximate the transformation itself by dropping terms at a given order
in $\delta H$.  Second, one can approximate the trajectory by discarding
specific operators; \eg, all or some
of the irrelevant operators.  Third, one
can employ coupling coherence to compute the Hamiltonian to a given
order in the running canonical coupling and drop higher orders.  One can
also employ combinations of these approximations.

I assume that Feynman perturbation theory is accurate, and estimate
the errors from each approximation by identifying the order in Feynman
perturbation theory at which an error first arises and then discussing
the magnitude of this error.  Such an analysis is of limited use if
Feynman perturbation theory is not valid for the computation of low
energy observables, which is exactly the case for which the light-front
renormalization group is being developed;
but this should give a rough guide to the problems we
should study.
Let us begin by approximating the transformation by
dropping terms starting at some given order in $\delta H$.  The simplest
comparison we can make that is of any interest is between the
$\order(\delta H)$ analysis and the $\order(\delta H^2)$ analysis.  The
linear analysis is trivial to complete.  We set the irrelevant variables
to arbitrary strengths, and find that they go to zero at the lower
cutoff as the upper cutoff is taken to infinity.
This leads to errors in Feynman perturbation theory that are
$\order(g^2)$, at least in the irrelevant variables.  This means for
example, that the real part of the boson-boson scattering amplitude,
computed in Eq. (6.53), contains errors of at least
$\order(g^2\;M^2/\Lambda_\calN^2)$.  As long as one is studying scattering
for states whose invariant mass is much less than the cutoff, these
errors are small.  This assumes that one inputs the correct marginal and
relevant variables at the lower cutoff, $\Lambda_\calN$.
For example, if one does not input the marginal
operators computed above to $\order(g^2)$ using coupling coherence,
there are logarithmic errors shown in Eq. (6.59) that are arbitrarily
large.  As $w \rightarrow 0$, these errors diverge like $g^2 \;log(w)$;
and as the longitudinal momenta of the outgoing bosons approaches those of
the incoming bosons, there are comparable logarithmic errors.  For small
$g$ these errors are suppressed relative to the leading term by one
power of $g$, but the relative error can be arbitrarily large for scattering
states of any invariant mass.

The lesson here is that arbitrarily large errors arise when functions of
longitudinal momenta diverge.  Longitudinal divergences forced us to
discard several candidate transformations, and we are seeing that the
invariant-mass transformations do not completely control the spectrum of
the longitudinal operators.  This
problem is so severe that its solution may require a completely
different renormalization group than has been developed in this paper.

If we improve the analysis by keeping $\order(\delta H^2)$
corrections to the
trajectory, we automatically generate the correct irrelevant operators to
$\order(g^2)$, even if we do not include the correct $\order(g^2)$
relevant and marginal counterterms at the lower cutoff.
This means that we obtain Eq. (6.59) instead of
Eq. (6.53) for the real part of the boson-boson scattering amplitude,
for example.  If the correct marginal operators are included in the
Hamiltonian at the lower cutoff, we obtain Feynman results to
$\order(g^2)$.  Therefore, if we approximate the transformation at
$\order(\delta H^2)$, and in addition we make a perturbative expansion
in terms of the canonical coupling to $\order(g^2)$, and we impose the
correct boundary conditions on the marginal and relevant operators at
the lower cutoff, which means fixing functions of longitudinal momenta,
we obtain the Feynman results to $\order(g^2)$.  Of course, we do
not need to make the additional perturbative expansion in powers of $g$,
but we must make some additional approximations, because the
second-order transformation generates an infinite number of vertices and
each contains an entire function of momenta.

When the transformation is approximated to $\order(\delta H^2)$,
$u_2$ does not affect any of the other functions in the Hamiltonian, as
discussed in Section \rnsix.  This means that the relevant operators
produced by the transformation have no effect on the other operators and
can be studied separately.  There is a large host of additional
approximations one might consider that are not perturbative in the
canonical coupling and may be of interest.  All of them produce
`errors' at $\order(g^3)$; so if Feynman perturbation theory in powers of
the canonical coupling is accurate, most of the additional
approximations one can make offer little or no improvement to the
perturbative approximation discussed in the last paragraph.  One
interesting approximation is to keep only the marginal part of $u_4$,
and complete an analysis that is a generalization of the analysis
leading up to Eq. (6.36) for the running canonical coupling.  Since the
boundary condition on this marginal operator includes functions of
longitudinal momenta, the analysis in Section \rnsix~ must be
generalized to include such functions.  This analysis reduces to the
study of coupled one-dimensional integral equations, because the
dependence of the marginal operator on transverse momentum is
fixed, making all transverse integrals the same.
Perhaps the most important
question one can address with such an investigation is whether functions
of longitudinal momenta with nonintegrable singularities are generated,
since the boundary functions include logarithmic divergences.  I believe
that it is relatively easy to show that in the $\order(\delta H^2)$
analysis no such singularities arise.  One can solve the coupled
equations for the marginal part of $u_4$ `backwards', towards larger
cutoffs, using the functions in Eqs. (7.21) and (7.22) as a boundary
condition.  In each step one simply convolutes the functions produced
from the previous step, which means that after any finite number of steps one
effectively considers a multidimensional convolution of logarithms, and
these are always finite.

Any
estimation of errors requires one to place bounds on the longitudinal
functions.  I have ignored this issue, except where the integrals over
longitudinal momentum produce divergences.  I have shown that this does
not happen to lowest orders in the running canonical coupling when one
uses the invariant-mass renormalization group; however, I have not shown
that this does not happen in higher orders.  Even if a perturbative
expansion in powers of the canonical coupling is finite to every order,
it is possible for some of the running variables in the invariant-mass
renormalization group to diverge; as long as
all such divergences cancel against one another when one re-expresses
any physical
result as a power series in terms of the canonical coupling.
If one performs only an expansion in powers of the canonical coupling,
and makes no further approximations, it is even possible to use the
transformations that run cutoffs on the transverse and longitudinal
momenta of individual particles.  One should be able to use coupling
coherence in this case to again obtain the Feynman results to any order
one desires.  The problem is that these results depend on
huge cancellations that must be precisely maintained.
I have not found any reason to believe that nonintegrable singularities
arise in the invariant-mass analysis; however, this is far from
satisfactory.

Suppose one uses coupling coherence and computes the Hamiltonian exactly
to a given order in $g$, and then computes low energy results using this
Hamiltonian.  How
large are the errors in perturbation theory?  Obviously the results are
exactly those of Feynman perturbation theory up to the order to which
the Hamiltonian is computed, but beyond this order one encounters
errors.  For example, the canonical $\phi^4$ Hamiltonian is accurate to
$\order(g)$; but if we compute the boson-boson scattering amplitude to
$\order(g^2)$ we obtain the result in Eq. (6.53), with the same errors
discussed above.  The errors in perturbation theory are of $\order(g^2)$,
but even when $g$ is small the errors can be arbitrarily large.  This
same type of error arises no matter how many orders in $g$ are
included in the calculation of $H$.  This problem may best be understood
by thinking of the coupling as running with longitudinal momentum.  The
second order corrections to the Hamiltonian in Eqs. (7.21) and (7.22)
show that the coupling decreases as the longitudinal momentum transferred
through the vertex decreases.  In fact, according to the perturbative
analysis, the coupling actually becomes negative at sufficiently small
longitudinal momentum transfer.  The perturbative analysis breaks down
when such logarithmic divergences arise, and one must find a method of
re-summing these corrections.  In this case, we are finding that even if
the coupling is small for all longitudinal momentum transfers, we cannot
expand the coupling for one momentum transfer in powers of the coupling
at some drastically different momentum transfer.  This is the sort of
problem the Euclidean renormalization group manages to avoid, but the
invariant-mass renormalization group does not treat all components of
the momentum on an equal footing, and this is a price that must be
paid.


\bigskip
\noindent {\bf \rneight. Conclusion}
\medskip

After defining several renormalization groups that might be of interest
in the study of light-front field theory, I have found the Gaussian
fixed points and completed the linear analysis about the Gaussian fixed
point of interest for relativistic field theory.  The subsequent
second-order analyses of these transformations show that a
perturbative expansion
of a transformation about the Gaussian fixed point,
in which some of the irrelevant operators are discarded,
can only converge for transformations
that remove states of higher `free' energy than all states retained,
where the free energy is determined by the Gaussian fixed point.  This
constraint naturally leads to transformations that employ
invariant-mass cutoffs.  The linear analysis of invariant-mass
transformations reveals that there are an infinite number of relevant
and marginal variables, because functions of longitudinal momenta do not
affect the invariant-mass scaling dimension of operators.

While the linear analysis of a light-front transformation about the
Gaussian fixed point is readily completed for arbitrary Hamiltonians,
the second-order analysis has a complicated dependence on the
interactions.  One can easily write a general expression for the
transformed Hamiltonian using Eq. (4.14).  The
relevant and marginal operators in $u_2$ do not affect other operators
in the second-order
light-front analysis when zero-modes are dropped, and this
simplifies the analysis considerably.  The
only marginal interaction that directly enters
the second-order analysis is
the marginal part of $u_4$.  If
all irrelevant operators are dropped, only the relevant and marginal parts of
$u_2$ and $u_4$ survive.  Since the marginal part of $u_4$
contains an arbitrary function of longitudinal momenta, the
second-order correction to the marginal part of $u_4$ involves a new
function of longitudinal momenta and a complete analysis
requires one to compute trajectories of such functions, in general.
While an analysis that allows arbitrary functions is not extremely
complicated, the simplest example is the trajectory generated when
the initial value of the marginal part of $u_4$ is a constant, as suggested
by canonical field theory.  In this case
the second-order correction leads to a new
constant and not to a function of longitudinal momentum.  This case
was studied in detail, and it was shown that the canonical coupling
decreases as the cutoff is lowered.  This analysis is readily improved
by retaining leading irrelevant operators, and allowing functions to
appear in the marginal operator, but this was not done in this paper.

The invariant-mass cutoff violates explicit Lorentz covariance and
cluster decomposition, so the Hamiltonians one must investigate do not
display these properties manifestly.
If one uses the canonical $\phi^4$ Hamiltonian with an invariant-mass
cutoff to compute Green's functions, it is not surprising to find that
covariance and cluster decomposition are violated.  If one arbitrarily
chooses a more complicated renormalized Hamiltonian, there is no reason
to expect that these properties are restored.  One must
select the correct functions of longitudinal momentum in the marginal
and relevant operators at the lowest cutoff, and use the perturbative
renormalization group to estimate the irrelevant operators at the lowest
cutoff, to restore Lorentz covariance and cluster decomposition to
observables.  While it is possible to adjust these functions until
correct results are obtained in perturbation theory, I showed that one
can also use coupling constant coherence \APrefmark{\rPERWIL}
to restore covariance and
cluster decomposition.  By insisting that the canonical coupling is the
only variable that explicitly runs with the cutoff, and that all other
variables depend on the cutoff through perturbative dependence on the
canonical coupling, one uniquely fixes the relevant and irrelevant
operators at the lowest cutoff to $\order(g^2)$.  The marginal variables
are not uniquely fixed until one completes a third-order calculation,
and the correct results are not obtained unless $\order(\delta H^3)$
terms are kept in the transformation.  When these third-order terms are
retained, it is possible to use coupling coherence to fix the functions
appearing in the marginal part of $u_4$ to $\order(g^2)$, and these
functions are exactly those required to restore Lorentz covariance and
cluster decomposition to observables computed in second-order
perturbation theory.

The appearance of entire functions in the relevant and marginal
operators severely complicates the development of a perturbative
light-front renormalization group.  While it is encouraging to find that
coupling coherence apparently fixes these functions without direct
reference to Lorentz covariance and cluster decomposition in the output
observables, the light-front renormalization group does not offer a
convenient method for performing perturbative calculations.  The
Euclidean renormalization group is far more convenient for perturbative
calculations, because there are a small number of relevant and marginal
operators \APrefmark{\rWILNINE}.
In gauge theories Euclidean cutoffs typically
violate gauge invariance, and thereby
covariance (with the lattice providing an example of
how one can take advantage of irrelevant operators to maintain
gauge-invariance \APrefmark{\rWILFORT});
but one should be able to use coupling
coherence to restore these properties in a Euclidean renormalization
group analysis, as was done for the light-front renormalization group in
this paper.  Thus, the appearance of functions in the relevant and
marginal operators leads to computational complication of a perturbative
light-front analysis in comparison to a Euclidean analysis at best.

Many of the basic problems one encounters in trying to apply the
light-front renormalization group to QCD have been discussed in this
paper; however, several very important problems have been completely
avoided.  The most immediate problems result from divergences in QCD that
are not regulated by the invariant-mass cutoff.  In second-order
perturbation theory, one finds that $u_2$ for both quarks and gluons
diverges if one uses only an invariant-mass cutoff. There are logarithmic
divergences that remain in one-loop longitudinal momentum integrals
after the invariant-mass cutoff is imposed, and these require an
additional regulator.  These infrared longitudinal divergences do not
appear in the one-loop corrections to the three-particle and
four-particle interactions in QCD if one uses the canonical Hamiltonian,
but they are avoided only because there are precise cancellations
maintained by the canonical Hamiltonian \APrefmark{\rTHORN,\rPERQCD}.
These cancellations require
sectors of Fock space with different parton content to contribute with
relative strengths given by perturbation theory, and there is no reason
to believe that such cancellations are maintained
nonperturbatively.


In addition to new infrared divergences, a perturbative analysis using
coupling coherence in QCD may reveal that the canonical coupling
increases in strength
for small longitudinal momentum transfer, rather than
decreasing as found in the scalar theory.  This would mean that one cannot
simply re-sum the logarithmic corrections to the canonical coupling and
produce a reasonable coupling that runs with longitudinal momentum transfer,
because the coupling one obtains in this fashion diverges for a
finite longitudinal momentum transfer.

At this point it is not even obvious that these problems are
unwelcome.  When dealing with QCD one must balance seemingly contradictory
goals.  Because of asymptotic freedom a perturbative renormalization
group analysis may enable one to lower the cutoff on invariant masses to
a few GeV.  At the same time, the theory should confine quarks and
gluons, and confinement certainly cannot result from any interaction that
can be treated perturbatively at all energies.
Ideally only a few operators are
required to accurately approximate confinement effects.  If a large
number of operators are required it is probably quite
difficult to find them in an approximate analysis.  Moreover, hopefully
these interactions do not change particle number.  If operators that
change particle number diverge in strength, it is difficult to
approximate the states.  A renormalization group analysis that produces
an accurate effective QCD Hamiltonian with a cutoff of a few GeV must
either be able to treat the confining interactions nonperturbatively
when eliminating high energy states, or these interactions must be
perturbative in strength for high energy states and diverge in strength
only for low energy states.

Another problem that has been avoided in this paper is symmetry
breaking.
Unless the cutoff violates the symmetry of interest, one
never finds symmetry breaking terms in a perturbative analysis.  They
must simply be added to the Hamiltonian by hand, and their strength
must be tuned to reproduce an observable that is computed using the
final Hamiltonian, or one must use coupling coherence.  It is also
possible for terms that do not violate any symmetry to arise in this
fashion.  Any effective interaction associated with a vacuum condensate,
or more generally any effective interaction that does not depend on the
canonical coupling analytically,
must be introduced directly in the Hamiltonian if one performs a
light-front calculation with the zero modes removed, because the vacuum
is forced to be trivial in this case.  Once again, there are two
attitudes one can take concerning vacuum-induced interactions.  One can
emphasize the fact that the light-front offers little or no advantage to
anyone interested in solving the QCD vacuum problem; or one can
emphasize the fact that light-front field theory may offer anyone who is
primarily interested in building hadrons a way to reformulate the vacuum
problem \APrefmark{\rGLAZFOUR}.

The best way to check whether a
cutoff QCD Hamiltonian resulting from a perturbative light-front
renormalization group analysis is
even crudely accurate is to
diagonalize it and determine whether it produces even a
crude description of low-lying hadrons.  One can use a trial wave
function analysis once the Hamiltonian is selected, which is a powerful
nonperturbative tool that is not usually available in a field theoretic
analysis.  It seems highly unlikely that a perturbative QCD Hamiltonian
will produce reasonable results, and one knows that it will not produce
the mass splittings associated with chiral symmetry breaking if zero
modes are discarded.  It is
necessary to introduce new operators to produce symmetry breaking in
the spectrum, and it is almost certainly necessary to introduce
additional vacuum-induced interactions to produce a reasonable
description of the mass splittings associated with confinement.

Many of the problems one
expects to encounter in such an endeavor are familiar from the histories
of the constituent quark model \APrefmark{\rKOKK}
and lattice QCD \APrefmark{\rREBBI}.
The constituent quark model indicates that it should be
possible to obtain accurate models of all low-lying hadrons with cutoff
Hamiltonians, and a few degrees of freedom.
A constituent picture may arise naturally from
light-front QCD if the parton number-conserving interactions
in the cutoff Hamiltonians produce mass gaps
between sectors with different parton content that are reasonably large
in comparison to the cutoff.
The easiest Hamiltonians to use in producing simple
constituent hadrons may
employ fairly low invariant-mass cutoffs, making them
analogous to strong-coupling lattice actions \APrefmark{\rWILFORT}.  To obtain
reasonable results with `strong-coupling' Hamiltonians, it may be
necessary to simply tune the strength of various operators by hand,
using phenomenology as a
guide when selecting candidate
operators.  It is essential to
show that if one makes such uncontrolled approximations it is possible
to reproduce low energy hadronic observables.  On the other hand, if one
wants to work with large cutoffs for which the canonical coupling remains
small, it may be possible to use somewhat simpler Hamiltonians;
but then the Fock space required
to diagonalize the Hamiltonian is large and there is no reason to
expect that a few constituents will yield even crude approximations.
Hopefully an intermediate ground exists in which low-lying
hadrons are adequately
approximated as few parton states, while the Hamiltonian is not forced
to be absurdly complicated by the low cutoff.

The light-front renormalization group may eventually lead to the
solution of some of the most
important and interesting problems encountered in the study of
low-energy QCD.  At this point it merely offers a new perspective on
these problems.  This perspective differs radically from those offered
by Euclidean field theory, and it is not surprising that difficult new
challenges appear.  While the primary accomplishment of this article is
to illustrate how coupling coherence may allow one to obtain unique
answers in perturbation theory without maintaining manifest covariance
and gauge invariance, I hope that this article also elucidates some of
the
challenges we must meet to
use light-front field theory to solve QCD.


\bigskip
\noindent{\bf Acknowledgments }
\medskip

\indent I want to thank Ken Wilson, without whose help I could
not have completed this work.  I also owe a great debt to Avaroth
Harindranath, with whom I have discussed light-front field theory for
several enjoyable years.  I have profitted from discussions of the
renormalization group with Edsel Ammons,
Richard Furnstahl
and Tim Walhout, who made many useful comments on the text.
In addition I have benefitted from discussions
with Armen Ezekelian, Stanislaw G{\/l}azek, Paul Griffin, Kent Hornbostel,
Yizhang Mo, Steve Pinsky,
Dieter Schuette, Junko Shigemitsu, Brett Van de Sande,
and Wei-Min Zhang.
This work was supported by the National Science Foundation under
Grant No. PHY-9102922 and the Presidential Young Investigator Program
through Grant PHY-8858250.

\vfill
\eject

\bigskip
\noindent {\bf Appendix A:  Canonical Light-Front Scalar Field Theory}
\medskip

Canonical light-front scalar field theory is discussed by Chang, Root
and Yan \APrefmark{\rCHAONE,\rCHATWO}, who derive many of the results below.
This Appendix is not intended as an introduction to light-front field
theory, but merely collects some useful formulae and establishes
notation.
In this paper I choose the light-front time variable to be
$$ x^{+}=x^{0}+x^{3}, \eqno(A.1)$$
and the light-front longitudinal space variable to be
$$ x^{-} = x^{0}-x^{3} . \eqno(A.2)$$

\noindent With these choices the scalar product is

$$a \cdot b = a^\mu b_\mu = {1 \over 2} a^+ b^- + {1 \over 2} a^- b^+ -
{\bf a}^\perp \cdot {\bf b}^\perp \;,\eqno(A.3)$$

\noindent the derivative operators are

$$\partial^\pm = 2 {\partial \over \partial x^\mp} \;,\eqno(A.4)$$

\noindent and the four-dimensional volume element is

$$d^4x = {1 \over 2} dx^+ dx^- d^2x^\perp \;.\eqno(A.5)$$

The Lagrangian is independent of the choice of variables,

$${\cal L} = {1 \over 2} \partial^\mu \phi \partial_\mu \phi - {1 \over
2} \mu^2 \phi^2 - {\lambda \over 4!} \phi^4 \;, \eqno(A.6)$$

\noindent for example.  The commutation relation for the boson
field is

$$ [\phi(x^{+}, x^{-},\vec x^\perp), \partial^{+} \phi(x^{+}, y^{-},
\vec y^\perp) ] =
i \delta^3(x-y). \eqno(A.7)$$

In order to derive the Hamiltonian and other Poincar\'e generators, one
typically begins with the energy-momentum
tensor \APrefmark{\rCHAONE,\rCHATWO}.
While it is certainly
possible to derive a formal, ill-defined expression for the complete
tensor, I merely
list the canonical Hamiltonian,

$$\eqalign{ H = \int & dx^-d^2x^\perp \biggl[{1 \over 2}
\;\;\phi(x) \biggl(-\partial^{\perp 2}+\mu^2\biggr) \phi(x) +
{\lambda \over 4!} \phi^4(x)
\biggr]
.} \eqno(A.8)$$


Eq. (A.8) provides a formal definition of the Hamiltonian.
The boson field can be expanded in terms of a free particle basis,

$$ \phi(x) = \int \dqq \;\theta(q^+)\;\biggl[ a(q) e^{-iq \cdot x} +
a^\dagger(q) e^{iq \cdot x}\biggr], \eqno(A.9)$$

\noindent with
$$ [a(q),a^\dagger(q')] = 16 \pi^3 q^+ \delta^3(q-q'). \eqno(A.10) $$

\noindent If we use

$$\phi(x)=\int{d^4k \over
(2\pi)^3}\;\delta(k^2-\mu^2)\;\phi(k)\;e^{-ik\cdot x} ,\eqno(A.11)$$

\noindent we find that,

$$\phi(q)=a(q)\;\;\;[q^+>0]\;\;\;\;,\;\;\;\;
\phi(q)=-a^\dagger(-q)\;\;[q^+<0] \;.
\eqno(A.12)$$

\noindent In the remainder of this Appendix I simply assume that
all longitudinal momenta are greater than zero.

The free part of the Hamiltonian in Eq. (A.8) is,

$$H_0=\int dx^-d^2x^\perp {1 \over 2} \;\phi(x) \biggl(-\partial^{\perp
2}+\mu^2\biggr) \phi(x) = \int \dqq \; \Bigl({q^{\perp
2}+\mu^2 \over q^+} \Bigr) \;a^\dagger(q) a(q) \;.\eqno(A.13)$$

\noindent The interaction term is more complicated and I do not expand
it in terms of creation and annihilation operators.  Fock space
eigenstates of the free Hamiltonian are,

$$|q_1,q_2,...\rangle = a^\dagger(q_1) a^\dagger(q_2) \cdot\cdot\cdot
|0\rangle \;,\eqno(A.14)$$

\noindent with the normalization being,

$$\langle k|q\rangle = 16 \pi^3 q^+ \delta(k^+-q^+)\delta^2(k^\perp -
q^\perp) \;.\eqno(A.15)$$

\noindent Completeness implies that

$$1= |0\rangle \langle 0| \;+\;\int \dqq |q\rangle \langle q| \;+\;
{1 \over 2!} \int \dqq \int \dkk |q,k\rangle \langle q,k| \;+\;
\cdot\cdot\cdot \;,\eqno(A.16)$$

\noindent so we can write the free Hamiltonian as

$$\eqalign{
H_0 &= \int \dqq \Bigl({q^{\perp 2}+\mu^2 \over q^+} \Bigr) |q\rangle
\langle q| \cr &+\; {1 \over 2!}
\int \dqq \int \dkk \Bigl({q^{\perp 2}+\mu^2 \over q^+}
+ {k^{\perp 2}+\mu^2 \over k^+} \Bigr) |q,k\rangle \langle q,k| \cr
&+\; \cdot\cdot\cdot \;.}\eqno(A.17)$$

In order to provide further orientation, let me consider the problem of
computing the connected Green's functions for this theory.  This
problem in equal-time field theory is discussed in many textbooks, and
has been reviewed for light-front field theory
\APrefmark{\rLEPAGE-\rBROONE}.
To find the Green's
functions of a theory, consider the overlap of a state $\vert i(0)
\rangle$ at light-front time $x^+=0$ with a second state $\vert
f(\tau) \rangle$ at time $x^+=\tau$.  We split the Hamiltonian into
a free part $H_0$ and an interaction $V$, and find that

$$\eqalign{
\langle f(\tau) \vert i(0) \rangle &= (16 \pi^3) \delta^3(P_f-P_i)\;
G(f,i;\tau) = \langle f \vert e^{-i H \tau/2} \vert i \rangle \cr
&= i \int {d\epsilon \over 2 \pi} e^{-i \epsilon \tau /2}
(16 \pi^3) \delta^3(P_f-P_i)\; G(f,i;\epsilon)
.}\eqno(A.18)$$

\noindent This definition differs slightly from that given in some other
places.
It is then straightforward to demonstrate that

$$\eqalign{(16 \pi^3) \delta^3(P_f-P_i)\;
G(f,i;\epsilon) &= \langle f \vert {1
\over \epsilon-H+i0_+} \vert i \rangle \cr
&= \langle f \vert {1 \over \epsilon-H_0+i0_+} + {1 \over \epsilon-H_0+i0_+}
V {1 \over \epsilon-H_0+i0_+} \cr
&+ {1 \over \epsilon-H_0+i0_+} V {1 \over
\epsilon-H_0+i0_+} V {1 \over \epsilon-H_0+i0_+} + \cdot \cdot \cdot
\vert i \rangle .}\eqno(A.19)$$

\noindent
Operator products are evaluated by inserting a complete set of eigenstates of
$H_0$ between interactions, using Eq. (A.16), and using

$$H_0 \vert q_1,q_2,...\rangle = \Bigl[\sum_i q_i^-\Bigr] \vert
q_1,q_2,...\rangle \;, $$

$$q_i^- = {q_i^{\perp 2}+\mu^2 \over q_i^+} \;,\eqno(A.20)$$

\noindent to replace operators occurring in the denominators with
c-numbers.

Divergences arise from high energy states that are created
and annihilated by adjacent $V$'s, for example.
All divergences in perturbation theory come from
intermediate states (internal lines) that have large free energy.
The free energy is given by Eq. (A.20),
so divergences occur in
diagrams containing internal lines that carry large perpendicular
momentum (`ultraviolet' divergences) and/or small longitudinal momentum
(`infrared' divergences).

Given a light-front Hamiltonian, $H=H_0+V$, one can
determine the rules for constructing time-ordered perturbation theory
diagrams.  These diagrams are similar to the Hamiltonian diagrams used
in the text, but there are important differences related to the energy
denominators.
The diagrammatic rules allow us to evaluate all terms that occur in the
expansion for the Green's functions given in eq. (A.19).  For $\phi^4$
canonical field theory the diagrammatic rules for time-ordered connected
Green's functions are:

\item{(1)} Draw all allowed time-ordered diagrams with the quantum numbers
of the specified initial and final states on the external legs.  Assign
a separate momentum $k^\mu$ to each internal and external line,
setting $k^-=(k_\perp^2+\mu^2)/k^+$ for each line.
The momenta are directed
in the direction time flows.

\item{(2)} For each intermediate state there is a factor
$\bigl(\epsilon-\sum_i k_i^-+i0_+\bigr)^{-1}$, where the sum is over all
particles in the intermediate state.

\item{(3)} Integrate $\int {dk^+ d^2k_\perp \over 16 \pi^3 k^+}\;
\theta(k^+)$ for each
internal momentum.

\item{(4)} For each vertex associate a factor of $\lambda \;
\delta(K^+_{in}-K^+_{out})\;\delta^2(K^\perp_{in}-K^\perp_{out})$, where
$K^+_{in}$ is the sum of momenta entering a vertex, etc.

\item{(5)} Multiply the contribution of each time-ordered
diagram by a symmetry
factor $1/S$, where
$S$ is the order of the permutation group of the
internal lines and vertices leaving the diagram unchanged
with the external lines fixed.

To obtain the Green's function defined in Eq. (A.19), propagators for
the incoming and outgoing states must be added, and one must divide by an
overall factor of $(16 \pi^3)\delta^3(P_f-P_i)$.

\vfill
\eject

\bigskip
\noindent {\bf Appendix B:  Example Calculation of $u_2$ in the
Effective Hamiltonian}
\medskip

In this Appendix the second-order change in the effective Hamiltonian
shown in figure 3a is computed.  Let the first term in the Hamiltonian
be

$$ H = \int \dqt_1 \; \dqt_2 \; (16 \pi^3)
\delta^3(q_1-q_2) \; u_2(-q_1,q_2)
\;a^\dagger(q_1) a(q_2)\;+\cdot\cdot\cdot \;.\eqno(B.1)$$

\noindent  Then the matrix element of this operator between single
particle states is

$$\langle p'|H|p\rangle = \langle 0|\;a(p')\; H\; a^\dagger(p)\;|0\rangle
= (16 \pi^3)
\delta^3(p'-p)\;u_2(-p,p) \;.\eqno(B.2)$$

\noindent Thus, we easily determine $u_2$ from the matrix element.  It
is easy to compute matrix elements between other states.  Eq.
(4.14) gives us the matrix elements of the effective Hamiltonian
generated when the cutoff is lowered, and we are interested in a
second-order term generated by the interactions in Eq. (3.6),

$$\eqalign{v=
&{1 \over 6} \int \dqt_1\; \dqt_2\; \dqt_3\; \dqt_4 \; (16 \pi^3)
\delta^3(q_1+q_2+q_3-q_4) \cr
&\qquad\qquad\qquad\qquad u_4(-q_1,-q_2,-q_3,q_4)\; a^\dagger(q_1)
a^\dagger(q_2) a^\dagger(q_3) a(q_4)  \;,}\eqno(B.3)$$

$$\eqalign{v^\dagger=
&{1 \over 6} \int \dqt_1\; \dqt_2\; \dqt_3\; \dqt_4 \; (16 \pi^3)
\delta^3(q_1-q_2-q_3-q_4) \cr
&\qquad\qquad\qquad\qquad\qquad u_4(-q_1,q_2,q_3,q_4)\; a^\dagger(q_1)
a(q_2) a(q_3) a(q_4)  \;.}\eqno(B.4)$$

Since we will find that the incoming and outgoing momenta must be the
same, implying that $\epsilon_a$ and $\epsilon_b$ in Eq. (4.14) are the
same in this case, I combine the two second-order terms and obtain,

$$\eqalign{
&(16 \pi^3)\delta^3(p'-p)\;\delta v_2(-p,p)= \cr
&\qquad\qquad\qquad\qquad {1 \over 3!} \langle p'|\;
v^\dagger \;\; \int \dkt_1 \dkt_2 \dkt_3 \; \Theta(k_1,k_2,k_3)\; {
|k_1,k_2,k_3\rangle \langle k_1,k_2,k_3| \over p^--k_1^--k_2^--k_3^-}
\;\;v\;|p\rangle \;;}\eqno(B.5)$$

\noindent where $p^-=\pp^2/p^+$, etc.  $\Theta(k_1,k_2,k_3)$ is the
appropriate cutoff, which is determined by the transformation employed.
One can readily verify that

$$\langle p'|v^\dagger|k_1,k_2,k_3\rangle = (16 \pi^3)
\delta^3(p'-k_1-k_2-k_3) \; u_4(-p',k_1,k_2,k_3) \;, \eqno(B.6)$$

$$\langle k_1,k_2,k_3|v|p\rangle= (16 \pi^3) \delta^3(p-k_1-k_2-k_3) \;
u_4(-k_1,-k_2,-k_3,p) \;. \eqno(B.7)$$

\noindent Substituting these results in Eq. (B.5) leads to the final
result,

$$\eqalign{
\delta v_2(-p,p)=&{1 \over 3!} \int \dkt_1 \dkt_2 \dkt_3 \;
\Theta(k_1,k_2,k_3)\; (16 \pi^3)\delta^3(p-k_1-k_2-k_3) \cr
&\qquad\qquad\qquad\qquad
{u_4(-p,k_1,k_2,k_3)\; u_4(-k_1,-k_2,-k_3,p) \over
p^--k_1^--k_2^--k_3^-} \;.}\eqno(B.8)$$

\noindent
To obtain $\delta u_2$ from $\delta v_2$ we must complete the set of
re-scalings appropriate to the transformation.  If we want to study
spectator dependence, we simply need to compute matrix elements between
states containing additional particles, which directly affect the cutoff
function $\Theta$ for some transformations.

\bigskip
\noindent {\bf Appendix C:  Physical Masses in the Light-Front
Renormalization Group}
\medskip

What happens if the physical particles are massive?  In this case one
expects to find at
least one mass parameter in the relevant mass operator that
is an
independent function of the cutoff.  I assume that this is the mass
term that produces the correct relativistic dispersion relation for free
particles (\ie, the part of $u_2$ that does not depend on either
transverse or longitudinal momenta),
and I call this the `physical mass', even though it is a
running mass in the Hamiltonian that should not be confused with the
mass of a physical particle. The complete `mass' operator contains an infinite
number of relevant operators (\ie, functions of longitudinal momentum
fractions that produce a complicated dispersion relation) even in the
critical theory, and one should expect new functions to appear in the
massive theory.  In the massive theory, one can again use the
conditions Wilson and I have proposed, selecting a single coupling and a
single mass to explicitly run with the cutoff, while all other operators
depend on the cutoff only through their dependence on this coupling and
mass.  In this Appendix I
briefly study a few of the consequences of adding a physical mass.
This study is both incomplete and preliminary.  I explicitly
consider only the second-order behavior of the transformation, and I
focus on the portion of the Hamiltonian trajectory near the
critical Gaussian fixed point where the physical mass is exponentially
small in comparison to the cutoff.

The physical mass is proportional to some
$\Lambda_\calN$, which sets the scale at which the physical mass can no
longer be regarded as small and must be treated nonperturbatively.
Since
$\Lambda_\calN=2^{-\calN}\Lambda_0$, as we construct a renormalized
trajectory we expect the physical mass in
$H^{\Lambda_0}_{\Lambda_0}$, $m_0$, to be
exponentially small in comparison to $\Lambda_0$.
In other words, on a
renormalized trajectory we should find an infinite number of
Hamiltonians near the critical fixed point in an asymptotically free
theory.  In a scalar theory, if we let $\Lambda_0 \rightarrow \infty$
while keeping the initial coupling small, we expect to find the
trajectory approach the critical fixed point and then asymptotically
approach the line between the critical and trivial Gaussian fixed
points, with the strength of the interaction going to zero.  For an
infinitely long trajectory there are interactions only over an
infinitesimal initial portion of the trajectory, after which the
trajectory misses the critical fixed point by an infinitesimal amount
and then follow infinitesimally close to the line of Gaussian
Hamiltonians joining the critical and trivial Gaussian fixed points.

After
adding a physical mass to the Hamiltonian, the analysis
that led to Eq. (6.11) yields

$$\eqalign{ \delta v_4 = {g^2 \over 4}\; \int
& \dsx  \Bigl[{\rp^2+m^2 \over y(1-y)}
- {\sp^2+m^2 \over x(1-x)}\Bigr]^{-1} \cr
&\theta\Bigl(\Lambda_0^2-{\sp^2 \over x(1-x)}\Bigr)
\theta\Bigl({\sp^2 \over x(1-x)} -
\Lambda_1^2\Bigr)
\;.}\eqno(C.1)$$

\noindent  Note that I have placed the physical mass in the denominator.
If we are considering perturbations about the critical Gaussian fixed
point, we should include the mass as an interaction; but we can find the
results of such a treatment here by expanding in powers of
the physical mass.  When the physical mass is an independent parameter in the
Hamiltonian, and it is small, we expect to find a well-defined
perturbative expansion in powers of the running coupling and in powers
of the running mass.  Here I seek a perturbative renormalization group
in which such expansions exist.

In Eq. (C.1) I have
used an invariant-mass cutoff that contains no mass, and
the same problem that plagued $T^\perp$ is encountered;
the denominator can change sign inside the integral.  There is nothing
to prevent $m^2/y/(1-y)-m^2/x/(1-x)$ from becoming large and positive
for small $y$.  If we expand the denominator in
powers of $\rp^2$, and complete the integrals,
the coefficients in the expansion become arbitrarily large, producing
divergences in subsequent transformations.

The only simple
solution that I have found is to abandon the first
invariant-mass transformation and use the transformation that employs
the cutoffs in Eq. (3.24).  Some of the peculiarities of this
transformation have already been discussed.  Here I want to focus on the
changes in the analysis of the massless theory
wrought by the physical mass terms in energy denominators and cutoffs.
Perhaps the
most important observation to make at this point is that the addition of
a small mass {\it may} lead to small changes in the analysis, because all
integrals are finite before the mass is added.  To see that this is
apparently the case, return to the correction to $u_4$ in Eq. (C.1), and use
the new invariant mass cutoffs to get

$$\eqalign{ \delta v_4 = 144 g^2\; \int
& \dsx  \Bigl[{\rp^2+m^2 \over y(1-y)}
- {\sp^2+m^2 \over x(1-x)}\Bigr]^{-1} \cr
&\theta\Bigl(\Lambda_0^2-{\sp^2+m^2 \over x(1-x)}\Bigr)
\theta\Bigl({\sp^2+m^2 \over x(1-x)} -
\Lambda_1^2\Bigr)
\theta\Bigl(\Lambda_1^2-{\rp^2+m^2 \over y(1-y)}\Bigr)
\;.}\eqno(C.2)$$

\noindent  Here I display the cutoff associated with the
incoming state, which prevents the energy denominator from going
through zero.
Let me show the
analytic result and use it to analyze $\delta u_4$ when there is a
physical mass.  Defining

$$E = {\rp^2+m^2 \over y(1-y)} \;, \eqno(C.3)$$

\noindent we obtain,

$$\eqalign{ & \qquad \qquad \delta v_4 = -{g^2 \over 64 \pi^2} \;
\theta\Bigl(\Lambda_1^2-E\Bigr)
\Biggl\{ \;ln \Biggl[ {1+\sqa \over 1-\sqa}\cdot{1-\sqb \over 1+\sqb}
\Biggr] \cr
&+ \; \sqc \; ln \Biggl[ {\sqa-\sqc \over \sqa+\sqc} \cdot
{\sqb+\sqc \over \sqb-\sqc} \Biggr] \Biggr\}
\;. } \eqno(C.4)$$

\noindent Without the cutoff on the external energy,
we would encounter negative arguments in the logarithm.  With the
cutoff the argument can still go to zero; however, I have discussed
this same basic issue following Eq. (6.15), where I argued that it is not
necessarily a serious problem for a perturbative analysis.

The marginal part of this correction is obtained by taking the
limit $\rp \rightarrow 0$ with $y$ and $m$ fixed, which leads to

$$\eqalign{ & \qquad \qquad \delta v_4 \rightarrow
-{g^2 \over 64 \pi^2} \;
\Biggl\{ \;ln \Biggl[ {1+\sqa \over 1-\sqa}\cdot{1-\sqb \over 1+\sqb}
\Biggr] \cr
&+ \; \sqd \; ln \Biggl[ {\sqa-\sqd \over \sqa+\sqd} \cdot \cr
&\qquad\qquad\qquad\qquad\qquad\qquad\qquad\qquad\qquad
{\sqb+\sqd \over \sqb-\sqd} \Biggr] \Biggr\}
\;. } \eqno(C.5)$$

\noindent  Unlike the ${\cal O}(g^2)$ correction encountered in the
theory with no physical mass, this marginal operator has a complicated
dependence on $y$.  One can continue to compute transformations, using
this entire marginal operator, or one can make an expansion of this
operator in powers of $m^2$.  Such an expansion should converge, as long
as $m^2$ is exponentially small in comparison to $\Lambda_0^2$.  After
setting $\Lambda_1=\Lambda_0/2$,
the expansion yields

$$\eqalign{ & \qquad \qquad \delta v_4 \rightarrow
-{9 g^2 \over \pi^2} \;\Biggl\{ \;ln(4) - {3 m^2 \over \Lambda_0^2}
{1-2y(1-y) \over y(1-y)}+\cdot\cdot\cdot \Biggr\}
\;. } \eqno(C.6)$$

The first term in the expansion is the result obtained for the massless
theory, while the second term must be considered further.  There are
cutoffs on the remaining states that prevent $m^2/y/(1-y)$ from becoming
larger than $\Lambda_0^2/4$, so the first correction to $ln(4)$ is at
most $3/4$.
Moreover, the cutoff actually
involves $\rp^2+m^2$, and not just $m^2$.  Thus, no individual term in
the expansion of $\delta v_4$ given in Eq. (C.5) diverges within the
limits imposed by the cutoffs, although the entire sum can diverge.  I
do not repeat the discussion following Eq. (6.15).

As long as the mass remains small, we can
make a perturbative expansion of every term in powers of the mass
divided by the cutoff.  In this case, the mass is treated as if it were
a transverse momentum.  This must be done with some care.  As we have
already seen in the critical theory, the step function cutoffs sometimes
lead to singular distributions when limits involving their arguments are
taken.  Just as taking transverse momenta to zero in the critical theory
led to delta functions involving longitudinal momentum fractions, taking
the mass to zero may produce such singular distributions, and it is
essential that this limit be taken exactly.  Thus, as long as
the mass is small, we can continue to expand the Hamiltonian about the
critical fixed point provided we carefully evaluate distributions.
We expand every
operator in powers of transverse momenta, because the linear analysis
reveals that such powers lead to increasingly irrelevant eigenoperators
of the linearized transformation.  While powers of the physical mass do
not lead to increasingly irrelevant operators, they do lead to
increasingly large powers of $m^2/\Lambda_0^2$.  The subsequent
rescaling replaces each power with $4\;m^2/\Lambda_0^2$; however, the
expansion should remain reasonable until we run the effective cutoff
down to the point where $m^2/\Lambda_n^2$ is not small.

To this point I have ignored spectators in the discussion of $\delta
v_4$.  Spectators have a drastic, potentially disastrous, effect on the
analysis.  When we studied the marginal part of $\delta u_4$ in the
massless theory, spectators had no effect, because their transverse
momenta were set to zero to find the marginal operator and they dropped
out of the cutoffs as a result.  However, if the spectators are massive,
they do not drop out of the cutoffs when their transverse momenta are
zero.  Their presence effectively lowers the cutoff.  If there are $n$
spectators, the cutoffs are effectively shifted,

$$\Lambda_0^2 \rightarrow \Lambda_0^2 - \sum_{i=1}^n { m^2 \over x_i} \;
\;\;,\;\;\;
\Lambda_1^2 \rightarrow \Lambda_1^2 - \sum_{i=1}^n { m^2 \over x_i} \;
, \eqno(C.7)$$

\noindent where $x_i$ are the longitudinal momentum fractions of the
spectators.  If the total longitudinal momentum fraction of the
spectators is $1-w$, the smallest shift that this
sum can introduce is $n m^2/(1-w)$.  For sufficiently large $n$, this
always becomes larger than $\Lambda_0^2$, and we find that the
invariant-mass cutoff is also a cutoff on particle number.  Moreover, in
the highest sectors of Fock space that survive this cutoff, the effective
cutoff is arbitrarily small and we must expect that any change in this
cutoff produces nonperturbative effects in the exact results.
I see no way around this
conclusion; however, the error made by treating the highest sectors of
Fock space perturbatively as the cutoff is changed may still be small if
one is interested only in states of much lower energy than the states
removed, which is always the case in a renormalization group analysis.

When the effective cutoff becomes very small, it is because
the spectator state is a high energy state.  Thus, the nonperturbative
problem occurs when we need to accurately approximate the part of the
Hamiltonian that governs states of high energy.  In this case, the
intermediate state has high energy because there are a large number of
massive particles.  If our ultimate interest is to study states with
such high energy, we do not expect to be able to lower the cutoff to this
scale without encountering a nonperturbative problem.  However, if our
ultimate interest is to study low energy states, we need only concern
ourselves with the error made in the matrix elements that ultimately
govern the
mixing of these many-body high energy intermediate states with the few-body
low energy states of interest.  It may be possible to accurately approximate
these matrix elements using perturbation theory even when large errors
are made in the complete wave function for the many-body state.
Since it is far easier to study such spectator effects quantitatively
when computing $\delta u_2$, I turn to this.

In order to compute $\delta u_2$ for the massive theory, we can follow
the same steps that led to Eq. (6.25) and simply add a mass to obtain

$$\eqalign{ \delta v_2 =  {g^2 \over 3!} \;w \Lambda_0^2
\; \int & \dqp \drp \dssp
\Bigl[{\tp^2 \over w \Lambda_0^2} -
{\qp'^2 \over x'} - {\rp'^2 \over y'} - {\sp'^2 \over z'} - {2 m^2 \over w
\Lambda_0^2}
\Bigr]^{-1} \cr
&(16 \pi^3) \delta(1-x'-y'-z') \delta^2({\tp \over \sqrt{w} \Lambda_0} -
\qp'-\rp'-\sp') \cr
&\theta\Bigl(1- {\tp^2+M^2 \over (1-w)\Lambda_0^2}- {3 m^2
\over w \Lambda_0^2} -
{\qp'^2 \over x'} - {\rp'^2 \over y'} - {\sp'^2
\over z'} \Bigr) \cr
&\theta\Bigl({\qp'^2 \over x'} + {\rp'^2 \over y'} + {\sp'^2 \over z'}
+ {\tp^2+M^2 \over (1-w)\Lambda_0^2}+ {3 m^2 \over w
\Lambda_0^2} - {1 \over 4} \Bigr) \cr
&\theta\Bigl(1/4-{\tp^2+m^2 \over w \Lambda_0^2} - {\tp^2+M^2
\over (1-w)\Lambda_0^2}\Bigr)
\;.}\eqno(C.8)$$

\noindent Here the invariant-mass-squared of the spectators,
$M^2$, is dependent on the individual
longitudinal momentum fractions of the spectators, and it no longer
vanishes when all spectator transverse momenta vanish.
I have displayed the
cutoff on the external state to make it clear that an expansion in
powers of $m^2$ may be reasonable even though one might fear that
$M^2$, $1/(1-w)$, or $1/w$ could become large.  When $\delta v_2$ is
expanded in powers of $m^2$, an infinite number of functions
of longitudinal momentum fractions appear.  Since this happens even when
all external transverse momenta are taken to zero, this means that
an infinite number of relevant operators appear.
Relevant operators usually
must be precisely controlled
because they grow at an exponential rate.  The exception is when their
initial strength is exponentially small, and remains exponentially small
over all but the final part of a trajectory of Hamiltonians.  In this
case, a small error in the coefficient of a relevant operator remains
small over the entire trajectory.  This error becomes exponentially
large only when the mass becomes exponentially large, and I am interested
in approximating the portion of the renormalization group trajectory
over which the mass is small.  There is a separate important question of
how precisely one must approximate this trajectory if one wants accurate
predictions for low energy observables to result from an exact
diagonalization of the final Hamiltonian on the trajectory.  I do not
address this issue here.

If we expand every term in the transformed Hamiltonian in powers of
$m^2$, we find that the second-order transformation is almost identical
to the massless case.  We can compute a trajectory of massive Hamiltonians,
using the second-order transformation and expanding every term in powers
of $m^2/\Lambda_0^2$.  We can choose $H^{\Lambda_0}_{\Lambda_0}$ to be

$$\eqalign{ H^{\Lambda_0}_{\Lambda_0} =&
\int {d^2\calp^\perp d\calp^+ \over 16\pi^3 \calp^+}\;\Biggl\{ \cr
&\qquad \; \int {d^2r^\perp_1 dx_1 \over 16 \pi^3 x_1 } \;
(16 \pi^3) \delta^2(\rp_1) \delta(1-x_1)
\;\Biggl[ {\calp^{\perp 2} \over \calp^+}+
(1+\xi_0) {\rp^2_1 +  m_0^2 \over x_1 \calp^+}+\mu_0^2 \Biggr]\;
|q_1 \rangle \langle q_1 |  \cr
&\qquad + \;\int {d^2r^\perp_1 dx_1 \over 16 \pi^3 x_1 }
\; \int {d^2r^\perp_2 dx_2 \over 16 \pi^3 x_2 } \;
(16 \pi^3) \delta^2(\rp_1+\rp_2) \delta(1-x_1-x_2) \cr
& \qquad \Biggl[{\calp^{\perp 2} \over \calp^+}+
(1+\xi_0) {\rp^2_1 +m_0^2 \over x_1 \calp^+} + (1+\xi_0) {\rp^2_2+m_0^2
\over x_2 \calp^+}+2\mu_0^2\Biggr]
\;|q_1,q_2 \rangle \langle q_1,q_2 | \;\;+\;\;\cdot\cdot\cdot \;\Biggr\}
\cr
+&{g_0 \over 6} \int \dqt_1\; \dqt_2\; \dqt_3\; \dqt_4 \; (16 \pi^3)
\delta^3(q_1+q_2+q_3-q_4)
\; a^\dagger(q_1) a^\dagger(q_2)
a^\dagger(q_3)
a(q_4)  \cr
+&{g_0 \over 4} \int \dqt_1\; \dqt_2\; \dqt_3\; \dqt_4 \; (16 \pi^3)
\delta^3(q_1+q_2-q_3-q_4)
\; a^\dagger(q_1)
a^\dagger(q_2) a(q_3) a(q_4)  \cr
+&{g_0 \over 6} \int \dqt_1\; \dqt_2\; \dqt_3\; \dqt_4 \; (16 \pi^3)
\delta^3(q_1-q_2-q_3-q_4)
\; a^\dagger(q_1)
a(q_2) a(q_3) a(q_4)
\;.}\eqno(C.9)
$$

\noindent This Hamiltonian contains a complete set of relevant and
marginal
operators through ${\cal O}(m^2/\Lambda_0^2)$.  When the
second-order invariant-mass transformation is applied to the Hamiltonian,
and all irrelevant operators are dropped,
the resultant Hamiltonian contains no new relevant or marginal operators
to these orders, and the only effect is to change the constants
$m_0$, $\xi_0$, $\mu_0$, and $g_0$.  Thus we can write the relevant
and marginal operators in $H^{\Lambda_0}_{\Lambda_n}$
as in Eq. (6.35), using the constants $m_n$,
$\xi_n$, $\mu_n$, and
$g_n$.

The complete second-order equations
for the evolution of $\xi_n$, $\mu_n$, and
$g_n$ are identical to Eqs. (6.36)-(6.38).  Moreover the equation
for the evolution of $m$ to this order is trivial,

$$m_{n+1}^2 = 4 m_n^2 \;.
\eqno(C.10)$$

\noindent
It is clear that the presence of a physical mass severely complicates
the higher-order analyses, but the qualitative features of
the analysis for the massless theory should
survive when an additional expansion in powers of $m^2/\Lambda_0^2$ is
made.

\vfill
\eject

\bigskip
\noindent{\bf References}
\medskip

\refitem {1.}
\obeyendofline \frenchspacing E. C. G. Stueckelberg and A. Peterman, {\it Helv.
Phys. Acta} {\bf  26} (1953), 499.
\ignoreendofline
\refitem {2.}
\obeyendofline \frenchspacing M. Gell-Mann and F. E. Low, {\it Phys. Rev.} {\bf
95} (1954), 1300.
\ignoreendofline
\refitem {3.}
\obeyendofline \frenchspacing N. N. Bogoliubov and D.V. Shirkov, ``Introduction
to the Theory of Quantized Fields'', Interscience, New York, 1959.
\ignoreendofline
\refitem {4.}
\obeyendofline \frenchspacing K. G. Wilson, {\it Phys. Rev.} {\bf 140} (1965),
B445.
\ignoreendofline
\refitem {5.}
\obeyendofline \frenchspacing K. G. Wilson, {\it Phys. Rev.} {\bf D2} (1970),
1438.
\ignoreendofline
\refitem {6.}
\obeyendofline \frenchspacing K. G. Wilson, {\it Phys. Rev.} {\bf D3} (1971),
1818.
\ignoreendofline
\refitem {7.}
\obeyendofline \frenchspacing K. G. Wilson, {\it Phys. Rev.} {\bf D6} (1972),
419.
\ignoreendofline
\refitem {8.}
\obeyendofline \frenchspacing K. G. Wilson, {\it Phys. Rev.} {\bf B4} (1971),
3174.
\ignoreendofline
\refitem {9.}
\obeyendofline \frenchspacing K. G. Wilson, {\it Phys. Rev.} {\bf B4} (1971),
3184.
\ignoreendofline
\refitem {10.}
\obeyendofline \frenchspacing K. G. Wilson and M. E. Fisher, {\it Phys. Rev.
Lett.} {\bf 28} (1972), 240.
\ignoreendofline
\refitem {11.}
\obeyendofline \frenchspacing K. G. Wilson, {\it Phys. Rev. Lett.} {\bf 28}
(1972), 548.
\ignoreendofline
\refitem {12.}
\obeyendofline \frenchspacing K. G. Wilson and J. B. Kogut, {\it Phys. Rep.}
{\bf 12C} (1974), 75.
\ignoreendofline
\refitem {13.}
\obeyendofline \frenchspacing K. G. Wilson, {\it Rev. Mod. Phys.} {\bf 47}
(1975), 773.
\ignoreendofline
\refitem {14.}
\obeyendofline \frenchspacing K. G. Wilson, {\it Rev. Mod. Phys.} {\bf 55}
(1983), 583.
\ignoreendofline
\refitem {15.}
\obeyendofline \frenchspacing K. G. Wilson, {\it Adv. Math.} {\bf 16} (1975),
170.
\ignoreendofline
\refitem {16.}
\obeyendofline \frenchspacing K. G. Wilson, {\it Scientific American} {\bf 241}
(1979), 158.
\ignoreendofline
\refitem {17.}
\obeyendofline \frenchspacing F. J. Wegner, {\it Phys. Rev.} {\bf B5} (1972),
4529.
\ignoreendofline
\refitem {18.}
\obeyendofline \frenchspacing F. J. Wegner, {\it Phys. Rev.} {\bf B6} (1972),
1891.
\ignoreendofline
\refitem {19.}
\obeyendofline F. J. Wegner, {\it in} ``Phase Transitions and Critical
Phenomena'' (C. Domb and M. S. Green, Eds.), Vol. 6,
Academic Press, London, 1976.
\ignoreendofline
\refitem {20.}
\obeyendofline \frenchspacing L. P. Kadanoff, {\it Physica} {\bf 2} (1965),
263.
\ignoreendofline
\refitem {21.}
\obeyendofline \frenchspacing C. Rebbi, ``Lattice Gauge Theories and Monte
Carlo Simulations'', World Scientific, Singapore, 1983.
\ignoreendofline
\refitem {22.}
\obeyendofline \frenchspacing C. G. Callan, {\it Phys. Rev.} {\bf D2} (1970),
1541.
\ignoreendofline
\refitem {23.}
\obeyendofline \frenchspacing K. Symanzik, {\it Comm. Math. Phys.} {\bf 18}
(1970), 227.
\ignoreendofline
\refitem {24.}
\obeyendofline \frenchspacing R. P. Feynman, {\it Rev. Mod. Phys.} {\bf 20}
(1948), 367.
\ignoreendofline
\refitem {25.}
\obeyendofline \frenchspacing I. Newton, ``Philosophiae Naturalis Principia
Mathematica'', S. Pepys, London, 1686.
\ignoreendofline
\refitem {26.}
\obeyendofline \frenchspacing S. K. Ma, {\it Rev. Mod. Phys.} {\bf 45} (1973),
589.
\ignoreendofline
\refitem {27.}
\obeyendofline \frenchspacing G. Toulouse and P. Pfeuty, ``Introduction to the
Renormalization Group and to Critical Phenomena'', Wiley, Chichester, 1977.
\ignoreendofline
\refitem {28.}
\obeyendofline \frenchspacing S. K. Ma, ``Modern Theory of Critical
Phenomena'', Benjamin, New York, 1976.
\ignoreendofline
\refitem {29.}
\obeyendofline \frenchspacing D. Amit, ``Field Theory, the Renormalization
Group, and Critical Phenomena'', Mc-Graw-Hill, New York, 1978.
\ignoreendofline
\refitem {30.}
\obeyendofline \frenchspacing J. Zinn-Justin, ``Quantum Field Theory and
Critical Phenomena'', Oxford, Oxford, 1989.
\ignoreendofline
\refitem {31.}
\obeyendofline \frenchspacing N. Goldenfeld, ``Lectures on Phase Transitions
and the Renormalization Group'', Add-ison-Wesley, Reading Mass., 1992.
\ignoreendofline
\refitem {32.}
\obeyendofline \frenchspacing P. A. M. Dirac, {\it Rev. Mod. Phys.} {\bf 21}
(1949), 392.
\ignoreendofline
\refitem {33.}
\obeyendofline \frenchspacing P. A. M. Dirac, ``Lectures on Quantum Field
Theory'',  Academic Press, New York, 1966.
\ignoreendofline
\refitem {34.}
\obeyendofline An extensive list of references on light-front physics
({\it light.tex}) is available via anonymous ftp from
public.mps.ohio-state.edu in the subdirectory pub/infolight.
\ignoreendofline
\refitem {35.}
\obeyendofline \frenchspacing S. Weinberg, {\it Phys. Rev.} {\bf 150} (1966),
1313.
\ignoreendofline
\refitem {36.}
\obeyendofline \frenchspacing A. Harindranath and J. P. Vary, {\it Phys. Rev.}
{\bf D36} (1987), 1141.
\ignoreendofline
\refitem {37.}
\obeyendofline \frenchspacing J. R. Hiller, {\it Phys. Rev.} {\bf D44} (1991),
2504.
\ignoreendofline
\refitem {38.}
\obeyendofline \frenchspacing J. B. Swenson and J. R. Hiller, {\it Phys. Rev.}
{\bf D48} (1993), 1774.
\ignoreendofline
\refitem {39.}
\obeyendofline \frenchspacing R. J. Perry and K. G. Wilson, {\it Nucl. Phys.}
{\bf B403} (1993), 587.
\ignoreendofline
\refitem {40.}
\obeyendofline \frenchspacing S. Fubini and G. Furlan, {\it Physics} {\bf 1}
(1964), 229.
\ignoreendofline
\refitem {41.}
\obeyendofline \frenchspacing R. Dashen and M. Gell-Mann, {\it Phys. Rev. Lett}
{\bf  17} (1966), 340.
\ignoreendofline
\refitem {42.}
\obeyendofline \frenchspacing J. D. Bjorken, {\it Phys. Rev.} {\bf 179} (1969),
1547.
\ignoreendofline
\refitem {43.}
\obeyendofline \frenchspacing R. P. Feynman, ``Photon-Hadron Interactions'',
Benjamin, Reading, Massachusetts, 1972.
\ignoreendofline
\refitem {44.}
\obeyendofline \frenchspacing J. B. Kogut and L. Susskind, {\it Phys. Rep.}
{\bf C8} (1973), 75.
\ignoreendofline
\refitem {45.}
\obeyendofline \frenchspacing S.-J. Chang and S.-K. Ma, {\it Phys. Rev.} {\bf
180} (1969), 1506.
\ignoreendofline
\refitem {46.}
\obeyendofline \frenchspacing J. B. Kogut and D. E. Soper, {\it Phys. Rev.}
{\bf D1} (1970), 2901.
\ignoreendofline
\refitem {47.}
\obeyendofline \frenchspacing J. D. Bjorken, J. B. Kogut, and D. E. Soper, {\it
Phys. Rev.} {\bf D3} (1971), 1382.
\ignoreendofline
\refitem {48.}
\obeyendofline \frenchspacing S.-J. Chang, R. G. Root and T.-M. Yan, {\it Phys.
Rev. } {\bf D7} (1973), 1133.
\ignoreendofline
\refitem {49.}
\obeyendofline \frenchspacing S.-J. Chang and T.-M. Yan, {\it Phys. Rev.} {\bf
D7} (1973), 1147.
\ignoreendofline
\refitem {50.}
\obeyendofline \frenchspacing T.-M. Yan, {\it Phys. Rev.} {\bf D7} (1973),
1760.
\ignoreendofline
\refitem {51.}
\obeyendofline \frenchspacing T.-M. Yan, {\it Phys. Rev.} {\bf D7} (1973),
1780.
\ignoreendofline
\refitem {52.}
\obeyendofline \frenchspacing S. J. Brodsky, R. Roskies and R. Suaya, {\it
Phys. Rev.} {\bf D8} (1973), 4574.
\ignoreendofline
\refitem {53.}
\obeyendofline \frenchspacing C. Bouchiat, P. Fayet, and N. Sourlas, {\it Lett.
Nuovo Cim.} {\bf  4} (1972), 9.
\ignoreendofline
\refitem {54.}
\obeyendofline \frenchspacing A. Harindranath and R. J. Perry, {\it Phys. Rev.}
{\bf D43} (1991), 492.
\ignoreendofline
\refitem {55.}
\obeyendofline \frenchspacing D. Mustaki, S. Pinsky, J. Shigemitsu, and K.
Wilson, {\it Phys. Rev.} {\bf D43} (1991), 3411.
\ignoreendofline
\refitem {56.}
\obeyendofline \frenchspacing M. Burkhardt and A. Langnau, {\it Phys. Rev.}
{\bf D44} (1991), 1187.
\ignoreendofline
\refitem {57.}
\obeyendofline \frenchspacing M. Burkhardt and A. Langnau, {\it Phys. Rev.}
{\bf D44} (1991), 3857.
\ignoreendofline
\refitem {58.}
\obeyendofline \frenchspacing D. G. Robertson and G. McCartor, {\it Z. Phys.}
{\bf C53} (1992), 661.
\ignoreendofline
\refitem {59.}
\obeyendofline \frenchspacing R. J. Perry, {\it Phys. Lett.} {\bf B300} (1993),
8.
\ignoreendofline
\refitem {60.}
\obeyendofline \frenchspacing I. Tamm, {\it J. Phys. (USSR)} {\bf 9} (1945),
449.
\ignoreendofline
\refitem {61.}
\obeyendofline \frenchspacing S. M. Dancoff, {\it Phys. Rev.} {\bf 78} (1950),
382.
\ignoreendofline
\refitem {62.}
\obeyendofline \frenchspacing H. A. Bethe and F. de Hoffmann, ``Mesons and
Fields, Vol. II'',  Row, Peterson and Company, Evanston, Illinois, 1955.
\ignoreendofline
\refitem {63.}
\obeyendofline \frenchspacing R. J. Perry, A. Harindranath and K. G. Wilson,
{\it  Phys. Rev. Lett.} {\bf 65} (1990), 2959.
\ignoreendofline
\refitem {64.}
\obeyendofline A. C. Tang, Ph.D thesis, Stanford University, SLAC-Report-351,
June (1990).
\ignoreendofline
\refitem {65.}
\obeyendofline \frenchspacing R. J. Perry and A. Harindranath, {\it Phys. Rev.}
{\bf D43} (1991), 4051.
\ignoreendofline
\refitem {66.}
\obeyendofline \frenchspacing A. C. Tang, S. J. Brodsky, and H. C. Pauli, {\it
Phys. Rev.} {\bf D44} (1991), 1842.
\ignoreendofline
\refitem {67.}
\obeyendofline \frenchspacing M. Kaluza and H. C. Pauli, {\it Phys. Rev.} {\bf
D45} (1992), 2968.
\ignoreendofline
\refitem {68.}
\obeyendofline \frenchspacing St. D. G{\l }azek and R.J. Perry, {\it Phys.
Rev.} {\bf D45} (1992), 3740.
\ignoreendofline
\refitem {69.}
\obeyendofline \frenchspacing A. Harindranath, R. J. Perry, and J. Shigemitsu,
{\it Phys. Rev.} {\bf D46} (1992), 4580.
\ignoreendofline
\refitem {70.}
\obeyendofline \frenchspacing P. M. Wort, {\it Phys. Rev.} {\bf D47} (1993),
608.
\ignoreendofline
\refitem {71.}
\obeyendofline \frenchspacing S. G{\l }azek, A. Harindranath, S. Pinsky, J.
Shigemitsu, and K. G. Wilson, {\it Phys. Rev.} {\bf D47} (1993), 1599.
\ignoreendofline
\refitem {72.}
\obeyendofline \frenchspacing H. H. Liu and D. E. Soper, {\it Phys. Rev.} {\bf
D48} (1993), 1841.
\ignoreendofline
\refitem {73.}
\obeyendofline \frenchspacing St. D. G{\l }azek and R.J. Perry, {\it Phys.
Rev.} {\bf D45} (1992), 3734.
\ignoreendofline
\refitem {74.}
\obeyendofline \frenchspacing B. van de Sande and S. S. Pinsky, {\it Phys.
Rev.} {\bf D46} (1992), 5479.
\ignoreendofline
\refitem {75.}
\obeyendofline S. D. G{\l }azek and K. G. Wilson, ``Renormalization of
Overlapping Transverse Divergences in a Model Light-Front Hamiltonian,''
Ohio State preprint (1992).
\ignoreendofline
\refitem {76.}
\obeyendofline \frenchspacing H. Bergknoff, {\it Nucl. Phys.} {\bf B122}
(1977), 215.
\ignoreendofline
\refitem {77.}
\obeyendofline \frenchspacing T. Eller, H. C. Pauli, and S. J. Brodsky, {\it
Phys. Rev.} {\bf D35} (1987), 1493.
\ignoreendofline
\refitem {78.}
\obeyendofline \frenchspacing Y. Ma and J. R. Hiller, {\it J. Comp. Phys.} {\bf
82} (1989), 229.
\ignoreendofline
\refitem {79.}
\obeyendofline \frenchspacing M. Burkhardt, {\it Nucl. Phys.} {\bf A504}
(1989), 762.
\ignoreendofline
\refitem {80.}
\obeyendofline \frenchspacing K. Hornbostel, S. J. Brodsky, and H. C. Pauli,
{\it Phys. Rev.} {\bf D41} (1990), 3814.
\ignoreendofline
\refitem {81.}
\obeyendofline \frenchspacing G. McCartor, {\it Zeit. Phys.} {\bf C52} (1991),
611.
\ignoreendofline
\refitem {82.}
\obeyendofline Y. Mo and R. J. Perry, ``Basis Function Calculations for the
Massive Schwinger Model in the Light-Front Tamm-Dancoff Approximation,''
to appear in J. Comp. Phys. (1993).
\ignoreendofline
\refitem {83.}
\obeyendofline G. P. Lepage, S. J. Brodsky, T. Huang and P. B. Mackenzie,
{\it in} ``Particles and Fields 2'' (A. Z. Capri and A. N. Kamal,
Eds.), Plenum Press, New York, 1983.
\ignoreendofline
\refitem {84.}
\obeyendofline \frenchspacing J. M. Namyslowski, {\it Prog. Part. Nuc. Phys.}
{\bf 74} (1984), 1.
\ignoreendofline
\refitem {85.}
\obeyendofline S. J. Brodsky and G. P. Lepage, {\it in} ``Perturbative
quantum
chromodynamics'' (A. H. Mueller, Ed.), World Scientific,
Singapore, 1989.
\ignoreendofline
\refitem {86.}
\obeyendofline S. Brodsky, H. C. Pauli, G. McCartor, and S. Pinsky, ``The
Challenge of Light-Cone Quantization of Gauge Field Theory,''
SLAC preprint no. SLAC-PUB-5811 and Ohio State preprint no.
OHSTPY-HEP-T-92-005 (1992).
\ignoreendofline
\refitem {87.}
\obeyendofline \frenchspacing W. Pauli and F. Villars, {\it Rev. Mod. Phys.}
{\bf 21} (1949), 434.
\ignoreendofline
\refitem {88.}
\obeyendofline \frenchspacing G. 't Hooft and M. Veltman, {\it Nucl. Phys.}
{\bf B44} (1972), 189.
\ignoreendofline
\refitem {89.}
\obeyendofline G. 't Hooft and M. Veltman, ``Diagrammar,'' CERN preprint
73-9 (1973).
\ignoreendofline
\refitem {90.}
\obeyendofline \frenchspacing C. Bloch and J. Horowitz, {\it Nucl. Phys.} {\bf
8} (1958), 91.
\ignoreendofline
\refitem {91.}
\obeyendofline \frenchspacing C. Bloch, {\it Nucl. Phys.} {\bf 6} (1958), 329.
\ignoreendofline
\refitem {92.}
\obeyendofline \frenchspacing B. H. Brandow, {\it Rev. Mod. Phys.} {\bf 39}
(1967), 771.
\ignoreendofline
\refitem {93.}
\obeyendofline \frenchspacing H. Leutwyler and J. Stern, {\it Ann. Phys. (New
York)} {\bf 112} (1978), 94.
\ignoreendofline
\refitem {94.}
\obeyendofline \frenchspacing R. Oehme, K. Sibold, and W. Zimmerman, {\it
Phys. Lett.} {\bf B147} (1984), 115.
\ignoreendofline
\refitem {95.}
\obeyendofline \frenchspacing R. Oehme and W. Zimmerman, {\it Commun. Math.
Phys.} {\bf 97} (1985), 569.
\ignoreendofline
\refitem {96.}
\obeyendofline \frenchspacing W. Zimmerman, {\it Commun. Math. Phys.} {\bf 95}
(1985), 211.
\ignoreendofline
\refitem {97.}
\obeyendofline \frenchspacing J. Kubo, K. Sibold, and W. Zimmerman, {\it Nuc.
Phys.} {\bf B259} (1985), 331.
\ignoreendofline
\refitem {98.}
\obeyendofline \frenchspacing R. Oehme, {\it Prog. Theor. Phys. Supp.} {\bf 86}
(1986), 215.
\ignoreendofline
\refitem {99.}
\obeyendofline \frenchspacing C. Lucchesi, O. Piguet, and K. Sibold, {\it Phys.
Lett.} {\bf B201} (1988), 241.
\ignoreendofline
\refitem {100.}
\obeyendofline \frenchspacing E. Kraus, {\it Nucl. Phys.} {\bf B349} (1991),
563.
\ignoreendofline
\refitem {101.}
\obeyendofline \frenchspacing A. Casher, {\it Phys. Rev.} {\bf D14} (1976),
452.
\ignoreendofline
\refitem {102.}
\obeyendofline \frenchspacing W.A. Bardeen and R.B. Pearson, {\it Phys. Rev.}
{\bf D14} (1976), 547.
\ignoreendofline
\refitem {103.}
\obeyendofline \frenchspacing W.A. Bardeen, R.B. Pearson and E. Rabinovici,
{\it Phys. Rev.} {\bf D21} (1980), 1037.
\ignoreendofline
\refitem {104.}
\obeyendofline \frenchspacing G.P. Lepage and S.J. Brodsky, {\it Phys. Rev.}
{\bf D22} (1980), 2157.
\ignoreendofline
\refitem {105.}
\obeyendofline See P. A. M. Dirac,
in `Perturbative Quantum Chromodynamics'' (D. W. Duke and J. F. Owens,
Eds.), Am. Inst. Phys., New York, 1981.
\ignoreendofline
\refitem {106.}
\obeyendofline K. G. Wilson, ``Light Front QCD,'' Ohio State internal
report, unpublished (1990).
\ignoreendofline
\refitem {107.}
\obeyendofline \frenchspacing E. Tomboulis, {\it Phys. Rev.} {\bf D8} (1973),
2736.
\ignoreendofline
\refitem {108.}
\obeyendofline \frenchspacing D. J. Gross and F. Wilczek, {\it Phys. Rev.
Lett.} {\bf 30} (1973), 1343.
\ignoreendofline
\refitem {109.}
\obeyendofline \frenchspacing H. D. Politzer, {\it Phys. Rev. Lett.} {\bf 30}
(1973), 1346.
\ignoreendofline
\refitem {110.}
\obeyendofline \frenchspacing J. Goldstone, {\it Proc. Roy. Soc. (London)} {\bf
A239} (1957), 267.
\ignoreendofline
\refitem {111.}
\obeyendofline \frenchspacing P. A. Griffin, {\it Nucl. Phys.} {\bf B372}
(1992), 270.
\ignoreendofline
\refitem {112.}
\obeyendofline \frenchspacing J. Schwinger, ``Quantum Electrodynamics'', Dover,
New York, 1958.
\ignoreendofline
\refitem {113.}
\obeyendofline \frenchspacing L. M. Brown, ``Renormalization'',
Springer-Verlag, New York, 1993.
\ignoreendofline
\refitem {114.}
\obeyendofline \frenchspacing K. G. Wilson, {\it Phys. Rev.} {\bf D10} (1974),
2445.
\ignoreendofline
\refitem {115.}
\obeyendofline \frenchspacing W. Buchm{\"u}ller, ``Quarkonia'', North Holland,
Amsterdam, 1992.
\ignoreendofline
\refitem {116.}
\obeyendofline \frenchspacing  C. B. Thorn, {\it Phys. Rev.} {\bf D20} (1979),
1934.
\ignoreendofline
\refitem {117.}
\obeyendofline \frenchspacing St. G{\l }azek, {\it Phys. Rev.} {\bf D38}
(1988), 3277.
\ignoreendofline
\refitem {118.}
\obeyendofline \frenchspacing J. J. J. Kokkedee, ``The Quark Model'', Benjamin,
New York, 1969.
\ignoreendofline

\vfill
\eject

\noindent {\bf Figure Captions}
\medskip

\item{1.} Wilson's triangle of renormalization.  The units are chosen so
that $\Lambda_\calN=1$.

\item{2.} Examples of Hamiltonian diagrams, (a) in which a single energy
denominator appears, and (b) in which two energy denominators appear.
The arrows indicate the energy differences found in these denominators.

\item{3.} Second-order corrections to: (a) $u_2$, (b) $u_4$, (c) $u_6$,
and (d) $u_8$.

\item{4.} Second-order correction to $u_2$ with spectators.

\item{5.} Third-order corrections to the one-boson to three-boson part
of $u_4$.

\item{6.} (a) A two-loop correction to $u_4$ paired with the appropriate
counterterm insertion in a one-loop correction to $u_4$. (b) The source of
the counterterm.

\item{7.} (a) A two-loop correction to $u_4$ paired with the appropriate
counterterm insertion in a one-loop correction to $u_4$. (b) The source of
the counterterm.

\item{8.} (a) A two-loop correction to $u_4$ paired with two appropriate
counterterm insertions in one-loop corrections to $u_4$. The sources of
these counterterms are shown in (b) and (c).

\item{9.} (a) A two-loop correction to $u_4$ paired with two appropriate
counterterm insertions in one-loop corrections to $u_4$. The sources of
these counterterms are shown in (b) and (c).

\item{10.} One-loop corrections to the two-boson to two-boson part of
$u_4$ with one marginal counterterm
vertex.

\item{11.} One-loop corrections to the one-boson to three-boson part of
$u_4$ with one marginal counterterm vertex.


\end